\documentclass[prd]{revtex4}
\usepackage{graphicx}
\usepackage{lscape}
\usepackage{amssymb}
\usepackage{mathptmx}

\usepackage{graphicx}
\usepackage{epsfig}
\usepackage{amssymb,epsfig}

\begin{document}
\def\ds{\displaystyle}
\def\beq{\begin{equation}}
\def\eeq{\end{equation}}
\def\bea{\begin{eqnarray}}
\def\eea{\end{eqnarray}}
\def\beeq{\begin{eqnarray}}
\def\eeeq{\end{eqnarray}}
\def\ve{\vert}
\def\vel{\left|}
\def\ver{\right|}
\def\nnb{\nonumber}
\def\ga{\left(}
\def\dr{\right)}
\def\aga{\left\{}
\def\adr{\right\}}
\def\lla{\left<}
\def\rra{\right>}
\def\rar{\rightarrow}
\def\nnb{\nonumber}
\def\la{\langle}
\def\ra{\rangle}
\def\ba{\begin{array}}
\def\ea{\end{array}}
\def\tr{\mbox{Tr}}
\def\ssp{{\Sigma^{*+}}}
\def\sso{{\Sigma^{*0}}}
\def\ssm{{\Sigma^{*-}}}
\def\xis0{{\Xi^{*0}}}
\def\xism{{\Xi^{*-}}}
\def\qs{\la \bar s s \ra}
\def\qu{\la \bar u u \ra}
\def\qd{\la \bar d d \ra}
\def\qq{\la \bar q q \ra}
\def\gGgG{\la g^2 G^2 \ra}
\def\q{\gamma_5 \not\!q}
\def\x{\gamma_5 \not\!x}
\def\g5{\gamma_5}
\def\sb{S_Q^{cf}}
\def\sd{S_d^{be}}
\def\su{S_u^{ad}}
\def\ss{S_s^{??}}
\def\ll{\Lambda}
\def\lb{\Lambda_b}
\def\sbp{{S}_Q^{'cf}}
\def\sdp{{S}_d^{'be}}
\def\sup{{S}_u^{'ad}}
\def\ssp{{S}_s^{'??}}
\def\sig{\sigma_{\mu \nu} \gamma_5 p^\mu q^\nu}
\def\fo{f_0(\frac{s_0}{M^2})}
\def\ffi{f_1(\frac{s_0}{M^2})}
\def\fii{f_2(\frac{s_0}{M^2})}
\def\O{{\cal O}}
\def\sl{{\Sigma^0 \Lambda}}
\def\ar{&+& \!\!\!}
\def\es{\!\!\! &=& \!\!\!}
\def\ek{&-& \!\!\!}
\def\cp{&\times& \!\!\!}
\def\se{\!\!\! &\simeq& \!\!\!}
\def\hml{\hat{m}_{\ell}}
\def\rr{\hat{r}_{\Lambda}}
\def\ss{\hat{s}}
\def\tep{$B \rightarrow K_0^{*}(1430) \ell^+ \ell^-$}
\def\tepm{$B \rightarrow K_0^{*}(1430) \mu^+ \mu^-$}
\def\tept{$B \rightarrow K_0^{*}(1430) \tau^+ \tau^-$}
\def\simlt{\stackrel{<}{{}_\sim}}
\def\simgt{\stackrel{>}{{}_\sim}}
\newcommand{\shat}{\hat{s}}
\newcommand{\barB}{\overline{B}}
\newcommand{\barK}{\overline{K}}
\newcommand{\kone}{{K_1}}
\newcommand{\barkone}{{\overline{K}_1}}
\newcommand{\leftu}{\gamma^\mu L}
\newcommand{\leftd}{\gamma_\mu L}
\newcommand{\rightu}{\gamma^\mu R}
\newcommand{\rightd}{\gamma_\mu R}
\newcommand{\pA}{p_{\kone}}
\newcommand{\konel}{K_1(1270)}
\newcommand{\koneh}{K_1(1400)}
\newcommand{\barkonel}{\barK_1(1270)}
\newcommand{\barkoneh}{\barK_1(1400)}
\newcommand{\konea}{K_{1A}}
\newcommand{\koneb}{K_{1B}}
\newcommand{\barkonea}{\barK_{1A}}
\newcommand{\barkoneb}{\barK_{1B}}
\newcommand{\mkone}{m_{\kone}}
\newcommand{\konep}{K_1^+}
\newcommand{\konem}{K_1^-}
\newcommand{\konelm}{K_1^-(1270)}
\newcommand{\konehm}{K_1^-(1400)}
\newcommand{\konelp}{K_1^+(1270)}
\newcommand{\konehp}{K_1^+(1400)}
\newcommand{\konelz}{\overline{K}{}^0_1(1270)}
\newcommand{\konehz}{\overline{K}{}^0_1(1400)}
\newcommand{\degree}{^\circ}
\newcommand{\ket}[1]{{#1} \rangle}
\newcommand{\bra}[1]{\langle {#1}}
\newcommand{\ebar}{{\bar{e}}}
\newcommand{\sbar}{\bar{s}}
\newcommand{\cbar}{\bar{c}}
\newcommand{\bbar}{\bar{b}}
\newcommand{\qbar}{\bar{q}}\renewcommand{\l}{\ell}
\newcommand{\lbar}{\bar{\ell}}
\newcommand{\psibar}{\bar{\psi}}
\newcommand{\lpm}{\l^+\l^-}
\newcommand{\epm}{e^+e^-}
\newcommand{\mupm}{\mu^+\mu^-}
\newcommand{\taupm}{\tau^+\tau^-}
\newcommand{\hats}{\hat{s}}
\newcommand{\hatp}{\hat{p}}
\newcommand{\hatq}{\hat{q}}
\newcommand{\hatm}{\hat{m}}
\newcommand{\hatu}{\hat{u}}
\newcommand{\alphaem}{\alpha_{\rm em}}
\newcommand{\AFB}{A_{\rm FB}}
\newcommand{\barAFB}{\overline{A}_{\rm FB}}
\newcommand{\A}{{\cal A}}
\newcommand{\B}{{\cal B}}
\newcommand{\C}{{\cal C}}
\newcommand{\D}{{\cal D}}
\newcommand{\Q}{{\cal Q}}
\newcommand{\N}{{\cal N}}
\newcommand{\E}{{\cal E}}
\newcommand{\F}{{\cal F}}
\newcommand{\G}{{\cal G}}
\renewcommand{\H}{{\cal H}}
\newcommand{\T}{{\cal T}}
\renewcommand{\Re}{\mathop{\mbox{Re}}}
\renewcommand{\Im}{\mathop{\mbox{Im}}}
\newcommand{\eff}{{\rm eff}}
\newcommand{\GeV}{{\,\mbox{GeV}}}
\newcommand{\MeV}{{\,\mbox{MeV}}}
\newcommand{\Bm}{B^-}
\newcommand{\Bz}{\overline{B}{}^0}
\providecommand{\dfrac}[2]{\frac{\displaystyle
{#1}}{\displaystyle{#2}}}
\newcommand{\Br}{{\cal B}}

\title{
         {\Large
                 {\bf
Lepton polarization and CP-violating effects  in  $\overline{B}\rightarrow \overline{K}_{0}^{*}(1430) \ell^+\ell^-$ Decay in Standard and
 Two Higgs Doublet Model
                  }
         }
      }

\author{F. Falahati$^{a}$\footnote{falahati@shirazu.ac.ir}, S. M. Zebarjad and S. Naeimipour}
\affiliation{$^{a}$Physics Department and Biruni Observatory,  College of Sciences,  Shiraz University, Shiraz 71454,
Iran}

\small{\begin{abstract}

In this paper we analyze the dilepton mass square $q^2$ dependency of single lepton polarization asymmetries and CP violation for $\overline{B}\rightarrow \overline{K}_0^{*}(1430) \ell^+\ell^-,  \ell=\mu,\tau)$ in  the  2HDM context. Also,  we study the averages of these asymmetries in the domain $4 m_{\ell}^2<q^2< (m_B-m_{{K}_0^{*}})^2$.  Our study manifests that the investigation of  the above-mentioned asymmetries  for  $\overline{B}\rightarrow \overline{K}_0^*(1430) \ell^+\ell^-$ processes
could provide useful information for probing new
Higgs bosons  in the future B-physics experiments.

\end{abstract}

\maketitle

Packs numbers: 12.60.-i, 13.30.-a, 14.20.Mr

\section{Introduction}
Now that  last missing ingredient of the Standard Model (SM) (SM Higgs particle)
 has been experimentally discovered at the LHC by the ATLAS \cite{ref1} and CMS \cite{ref2,ref3} collaborations, with a mass $m_H \simeq 125$ GeV, the possibility of discovery of an
enlarged scalar sector becomes very plausible. On the other hand, between the spectrum of extensions of the SM, there are predictions that anticipate more than one scalar Higgs doublet; for instance,
the case of the Minimal- Supersymmetric-Standard-Model (MSSM)). Based on this we can consider a prototype of  extensions of
the SM which are including a larger scalar sector, called generically the Two- Higgs-Doublet-Model (2HDM). There
are different types of such 2HDM models. In the model called type I, one Higgs doublet generates masses for the up and
down quarks, simultaneously. In the model type II, one Higgs doublet gives masses to the up-type quarks and the
other one to the down-type quarks. These two models include a discrete symmetry to prevent flavour changing neutral
currents (FCNC) at tree level. However, the addition of these discrete symmetries is not required and in this
case both doublets are contributing to provide the masses to up-type and down-type quarks. In the literature, such
a model is known as the 2HDM type III . It has been used to search for physics beyond the SM and specifically
for FCNC at tree level. In general, both doublets can acquire a vacuum expectation value (VEV), but one
of them can be absorbed redefining the Higgs boson fields properly. Nevertheless, other studies on 2HDM-III using
different basis have been done and there is a case where both doublets get VEVs that allows to study the models type
I and II in a specific limit\cite{ref4,r23}.

In the 2HDM models, the two complex Higgs doublets include eigth scalar states. Spontanoues Symmetry
breaking procedure generates  five Higgs fields: two neutral CP-even scalars $h^0$ and $H^0$, a neutral CP-odd scalar $A^0$,
and two charged scalars $H^{±}$. While the neutral Higgs bosons may be difficult to distinguish from the one of the SM,
the charged Higgs bosons would have a distinctive signal for physics beyond the SM. Therefore the direct or indirect
effect of a charged Higgs boson would play an important role in the discovery of an extended Higgs model. The limitations which come from the experimental results of $B-\bar{B}$ mixing, $\Gamma (b\to s\gamma)$, $\Gamma (b \to
c\tau\bar{\nu}_{\tau})$, $\rho_0,R_b$ and the electric dipole
moments (EDMS) of the electron and
neutron\cite{r21, r23, r28, r29} could constrain the range of variation of  masses of Higgs bosons and that of the other related parameters such as vertex parameters, $\lambda_{tt}$ and $\lambda_{bb}$.

FCNC and CP-violating are indeed the most sensitive probes of NP contributions to
penguin operators. Rare decays, induced by FCNC of $b\to s \ell^+
\ell^-(\ell=e,\mu,\tau)$ transitions are at the
forefront of our quest to understand flavor and the origins of CP violation asymmetry
(CPV), offering one of the best probes for NP beyond the SM, in particular to explore 2HDM.

Although the branching ratios of FCNC decays are small in the SM,  interesting
results are yielded in developing experiments. The inclusive $b\to X_s \ell^+
\ell^-$ decay is observed
in BaBaR \cite{r3} and Belle collaborations. These collaborations also measured exclusive modes
$B\to K \ell^+\ell^-$ \cite{r4,r5,r6} and $B\to K^* \ell^+\ell^-$ \cite{r7}. These experimental results show
high agreement with theoretical predictions \cite{r8,r9,r10}.

There exists another  group of rare decays induced by $b\to s$ transition, such as $B\to K_2^*(1430) \ell^+\ell^-$ and $B\to K_0^*(1430) \ell^+\ell^-$
in which B meson decays into a tensor or  scalar meson, respectively. These decays are deeply investigated in SM in \cite{r11,r12} and the related transition form factors  are formulated within the framework of light front quark model \cite{r12,r13,r14} and  QCD sum rules method \cite{r15,r16}, respectively.

In this paper, we will investigate the exclusive decay $\overline{B}\rightarrow \overline{K}_0^{*}(1430) \ell^+\ell^-(\ell=\mu,\tau)$, where $\overline{K}_0^{*}(1430)$ is a scalar meson, both in the SM and  2HDM.  We evaluate the  single lepton polarization asymmetries and CP violating effects with special
emphasis on the model III of 2HDM.

The  paper is organized as follows. In Section II, we describe the content of the general 2HDM and write down the Yukawa Lagrangian for model III. In Section III, the effective Hamiltonian and  matrix elements of  $\overline{B} \to \overline{K}_0^*(1430)  \ell^{+}\ell^{-}$ transition in SM and 2HDM are presented. Then the general expressions for  single lepton polarization asymmetries and CP violation have been extracted out. Section IV is devoted to discussion and our conclusions. In the final section a brief summery of our results is presented.

\section{The General Two-Higgs-Doublet Model}

In a general two-Higgs-doublet model, both the doublets can couple to the
up-type and down-type quarks.  Without missing any thing, we use a basis
such that the first doublet produces the masses of all the gauge-bosons and fermions\cite{r23}:
\begin{equation}
\langle \phi_1 \rangle = \left( \begin{array}{c} 0 \\
                               \frac{v}{\sqrt{2}} \end{array} \right )
\;, \qquad
\langle \phi_2 \rangle = 0
\end{equation}
where $v$ is due to the $W$ mass by $M_W={g\over2}v$.
Based on this, the first doublet $\phi_1$ is the same as the SM doublet, whereas
all the new Higgs fields originate  from the second doublet $\phi_2$.  They are
written as
\begin{equation}
\phi_1=\frac{1}{\sqrt 2} \left( \begin{array}{c}
                             \sqrt{2} G^+ \\
                             v + \chi_1^0 + i G^0 \end{array} \right )
\; , \qquad
\phi_2=\frac{1}{\sqrt 2} \left( \begin{array}{c}
                               \sqrt{2} H^+ \\
                               \chi_2^0 + i A^0 \end{array} \right ) \;,
\end{equation}
where $G^0$ and $G^\pm$ are the Goldstone bosons that would be absorbed
in the Higgs mechanism to provide the longitudinal components of the weak
gauge bosons.  The $H^\pm$ are the physical charged-Higgs bosons and $A^0$ is
the physical CP-odd neutral Higgs boson.  The $\chi_1^0$ and $\chi_2^0$ are
not physical mass eigenstates but are written as linear combinations of the CP-even
neutral Higgs bosons:
\begin{eqnarray}
\chi_1^0 &=& H^0 \cos\alpha  - h^0 \sin \alpha  \\
\chi_2^0 &=& H^0\sin\alpha  + h^0 \cos \alpha  \;,
\end{eqnarray}
where $\alpha$ is the mixing angle.  Using this basis, the couplings
of $\chi_2^0 ZZ$ and $\chi_2^0 W^+ W^-$ are disappeared.
We can present\cite{r25} the Yukawa Lagrangian for model III as
\begin{equation}
-{\cal L}_Y = \eta_{ij}^U \overline{Q_{iL}} \tilde{\phi_1} U_{jR} +
              \eta_{ij}^D \overline{Q_{iL}} \phi_1 D_{jR} +
       \xi_{ij}^U \overline{Q_{iL}} \tilde{\phi_2} U_{jR} +
       \xi_{ij}^D \overline{Q_{iL}} \phi_2 D_{jR}   \quad \; + {\rm h.c.} \;,
\end{equation}
where $i,j$ are generation indices, $\tilde{\phi}_{1,2} = i\sigma_2
\phi_{1,2} \;$, $\eta_{ij}^{U,D}$ and $\xi_{ij}^{U,D}$
are, in general, nondiagonal coupling matrices, and $Q_{iL}$ is the
left-handed fermion doublet and $U_{jR}$ and $D_{jR}$ are the right-handed
singlets.  Note that  these $Q_{iL}$, $U_{jR}$, and $D_{jR}$ are weak
eigenstates, which can be expanded by mass eigenstates.
As we have mentioned above, $\phi_1$ provides all the
fermion masses and, therefore, $\frac{v}{\sqrt{2}}\eta^{U,D}$ will become
the up- and down-type quark-mass matrices after a bi-unitary transformation.
Applying  the transformation the Yukawa Lagrangian becomes
\begin{eqnarray}
{\cal L}_Y &=& -\overline{U} M_U U - \overline{D} M_D D
   - \frac{g}{2M_W} (H^0\cos\alpha - h^0 \sin\alpha)
     \biggr(\overline{U} M_U U + \overline{D} M_D D \biggr ) \nonumber \\
&+& \frac{ig}{2 M_W} G^0 \left(\overline{U} M_U \gamma^5 U
                             - \overline{D} M_D \gamma^5 D \right)\nonumber \\
&+& \frac{g}{\sqrt{2}M_W} G^- \overline{D} V^\dagger_{\rm CKM} \biggr [
     M_U \hbox{$1\over2$}(1+\gamma^5) - M_D \hbox{$1\over2$}(1-\gamma^5) \biggr]
 U \nonumber \\
&-& \frac{g}{\sqrt{2}M_W} G^+ \overline{U} V_{\rm CKM} \biggr [
     M_D \hbox{$1\over2$}(1+\gamma^5) - M_U \hbox{$1\over2$}(1-\gamma^5) \biggr]
 D  \nonumber \\
&-&
\frac{H^0 \sin\alpha + h^0\cos\alpha }{\sqrt{2}} \Biggr[
  \overline{U} \biggr( {\hat\xi}^U \hbox{$1\over2$}(1+\gamma^5) + {\hat\xi}^{U\dagger}
 \hbox{$1\over2$}(1-\gamma^5)
         \biggr ) U  \nonumber \\
&& +\overline{D} \biggr( {\hat\xi}^D \hbox{$1\over2$}(1+\gamma^5) +
 {\hat\xi}^{D\dagger} \hbox{$1\over2$}(1-\gamma^5)
         \biggr ) D
  \Biggr ]  \nonumber \\
&+& \frac{i A^0}{\sqrt{2}} \Biggr [
    \overline{U} \biggr( {\hat\xi}^U
           \hbox{$1\over2$}(1+\gamma^5) -{\hat\xi}^{U\dagger}
           \hbox{$1\over2$}(1-\gamma^5)
        \biggr ) U
   -\overline{D} \biggr( {\hat\xi}^D
           \hbox{$1\over2$}(1+\gamma^5) - {\hat\xi}^{D\dagger}
           \hbox{$1\over2$}(1-\gamma^5)
        \biggr ) D
     \Biggr ] \nonumber \\
&-&  H^+ \overline{U} \biggr[ V_{\rm CKM} {\hat \xi}^D
             \hbox{$1\over2$}(1+\gamma^5) -
   {\hat\xi}^{U\dagger} V_{\rm CKM}
             \hbox{$1\over2$}(1-\gamma^5) \biggr] D
  \nonumber \\
&-&  H^- \overline{D} \biggr[ {\hat \xi}^{D\dagger} V^\dagger_{\rm CKM}
   \hbox{$1\over2$}(1-\gamma^5) -
      V^\dagger_{\rm CKM} {\hat\xi}^U \hbox{$1\over2$}(1+\gamma^5) \biggr] U
 \;\;,
\label{rule}
\end{eqnarray}
where
$U$ is a symbol for the mass eigenstates of $u,c,t$ quarks and
$D$ is a symbol for  the mass eigenstates of $d,s,b$ quarks.
The diagonal mass matrices  are defined by
$M_{U,D}=\hbox{diag}(m_{u,d}, m_{c,s}, m_{t,b})= {v\over\sqrt{2}}
({\cal L}_{U,D})^{\dagger} \eta^{U,D} ({\cal R}_{U,D})$,
$\hat\xi^{U,D} = ({\cal L}_{U,D})^{\dagger} \xi^{U,D} ({\cal R}_{U,D})$.
The Cabibbo-Kobayashi-Maskawa matrix \cite{ref5} is given by
$V_{\rm CKM}=({\cal L}_U)^\dagger({\cal L}_D)  $.

The  matrices $\hat\xi^{U,D} $ contain the FCNC couplings.  These matrices would be given as \cite{ref6}:
\begin{equation}
\label{anat}
\hat\xi^{U,D}_{ij} = \lambda_{ij} \frac{g\sqrt{m_i m_j}}{\sqrt{2}M_W}
\end{equation}
by which the quark-mass hierarchy is ensured  while the FCNC for the
first two generations are suppressed by the small quark
masses,  is allowed for  the third generation.
\section{Analytic Formulas }
\subsection{The Effective Hamiltonian for  $\overline{B} \to \overline{K}_0^*(1430)  \ell^{+}\ell^{-}$ transition in SM and 2HDM}
The exclusive  decay $\overline{B} \to \overline{K}_0^*(1430)  \ell^{+}\ell^{-}$ is described at quark level by $b\rar s \ell^{+}\ell^{-} $ transition.
Taking into account the additional Higgs boson exchange diagrams, the effective Hamiltonian is calculated in 2HDM as:
\small{\bea {\cal
H}_{eff}(b\to s \ell^{+}\ell^{-}) = -\frac{4 G_F}{\sqrt 2} V_{tb} V^*_{ts}
\left\{\sum_{i=1}^{10} C_i( \mu ) O_i( \mu ) + \sum_{i=1}^{10}
C_{Q_i}( \mu ) Q_i( \mu ) \right\} ~, \eea}where the first set of operators in the brackets are due to  the  SM effective Hamiltonian. Also note that the contributions of charged Higgs diagrams are taken into account in the aforementioned set of operators by modifying the corresponding Wilson coefficients.  The second part which includes new operators is extracted from contribution of the  massive neutral Higgs bosons to this decay.   All operators as well as the  related  Wilson coefficients   are given in ~\cite{r24, r25,r26}.
Now, using the
above  effective Hamiltonian, the one-loop matrix elements
of $b \rightarrow s \ell^+ \ell^-$   can be given as:
\bea {\cal M} &=& <s \ell^+  \ell^- |{\cal
H}_{\rm eff}|b> \nonumber\\
&=& -\frac{G_F \alpha}{2\sqrt 2 \pi} V_{tb} V^*_{ts}
\Bigg\{\tilde{C}_{9}^{\rm eff} \bar s \gamma_\mu (1- \gamma_5) b \,
\bar \ell \gamma^\mu \ell + \tilde{C}_{10} \bar s
\gamma_\mu (1- \gamma_5) b \, \bar \ell \gamma^\mu \gamma_5 \ell   \nnb \\
&-& 2C_7^{\rm eff}\frac{m_b}{q^2}\bar s i \sigma_{\mu \nu}q^\nu
(1+\gamma_5) b\,\bar \ell \gamma^\mu \ell-2C_7^{\rm
eff}\frac{m_s}{q^2}\bar s i \sigma_{\mu
\nu}q^\nu (1-\gamma_5) b \, \bar \ell \gamma^\mu \ell\nnb\\
 &+&C_{Q_1} \bar s (1 + \gamma_5) b \,\bar \ell \ell + C_{Q_2} \bar s
(1+\gamma_5) b \, \bar \ell \gamma_5 \ell \Bigg\}~. \label{me} \eea
The  Wilson
coefficients $C_7^{\rm eff},~\tilde{C}_{9}^{\rm eff}$, $\tilde{C}_{10}$ are obtained from their SM values by adding the contributions due to the charged Higgs bosons exchange diagrams. Note that this addition is performed at high $m_W$ scale, and then using the renormalization group equations, the coefficients are  calculated at lower $m_b$ scale. Coefficients $C_{Q_1}$ and $C_{Q_2}$ describe the neutral Higgs boson exchange diagrams' contributions. The operators $O_i (i=1,\cdots,10)$ do not mix with $Q_1$ and $Q_2$ and there is no mixing between $Q_1$ and $Q_2$. For this reason the evolutions of the coefficients $C_{Q_1}$ and
$C_{Q_2}$ are controlled by the anomalous dimensions of $Q_1$ and $Q_2$
respectively\cite{r26}:
$$C_{Q_i} (m_b) = \eta^{-\gamma_Q / \beta_0} C_{Q_i} (m_W)~,~~~i=1,~2 ,$$
where $\gamma_Q = -4 $ is the anomalous dimension of the operator $\bar s_L b_R$.

The coefficients $C_i (m_W)$ ($i=7,9$ and $10$) and $C_{Q_1}(m_W)$ and $C_{Q_2}(m_W)$ are given by:
\bea
C_7(m_W) &=&
x \, \frac{(7-5 x - 8 x^2)}{24 (x-1)^3} +
\frac{x^2 (3 x - 2)}{4 (x-1)^4} \, \ln x \nnb \\
&+& \vel \lambda_{tt} \ver^2 \Bigg( \frac{y(7-5 y - 8 y^2)}
{72 (y-1)^3} + \frac{y^2 ( 3 y - 2)}{12 (y-1)^4} \, \ln y \Bigg) \nnb \\
&+& \lambda_{tt} \lambda_{bb} \Bigg( \frac{y(3-5 y)}{12 (y-1)^2} +
\frac{y (3 y - 2)}{6 (y-1)^3} \, \ln y \Bigg)~, \\ \nnb \\ \nnb \\
C_9(m_W) &=& - \frac{1}{sin^2 \theta_W} \, B(m_W) +
\frac{1 - 4 sin^2 \theta_W}{sin^2 \theta_W} \, C(m_W) \nnb \\
&+& \frac{x^2(25-19 x)}{36 (x-1)^3} +
\frac{-3 x^4 + 30 x^3 - 54 x^2 + 32 x -8}{18 (x-1)^4} \, \ln x
+ \frac{4}{9} \nnb \\
&+& \vel \lambda_{tt} \ver^2 \Bigg[
\frac{1 - 4 sin^2 \theta_W}{sin^2 \theta_W} \, \frac{x y}{8} \Bigg(
\frac{1}{y-1} - \frac{1}{(y-1)^2} \, \ln y \Bigg)\nnb \\
&-& y \Bigg( \frac{47 y^2 - 79 y + 38}{108 (y-1)^3}
-\frac{3 y^3 - 6 y^3 + 4}{18 (y-1)^4} \, \ln y \Bigg) \Bigg]~,
\\ \nnb \\ \nnb \\
C_{10}(m_W) &=& \frac{1}{sin^2 \theta_W} \Big( B(m_W) -
C(m_W) \Big) \nnb \\
&+& \vel \lambda_{tt} \ver^2 \frac{1}{sin^2 \theta_W} \,\frac{x y}{8}
\Bigg( - \frac{1}{y-1} + \frac{1}{(y-1)^2} \, \ln y \Bigg)~,\\ \nnb \\ \nnb \\
C_{Q_1} ( m_W ) &=& \frac{m_b m_\ell}{m_{h^0}^2}
\frac{1}{\vel \lambda_{tt} \ver^2}
\frac{1}{sin^2 \theta_W} \frac{x}{4} \Bigg\{ \left(sin^2 \alpha + h\, cos^2
\alpha \right) f_1 (x,y) + \nnb \\
&+& \left[ \frac{m_{h^0}^2}{m_W^2} + \left(sin^2 \alpha + h\,
cos^2\alpha \right)(1-z) \right] f_2(x,y) +  \nnb \\
&+& \frac{sin^2 2 \alpha}{2 m_{H^\pm}^2} \left[m_{h^0}^2 -
\frac{(m_{h^0}^2 + m_{H^0}^2)^2}{2 m_{H^0}^2} \right] f_3 (y) \Bigg\}~,\\ \nnb \\ \nnb \\
C_{Q_2} (m_W) &=& -\frac{m_b m_\ell}{m_{A^0}^2} \frac{1}{\vel \lambda_{tt} \ver^2} \left\{
f_1(x,y) +
\left[1+ \frac{m_{H^\pm}^2 - m_{A^0}^2}{m_W^2} \right] f_2(x,y) \right\}~,
\eea
where
\bea
x &=& \frac{m_t^2}{m_W^2}~,~~~~y=\frac{m_t^2}{m_{H^\pm}^2}~,
~~~~z=\frac{x}{y}~,~~~~h=\frac{m_{h^0}^2}{m_{H^0}^2}~, \nnb \\
B(x) &=& - \frac{x}{4 (x-1)} + \frac{x}{4 (x-1)^2} \, \ln x ~, \nnb \\
C(x) &=& \frac{x}{4} \Bigg( \frac{x-6}{2 (x-1)} +
\frac{3 x +2 }{2 (x-1)^2} \ln x \Bigg)~,\nnb \\
f_1 (x,y) &=& \frac{x\, \ln x}{x-1} - \frac{y\, \ln y}{y-1}~,\nnb \\
f_2(x,y) &=& \frac{x\, \ln y}{(z-x)(x-1)} + \frac{\ln z}{(z-1)(x-1)}~,\nnb \\
f_3(y) &=& \frac{1 -y + y\, \ln y}{(y-1)^2}~,
\eea

It should be noted that  the coefficient $\tilde{C}_9^{\rm eff}(\mu)$ can be written by three parts:
\small{\bea
\tilde{C}_9^{\rm eff}(\mu)=\tilde{C}_9(\mu)+Y_{SD}(\hat{m}_c, \hat{s})+Y_{LD}(\hat{m}_c, \hat{s})~,
\eea}
where the parameters $\hat{m}_c$ and $\hat{s}$ are defined as $\hat{m}_c=m_c/m_b$, $\hat{s}=q^2/m_b^{2}$.
$Y_{SD}(\hat{m}_c, \hat{s})$ describes the short-distance
contributions from four-quark operators  which can be calculated in
the perturbative theory. The function $Y_{SD}(\hat{m}_c, \hat{s})$ is
given by:\small{\begin{eqnarray} Y_{SD}&=&
g(\hat{m}_c, \hat{s}) (3 C_1 + C_2 + 3 C_3 + C_4 + 3 C_5 + C_6)
\nonumber
\\&-& \frac{1}{2} g(1, \hat{s}) (4 C_3 + 4 C_4 + 3 C_5 + C_6) \nonumber
\\&-& \frac{1}{2} g(0, \hat{s}) (C_3 + 3 C_4) + \frac{2}{9} (3 C_3 +
C_4 + 3 C_5 + C_6), \end{eqnarray}} where the  explicit expressions
for the $g$ functions can be found  in ~\cite{r24}.
 The long-distance contributions $Y_{LD}(\hat{m}_c, \hat{s})$ originate from the real $c\bar{c}$ intermediate states, i.e., $J/\psi$, $\psi^\prime$
 $\cdots$. The $J/\psi$ family is
introduced by the Breit--Wigner distribution for the resonances
through the following function\cite{ref7,ref8}:
\small{\bea Y_{\rm LD} = {3\pi \over \alpha^2} C^{(0)} \sum_{V_i=\psi, \psi',
\cdots } k_i \, {\Gamma(V_i \rar \ell^+ \ell^-) m_{V_i}
\over m_{V_i}^2- q^2-im_{V_i}\Gamma_{V_i}} ~, \nnb \eea}
where
$\alpha$ is the fine structure constant and $C^{(0)}=(3 C_1 + C_2 + 3 C_3 + C_4 + 3 C_5 + C_6)$.
The phenomenological parameters $k_i$ for the
$\overline{B} \rar \overline{K}_0^*(1430)
\ell^+ \ell^-$ decay can be fixed from ${Br} (\overline{B}
\rar J/\psi \overline{K}_0^*(1430) \rar \overline{K}_0^*(1430) \ell^+ \ell^-) = {Br} (\overline{B}
\rar J/\psi \overline{K}_0^*(1430)) {Br} (J/\psi \rar \ell^+ \ell^-)$. However, since
the branching ratio of $\overline{B}
\rar J/\psi \overline{K}_0^*(1430)$ decay has not been measured yet, we assume that the values of
$k_i$ are in the order of one. Therefore, we use $k_1=k_2=1$  in the following numerical calculations\cite{ref8}.

\subsection{Form factors for $\overline{B} \to \overline{K}_0^*(1430) \ell^{+}\ell^{-}$ transition}
 The exclusive $\overline{B} \to \overline{K}_0^*(1430) \ell^{+}\ell^{-}$ decay is described in terms of
the matrix elements of the quark operators in eq. (\ref{me}) over meson
states, which can be parameterized in terms of the form factors.
The needed  matrix elements for the calculation of  $\overline{B} \to \overline{K}_0^*(1430) \ell^{+}\ell^{-}$ decay are:

 \small {\bea \label{e90} \lla
\overline{K}_0^\ast(1430)(p_{K_0^\ast}) \vel \bar{s}\gamma_\mu (1 \pm \gamma_5) b \ver
\overline {B}(p_B) \rra \es \pm\big[f_+ (q^2) {(p_B + p_{K_0^\ast})}_\mu + f_-(q^2) q_\mu\big], \\
\label{e91} \lla  \overline{K}_0^\ast(1430)(p_{K_0^\ast}) \vel \bar{s}i
\sigma_{\mu\nu} q^\nu (1 \pm \gamma_5) b  \ver \overline{B}(p_B) \rra \es {\pm f_T (q^2)
\over m_B + m_{K_0^\ast}} \big[ {(p_B + p_{K_0^\ast})}_\mu q^2 - (m_B^2 -
m_{K_0^\ast}^2 ) q_\mu \big],\\
\label{e92}
\lla \overline{K}_0^\ast(1430)(p_{K_0^\ast}) \vel \bar s (1 \pm \gamma_5) b \ver
\barB(p_B) \rra &=&\pm\lla \overline{K}_0^\ast(1430)(p_{K_0^\ast}) \vel \bar s   \gamma_5 b \ver
\barB(p_B) \rra =
\mp\frac{1}{m_{b}+m_{s}}[f_+ (q^2) (p_B + p_{K_0^\ast}).q +
f_-(q^2) q^{2}]\\\nonumber\es\mp \frac{f_0 (q^2)}{m_{b}+m_{s}}(m_B^2-m_{K_0^*}^2) , \\
\label{e93}
\lla \overline{K}_0^\ast(1430)(p_{K_0^\ast}) \vel \bar s    b \ver
\barB(p_B) \rra&=&0~.
 \eea}
where $q =p_B -p_{K_0^\ast}$ and the function  $f_0(q^2)$ has been extracted from
\small{\bea
\label{e8404} f_-(q^2)={(m_B^2-m_{K_0^\ast}^2)\over{q^2}} \big[f_0(q^2)-f_+(q^2)\big].
\eea}
For the form factors we have used the results of three-point QCD sum rules method
~\cite{r15} in which
the $q^2$ dependence of  all form factors is given by
\small{\bea
F(q^2)=\,{F(0)\over 1-a_F(q^2/m_{B}^2)+b_F(q^2/m_{B}^2)^2},
\eea}
where the values of parameters $F(0)$, $a_F$ and $b_F$ for the
$\overline{B} \to \overline{K}_0^*(1430) \ell^{+}\ell^{-}$ decay are exhibited in table \ref{tab:BK0star}.

\begin{table}[t]
\begin{center}
\caption{Form factors for $\overline{B} \to \overline{K}_0^*(1430)$ transition obtained
within three-point QCD sum rules are fitted to the
3-parameter form.} \label{tab:BK0star}
\begin{tabular}{clll}
\hline\hline
       $F$
    & $F(0)$
    & $a_F$
    & $b_F$
    \\
    \hline
$f_+^{\overline{B} \to \overline{K}_0^*}$ &$\phantom{-}0.31\pm 0.08$ & $0.81$ & $-0.21$ \\
$f_-^{\overline{B} \to \overline{K}_0^*}$ &$-0.31 \pm 0.07 $ & $0.80$ & $-0.36$\\
$f_T^{\overline{B} \to \overline{K}_0^*}$ &$-0.26 \pm 0.07 $ & $0.41$ & $-0.32$\\
\end{tabular}
\end{center}
\end{table}

\subsection{The  lepton polarization asymmetries and the CP-violating asymmetry of $\overline{B} \to \overline{K}_0^{*}(1430) \ell^{+}\ell^{-}$}
Making use of  eq.(\ref{me}) and the  definitions of form factors, the matrix element of the  $\overline{B} \to \overline{K}_0^{*}(1430) \ell^{+}\ell^{-}$
decay can be written as follows:
\begin{eqnarray}\label{ampl}
{\cal M} &=& \frac{G_F \alpha_{\rm em}}{4\sqrt{2}\pi} V_{ts}^*
V_{tb}^{}\, m_B \nnb \\ &&\Bigg\{
  [{\cal A}(p_B+p_{{K}_0^{*}}+{\cal B}q)_{\mu}] \lbar \gamma^\mu \l
 + [{\cal C}(p_B+p_{{K}_0^{*}}+{\cal D}q)_{\mu}]\lbar \gamma^\mu \gamma_5 \l
 +[{\cal Q}]\lbar \l+[{\cal N}]\lbar  \gamma_5 \l\Bigg\},
\end{eqnarray}
where the auxiliary functions $\A,
\cdots, {\cal Q}$ are listed  in the following:
\begin{eqnarray}
\A&=& -2\tilde{C}_{9}^{\rm eff}f_{+}(q^2)-4(m_b+m_s){C}_{7}^{\rm eff}{{f_T(q^2) }\over{m_B+m_{{K}_0^{*}}}},\label{fmA}\\
\B&=&  -2\tilde{C}_{9}^{\rm eff}f_{-}(q^2)+4(m_b+m_s){C}_{7}^{\rm eff}{{f_T(q^2) }\over{(m_B+m_{{K}_0^{*}}})q^2}(m_B^2-m_{{K}_0^{*}}^2),\\
\C&=&-2\tilde{C}_{10}f_{+}(q^2),\\
\D&=&-2\tilde{C}_{10}f_{-}(q^2),\\
{\cal Q}&=&-2C_{Q_{1}}f_{0}(q^2){(m_B^2-m_{{K}_0^{*}}^2)\over{m_b+m_s}},\\
{\cal N}&=&-2C_{Q_{2}}f_{0}(q^2){(m_B^2-m_{{K}_0^{*}}^2)\over{m_b+m_s}},\label{fmN}
\end{eqnarray}
with $q = p_B - p_{{K}_0^{*}} = p_{\ell^+} + p_{\ell^-} $.

The unpolarized differential decay rate  for the
$\barB\to \overline{K}_0^{*}(1430)\lpm$ decay in the rest frame of $B$ meson is given by:
\begin{eqnarray}\label{dg1}
\frac{d \Gamma(\barB\to{{K}_0^{*}}\lpm)}{d \hats} = -\frac{G_F^2
\alphaem^2 m_B}{2^{14}\pi^5}
 \left|V_{tb}V_{ts}^*\right|^2 v\sqrt{\lambda}\Delta,
\end{eqnarray}
with
\bea\label{dgds1}\nnb \Delta &=&16 m_{\ell} m_{B}^2 (1-\hat{r}_{{{K}_0^{*}}})\mbox{\rm Re}[\C{\cal N^*}]+4\hats m_B^2 v^2|{\cal Q}|^2+16 \hats m_{\ell}^2 m_B^2  |\D|^2+32 m_{\ell}^2 m_B^2 (1-\hat{r}_{{{K}_0^{*}}})\mbox{\rm Re}[\C{\cal D^*}]\nnb\\
&+&16\hats m_{\ell} m_B^2\mbox{\rm Re}[\D{\cal N^*}] +2 \hats m_B^2 |{\cal N}|^2+\frac{4}{3} m_B^4 \lambda (3-v^2)|\A|^2\nnb\\
&+&\frac{4}{3}m_B^4 |\C|^2\{2\lambda-(1-v^2)(2\lambda-3(1-\hat{r}_{{{K}_0^{*}}})^2)\},
\end{eqnarray}
where $v=\sqrt{1-4{m}_\ell^2/q^2}$, $\hats=q^2/m_B^2$, $\hat{r}_{{K}_0^{*}}=m_{{K}_0^{*}}^2/m_B^2$ and $\lambda = 1 +
\hat{r}_{{K}_0^{*}}^2 + \hats^2 - 2\hats - 2\hat{r}_{{K}_0^{*}} (1+\hats)$.

The CP-violating asymmetry of the
$\barB\to\overline{K}_0^{*}(1430\lpm$ decay is defined by:

\bea
\label{e6312}
{\cal A}_{CP}(\hats) = \frac{\ds \frac{d\Gamma}{d\hat{s} } -\frac{d\overline{\Gamma}}{d\hat{s} }
}
{\ds \frac{d\Gamma}{d\hat{s} } +\frac{d\overline{\Gamma}}{d\hat{s} }}~,
\eea
where $\frac{d\Gamma}{d\hat{s} }$ is the unpolarized differential decay rate given by eq.(\ref{dg1}) and $\frac{d\overline{\Gamma}}{d\hat{s} }$ is the unpolarized differential decay rate for the antiparticle channel. In  order to obtain the latter one we should change the parameters $V_{ts}^*V_{tb}$, $\lambda_{tt}$ and $\lambda_{tt}$ of the former one into $V_{ts}V_{tb}^*$, $\lambda_{tt}^*$ and $\lambda_{tt}^*$.

Having obtained the CP-violation asymmetry, let us now consider the
single lepton polarization  asymmetries associated with the polarized leptons.
 For this
purpose, we first define the following orthogonal unit vectors $s_i^{\pm\mu}$ in
the rest frame of $\ell^\pm$, where $i=L,N$ or $T$ are the abbreviations of
the longitudinal, normal and transversal spin projections, respectively:
\bea
\label{e6310}
s^{-\mu}_L \es \ga 0,\vec{e}_L^{\,-}\dr =
\ga 0,\frac{\vec{p}_{\ell^-}}{\vel\vec{p}_{\ell^-} \ver}\dr~, \nnb \\
s^{-\mu}_N \es \ga 0,\vec{e}_N^{\,-}\dr = \ga 0,\frac{\vec{p}_{{{K}_0^{*}}}\times
\vec{p}_{\ell^-}}{\vel \vec{p}_{{{K}_0^{*}}}\times \vec{p}_{\ell^-} \ver}\dr~, \nnb \\
s^{-\mu}_T \es \ga 0,\vec{e}_T^{\,-}\dr = \ga 0,\vec{e}_N^{\,-}
\times \vec{e}_L^{\,-} \dr~, \nnb \\
s^{+\mu}_L \es \ga 0,\vec{e}_L^{\,+}\dr =
\ga 0,\frac{\vec{p}_{\ell^+}}{\vel\vec{p}_{\ell^+} \ver}\dr~, \nnb \\
s^{+\mu}_N \es \ga 0,\vec{e}_N^{\,+}\dr = \ga 0,\frac{\vec{p}_{{{K}_0^{*}}}\times
\vec{p}_{\ell^+}}{\vel \vec{p}_{{{K}_0^{*}}}\times \vec{p}_{\ell^+} \ver}\dr~, \nnb \\
s^{+\mu}_T \es \ga 0,\vec{e}_T^{\,+}\dr = \ga 0,\vec{e}_N^{\,+}
\times \vec{e}_L^{\,+}\dr~,
\eea
where $\vec{p}_{\ell^{\mp}}$ and $\vec{p}_{K_0^{*}}$ are in the
CM frame of $\ell^- \,\ell^+$ system, respectively. Lorentz transformation is used to boost the  components of the lepton polarization
to the CM frame of the lepton pair as:
\bea
\label{e6311}
\ga s^{\mp\mu}_L \dr_{CM} \es \ga \frac{\vel\vec{p}_{\ell^{\mp}}\ver}{m_\ell}~,
\frac{E_{\ell} \vec{p}_{\ell^\mp}}{m_\ell \vel\vec{p}_{\ell^{\mp}} \ver}\dr~,\nnb\\
\ga s^{\mp\mu}_N \dr_{CM} \es \ga s^{\mp\mu}_N \dr_{RF}~,\nnb\\
\ga s^{\mp\mu}_T \dr_{CM} \es \ga s^{\mp\mu}_T \dr_{RF}~,
\eea
where  $RF$ refers to the rest frame of the corresponding lepton as well as $\vec{p}_{\ell^+} = - \vec{p}_{\ell^-}$ and $E_\ell$ and $m_\ell$ are the energy and mass
of leptons in the CM frame, respectively.

The single lepton polarization  asymmetries can be defined as:
\bea
\label{e6313}
{\cal P}^{-}_{i}(\hat{s}) \es
\frac{\bigg(\ds \frac{d\Gamma}{d\hat{s} }(s_i^-,s_i^+)+ \frac{d\Gamma}{d\hat{s} }(s_i^-,-s_i^+)\bigg) -\bigg(\ds \frac{d\Gamma}{d\hat{s} }(-s_i^-,s_i^+)+ \frac{d\Gamma}{d\hat{s} }(-s_i^-,-s_i^+)\bigg) }
{\bigg(\ds \frac{d\Gamma}{d\hat{s} }(s_i^-,s_i^+)+ \frac{d\Gamma}{d\hat{s} }(s_i^-,-s_i^+)\bigg) +\bigg(\ds \frac{d\Gamma}{d\hat{s} }(-s_i^-,s_i^+)+ \frac{d\Gamma}{d\hat{s} }(-s_i^-,-s_i^+)\bigg) }~,\\
{\cal P}^{+}_{i}(\hat{s}) \es
\frac{\bigg(\ds \frac{d\Gamma}{d\hat{s} }(s_i^-,s_i^+)+ \frac{d\Gamma}{d\hat{s} }(-s_i^-,s_i^+)\bigg) -\bigg(\ds \frac{d\Gamma}{d\hat{s} }(s_i^-,-s_i^+)+ \frac{d\Gamma}{d\hat{s} }(-s_i^-,-s_i^+)\bigg) }
{\bigg(\ds \frac{d\Gamma}{d\hat{s} }(s_i^-,s_i^+)+ \frac{d\Gamma}{d\hat{s} }(s_i^-,-s_i^+)\bigg) +\bigg(\ds \frac{d\Gamma}{d\hat{s} }(-s_i^-,s_i^+)+ \frac{d\Gamma}{d\hat{s} }(-s_i^-,-s_i^+)\bigg) }~,\\
\eea
where $\frac{d\Gamma(\hat{s})}{d\hat{s}}$'s are calculated in the CM frame. Using these definitions for the single lepton polarization asymmetries, the following explicit forms for ${\cal P}_{i}$'s  are obtained:
\bea
\label{e6314}
{\cal P}_{L}^{\mp}&=&\frac{4vm_B^2}{\Delta}\bigg\{\pm\frac{4}{3}\lambda m_B^2 \mbox{\rm Re}[\A\C^*]-4m_{\ell}(1-\hat{r}_{{{K}_0^{*}}})\mbox{\rm Re}[\C\Q^*]-4m_{\ell}\hat{s}\mbox{\rm Re}[\D\Q^*]-2\hat{s}\mbox{\rm Re}[\N\Q^*]\bigg\}~,\\
\label{e6315}
{\cal P}_{N}^{\mp}&=&\frac{2\pi v \sqrt{\lambda\hat{s}} m_B^3}{\Delta}\bigg\{+2m_{\ell} \mbox{\rm Im}[\D\C^*]+ \mbox{\rm Im}[\N\C^*] \mp\mbox{\rm Im}[\A\Q^*]\bigg\}~,\\
\label{e6316}
{\cal P}_{T}^{\mp}&=&\frac{\pi \sqrt{\lambda\hat{s}} m_B^3}{\Delta}\bigg\{\pm 2 \mbox{\rm Re}[\A\N^*]\pm \frac{4 m_{\ell}}{\hat{s}}(1-\hat{r}_{{{K}_0^{*}}})\mbox{\rm Re}[\A\C^*] \pm4 m_{\ell}\mbox{\rm Re}[\A\D^*]+2v^2\mbox{\rm Re}[\C\Q^*]\bigg\}~.
\eea

\section{Numerical Analysis}
In this section, we  would like to study  the asymmetries ${\cal A}_{CP}$ and ${\cal P}_{i}^{\pm}$'s and their averages for the exclusive decay $\overline{B}\rightarrow \overline{K}_{0}^*(1430) \ell^+\ell^-$ in SM and  model III of 2HDM. The constraints on 2HDM parameters come from the experimental limits of the  electric dipole
moments of  neutron(NEDM), $B^0 - \bar B^0$ mixing, $\rho
\,_0$, $R_b$ and $Br(b \rar s \gamma)$\cite{r21, r23, r28, r29}. A simple ansatz for $\lambda_{tt}\lambda_{bb}$ would be:
 \beq
\lambda_{tt}\lambda_{bb} =
|\lambda_{tt}\lambda_{bb}|\,e^{i\theta}. \eeq
Considering the restrictions of above references  on  the parameters of  model III of 2HDM and taking $\theta = \pi/2$, we use the following three classes of parameters  throughout the
numerical analysis\cite{r23}:
\bea
\rm{ \, Case A}: \,\,\,|\lambda_{tt}|&=& 0.03; \,\,\,\,\,|\lambda_{bb}|=
100,\nnb\\ \rm{\, Case B}: \,\,\,|\lambda_{tt}|&=& 0.15; \,\,\,\,\,|\lambda_{bb}|=
50,\nnb\\ \rm{\, Case C}: \,\,\,|\lambda_{tt}|&=& 0.3; \,\,\,\,\,\,\,\,|\lambda_{bb}|=
30.
\eea

In addition, in  this study we have applied four sets of masses of Higgs bosons which are displayed in table\ref{tabmassset}\cite{r23}.
\begin{table}[ht]
\begin{center}
\caption{List of the values for the masses of the Higgs particles.} \label{tabmassset}
\begin{tabular}{cllll}
\hline\hline

    & $\rm{ m_{H^{\pm}}}$
    & $\rm{ m_{A^0}}$
    & $\rm{ m_{h^0}}$
    & $\rm{ m_{H^0}}$
    \\
    \hline
$\rm{mass}~\rm{ set-1}$ &$200 \rm{Gev} $ & $125 \rm{Gev}$ & $125 \rm{Gev}$ & $160 \rm{Gev}$ \\
$\rm{mass}~\rm{ set-2}$ &$160 \rm{Gev} $ & $125 \rm{Gev}$ & $125 \rm{Gev}$ & $160 \rm{Gev}$ \\
$\rm{mass}~\rm{ set-3}$ &$200 \rm{Gev} $ & $125 \rm{Gev}$ & $125 \rm{Gev}$ & $125 \rm{Gev}$ \\
$\rm{mass}~\rm{ set-4}$ &$160 \rm{Gev} $ & $125 \rm{Gev}$ & $125 \rm{Gev}$ & $125 \rm{Gev}$ \\
 \hline\hline
\end{tabular}
\end{center}
\end{table}

The  corresponding averages are defined by the following equation \cite{ref9}: \bea \la {\cal A} \ra = \frac{\ds
\int_{4 \hat{m}_\ell^2}^{(1-\sqrt{\hat{r}_{M}})^2} {\cal A}
\frac{d{\cal B}}{d \hat{s}} d \hat{s}} {\ds \int_{4
\hat{m}_\ell^2}^{(1-\sqrt{\hat{r}_{M}})^2} \frac{d{\cal B}}{d
\hat{s}} d \hat{s}}~, \eea
where the subscript M refers to $\overline{K}_{0}^*(1430)$  meson and the subscript $\cal{A}$ refers to the asymmetries ${\cal A}_{CP}$ and ${\cal P}_{i}^{\pm}$'s.
The full kinematical interval of the dilepton invariant mass $q^2$ is $4
m_\ell^2 \le q^2 \le (m_B - m_M)^2$ for which the long
distance contributions (the charmonium resonances) can give substantial
effects by considering  the two
low lying resonances $J/\psi$ and $\psi^\prime$, in the interval of $8~GeV^2\le q^2 \le
14~GeV^2$. In order to decrease the hadronic uncertainties we
use the kinematical region of $q^2$ for muon as \cite{ref8}:
\bea
\begin{array}{cl}
\mbox{\rm I} & 4 m_\ell^2 \le q^2 \le (m_{J\psi} - 0.02~GeV)^2~,\\ \\
\mbox{\rm II} & (m_{J\psi} + 0.02~GeV)^2 \le q^2 \le
(m_{\psi^\prime} - 0.02~GeV)^2~, \\ \\
\mbox{\rm III} & (m_{\psi^\prime} + 0.02~GeV)^2 \le q^2 \le
(m_B-m_M)^2~, \nnb
\end{array} \nnb
\eea and for tau as: \bea
\begin{array}{cl}
\mbox{\rm I} & 4 m_\ell^2 \le q^2 \le (m_{\psi'} - 0.02~GeV)^2~,\\ \\
\mbox{\rm II} & (m_{\psi'} + 0.02~GeV)^2 \le q^2 \le
(m_B-m_M)^2. \nnb
\end{array} \nnb
\eea

We continue our analysis regarding the ${\cal A}$'s and their averages by plotting a  set of figures (\ref{ACPmKstar}-\ref{PTmptKstar}) and presentation of a  class of tables (\ref{masssetBK0smu12}-\ref{masssetBK0stau34}).
In these tables the theoretical and experimental
uncertainties corresponding to the SM averages  have been evaluated. In such a manner the theoretical uncertainties are extracted from the hadronic uncertainties
related to the form factors and the experimental uncertainties
originate from the mass of quarks and hadrons and Wolfenstein
parameters.

\begin{itemize}
\item \textbf { Analysis of ${\cal A}_{CP}$ asymmetry for $\overline{B}\to \overline{K}_0^{*} \mu^+ \mu^-$ decay}:
The relevant plots in figure(\ref{ACPmKstar}) show that while the SM prediction of this asymmetry is zero, it is quite sensitive  to the variation of the parameters  $\lambda_{tt}$ and $\lambda_{bb}$. For example, by enhancing the magnitude of $|\lambda_{tt}\lambda_{bb}|$ the deviation from the SM value is increased.
Also,  this asymmetry is quite sensitive  to the variation of mass of  $H^{\pm}$, this happens due to the reduction  of mass of $H^{\pm}$,  such that the deviations from the SM value in mass sets 2 and 4 are more than those in mass sets 1 and 3.  By combining the above analyses it is understood that the most deviations from the SM prediction occur in the case C of mass sets 2 and 4.  Next to $q^2=m_{{\psi}'}^2$ in the afore-mentioned case and mass sets,  a deviation around +0.05  is possible as compared to  the zero expectation of SM. In addition, it is found out through the corresponding tables (\ref{masssetBK0smu12} and \ref{masssetBK0smu34})  that the values of averages  show  ignorable sensitivities  to the presence of new Higgs bosons.

\item \textbf { Analysis of ${\cal P}_{L}^{\mp}$ asymmetries for $\overline{B}\to \overline{K}_0^{*} \mu^+ \mu^-$ decay}:
 As it is obvious from figure (\ref{PLmmKstar})  the predictions of all of mass sets throughout the domain $4 m_{\mu}^2\leq q^2< (m_B-m_{{K}_0^{*}})^2$  apart from  $q^2=(m_B-m_{{K}_0^{*}})^2$ are  the same and highly coincide with the SM prediction.  At $q^2=(m_B-m_{{K}_0^{*}})^2$ the deviation from the SM value  in the case A of mass set 3 is more than the others which is +1. At such point the SM prediction is zero. Moreover, it is seen from  the tables  \ref{masssetBK0smu12} and \ref{masssetBK0smu34} that the most deviations of $\lla {\cal P}_{L}^{-} \rra$ from the calculated  SM value happen in the case C  of mass sets 2 and 4 which are very small compared to SM prediction ( $- 3.2\% $ SM). Also it is clear from equation (\ref{e6314})  while by ignoring the signs of ${\cal P}_{L}^{-}$ and ${\cal P}_{L}^{+}$  in SM the magnitudes of them are the same (${\cal P}_{L}^{+}=-{\cal P}_{L}^{-}$ in SM), those asymmetries do not have any symmetrical relationship with each other in 2HDM.  As it is obvious from figure (\ref{PLpmKstar}) as well as tables  \ref{masssetBK0smu12} and \ref{masssetBK0smu34} that the predictions of all of mass sets and cases throughout the interval $4 m_{\mu}^2\leq q^2\leq (m_B-m_{{K}_0^{*}})^2$  coincide with that of  SM  very much. The most deviations of $\lla {\cal P}_{L}^{+} \rra$ from  the calculated SM value  happen in the case C  of mass sets 2 and 4 which are $- 3.2\% $ SM. Ignoring $q^2=(m_B-m_{{K}_0^{*}})^2$ and using the mentioned parameter space for 2HDM, it is found out that ${\cal P}_{L}^{+}=-{\cal P}_{L}^{-}$ in both SM and 2HDM.

\item \textbf { Analysis of ${\cal P}_{N}^{\mp}$ asymmetries for $\overline{B}\to \overline{K}_0^{*} \mu^+ \mu^-$ decay}:   The relevant plots in figure (\ref{PNmmKstar}) show that this asymmetry is quite sensitive  to the variation of the parameters  $\lambda_{tt}$ and $\lambda_{bb}$. For example, by decreasing the magnitude of $|\lambda_{tt}\lambda_{bb}|$ the deviation from the SM value is increased.
Also,  this asymmetry is quite sensitive  to the variation of masses of $H^0$ and $H^{\pm}$, this happens due to the reduction  of mass of $H^0$ and the increment of mass of $H^{\pm}$, such that the deviations from the SM value in mass sets 3 and 4 are more than those in mass sets 1 and 2.  By combining the above analyses it is understood that the most deviation from the SM prediction occurs in the case A of mass set 3.  Next to $q^2=(m_B-m_{{K}_0^{*}})^2$ in the afore-mentioned case and mass set,  a deviation around -0.09  is possible as compared to  SM expectation of zero asymmetry. In addition, it is found out through the corresponding tables  that the values of averages  show  ignorable dependencies  to the existence of new Higgs bosons.
Also  it is clear from equation (\ref{e6315})  whereas   in SM  ${\cal P}_{N}^{+}={\cal P}_{N}^{-}=0$, in 2HDM ${\cal P}_{N}^{+}=-{\cal P}_{N}^{-}$.

\item \textbf { Analysis of ${\cal P}_{T}^{\mp}$ asymmetries for $\overline{B}\to \overline{K}_0^{*} \mu^+ \mu^-$ decay}: It is found out from figures (\ref{PTmmKstar}), (\ref{PTpmKstar}) and (\ref{PNmmKstar}) that the asymmetries ${\cal P}_{T}^{\mp}$ and ${\cal P}_{N}^{\mp}$ show similar sensitivities to the variations of mass sets and cases. For example, in these asymmetries by decreasing the magnitude of $|\lambda_{tt}\lambda_{bb}|$ or the mass of $H^0$ and  increasing the of mass of $H^{\pm}$ the deviations from the SM predictions increase. According to this the largest deviations from the SM predictions arise in the case A of mass set 3. Next to $q^2=(m_B-m_{{K}_0^{*}})^2$ in the mentioned case and mass set,   deviations around +50\% SM  and -100\% SM are possible for ${\cal P}_{T}^{-}$ and ${\cal P}_{T}^{+}$, respectively. In addition, it is found out through the corresponding tables  that  the most deviations of $\lla {\cal P}_{T}^{-} \rra$ and $\lla {\cal P}_{T}^{+} \rra$ from the calculated SM values which happen in the case A  of mass set 3   are $+24\% $ SM and $-22\% $ SM, respectively.
Moreover it is clear from equation (\ref{e6316})  while   in SM  ${\cal P}_{T}^{-}=-{\cal P}_{T}^{+}$, there not exist any symmetrical relationship among them in 2HDM. Nevertheless, it is evident from the relevant figures  and  tables that in cases B and C to a large extent ${\cal P}_{T}^{+}=-{\cal P}_{T}^{-}$.

\item \textbf { Analysis of ${\cal A}_{CP}$ asymmetry for $\overline{B}\to \overline{K}_0^{*} \tau^+ \tau^-$ decay}:
The relevant plots in figure (\ref{ACPtKstar}) show that while the SM prediction of this asymmetry is zero, it is quite sensitive  to the variation of the parameters  $\lambda_{tt}$ and $\lambda_{bb}$. For example, by enhancing the magnitude of $|\lambda_{tt}\lambda_{bb}|$ the deviation from the SM value is increased.
Also,  this asymmetry is quite sensitive  to the variation of mass of  $H^{\pm}$, this happens due to the reduction  of mass of $H^{\pm}$  such that the deviations from the SM value in mass sets 2 and 4 are more than those in mass sets 1 and 3.  By combining the above analyses it is understood that the most deviations from the SM prediction occur in the case C of mass sets 2 and 4.  Next to $q^2=m_{{\psi}'}^2$ in the afore-mentioned case and mass sets,   deviations around +0.016  are possible as compared to  the zero expectation of SM. In addition, it is found out through the corresponding tables (\ref{masssetBK0stau12} and \ref{masssetBK0stau34} )  that the values of averages  show  ignorable sensitivities  to the presence of new Higgs bosons.

\item \textbf { Analysis of ${\cal P}_{L}^{\mp}$ asymmetries for $\overline{B}\to \overline{K}_0^{*} \tau^+ \tau^-$ decay}:
The relevant plots in figure (\ref{PLmtKstar}) show that this asymmetry is quite sensitive  to the variation of the parameters  $\lambda_{tt}$ and $\lambda_{bb}$. For example, by decreasing the magnitude of $|\lambda_{tt}\lambda_{bb}|$ the deviation from the SM value is increased.
Also,  this asymmetry is quite sensitive  to the variation of masses of $H^0$ and $H^{\pm}$, this happens due to the decrease  of mass of $H^0$ and the increase of mass of $H^{\pm}$ such that the deviations from the SM value in mass sets 3 and 4 are more than those in mass sets 1 and 2.
By gathering the above analyses it is understood that the most deviation from the SM prediction occurs in the case A of mass set 3.  Whereas the SM prediction is zero at $q^2=(m_B-m_{{K}_0^{*}})^2$ a deviation around +0.7  is possible at that point. Besides, it is found out through the corresponding tables  that  a deviation around -4.9 times of that of SM arises in the case A of mass set 3 at most. Also it is clear from equation (\ref{e6314})  while by ignoring the signs of ${\cal P}_{L}^{-}$ and ${\cal P}_{L}^{+}$  in SM the magnitudes of them are the same (${\cal P}_{L}^{+}=-{\cal P}_{L}^{-}$ in SM), those asymmetries do not have any symmetrical relationship with each other in 2HDM.  Nevertheless, it is evident from the corresponding figures (\ref{PLmtKstar}) and (\ref{PLptKstar})  and  tables that in cases B and C to a large extent ${\cal P}_{L}^{+}=-{\cal P}_{L}^{-}$.  The maximum deviations of ${\cal P}_{L}^{+}$ relative to the SM predictions which are observed in the respective diagrams and tables  take place in the case A of mass set 3 which are around +0.7  as compared to  SM expectation of zero asymmetry
at $q^2=(m_B-m_{{K}_0^{*}})^2$ and $ +6.6$  times  of the calculated SM prediction for the related averages.

\item \textbf { Analysis of ${\cal P}_{N}^{\mp}$ asymmetries for $\overline{B}\to \overline{K}_0^{*} \tau^+ \tau^-$ decay}:
 It is clear from figures (\ref{PNmtKstar}) and (\ref{PLmtKstar})  that the asymmetries ${\cal P}_{N}^{-}$ and ${\cal P}_{L}^{-}$ show the same sensitivities to the variations of mass sets and cases. For instance, in these asymmetries by reducing the magnitude of $|\lambda_{tt}\lambda_{bb}|$ or the mass of $H^0$ and  enhancing the of mass of $H^{\pm}$ the deviations from the SM predictions increase. According to this the largest deviation of ${\cal P}_{N}^{-}$ from the SM prediction arises in the case A of mass set 3. Next to $q^2=m_{\psi^\prime}^2$ in the mentioned case and mass set,   a deviation around -0.04  compared to the zero prediction of SM is possible for ${\cal P}_{N}^{-}$. In addition, it is obvious through the respective tables  that  the most deviation of $\lla {\cal P}_{N}^{-} \rra$  from the calculated SM value   is $-0.024 $   which happens in the case A  of mass set 3.
Moreover it is clear from equation (\ref{e6315})  while   in SM  ${\cal P}_{N}^{+}={\cal P}_{N}^{-}=0$, in 2HDM ${\cal P}_{N}^{+}=-{\cal P}_{N}^{-}$.

\item \textbf { Analysis of ${\cal P}_{T}^{\mp}$ asymmetries for $\overline{B}\to \overline{K}_0^{*} \tau^+ \tau^-$ decay}:
Since our analyses for the afore-mentioned  all mass sets show that   ${{\cal P}_{T}^{-}}_{2HDM}={{\cal P}_{T}^{-}}_{SM}  $ in all cases  and ${{\cal P}_{T}^{+}}_{2HDM}={{\cal P}_{T}^{+}}_{SM}$ in cases B and C  we have only presented the plots of mass set 3 for ${\cal P}_{T}^{\mp}$ in figure (\ref{PTmptKstar}). In this mass set the most deviation from the SM value for ${\cal P}_{T}^{+}$ arises somehow in the case A of the range $ m_{{\psi}'}^2<q^2< (m_B-m_{{K}_0^{*}})^2$ a discrepancy about $-25\%$SM is seen. Also it is clear from the corresponding tables that the largest deviation from the calculated SM anticipation for $\lla {\cal P}_{T}^{+} \rra$ is $-15\%$SM which occurs in the mentioned case and mass set. Moreover it is clear from equation (\ref{e6316})that ${\cal P}_{T}^{+}=-{\cal P}_{T}^{-}$ in SM.

\end{itemize}
Finally, let us see briefly  whether
 the lepton polarization asymmetries are visitable  or
 not. To measure  an asymmetry $\lla {\cal A}\rra$ of the decay with
 branching ratio $\cal{B}$ at $n\sigma$ level in experiment, the required number
 of events (i.e., the number of $B\bar{B}$) is given by the
 relation \bea N = \frac{n^2}{{\cal
B} s_1 s_2 \la {\cal A} \ra^2}~,\nnb \eea  where $s_1$ and $s_2$ are
the efficiencies of the leptons. The values of the efficiencies of
the $\tau$--leptons differ from $50\%$ to $90\%$ for their different
decay modes\cite{r30} and the error in $\tau$--lepton polarization
is nearly  $(10 - 15)\%$ \cite{r31}. So, the error in
measurements of the $\tau$--lepton asymmetries is estimated to be
about $(20 - 30)\%$, and the error in obtaining the number of events
is about $50\%$.

According to the above expression for  $N$, in order to measure the
single lepton polarization asymmetries in the $\mu$ and $\tau$ channels at
$3\sigma$ level, the lowest limit of required number of events are given
by(the efficiency of $\tau$--lepton is considered $0.5$):
\begin{itemize}
\item for $\overline{B} \rar \overline{K}_0^{*}(1430) \mu^+ \mu^-$ decay \bea N \sim \left\{
\begin{array}{llll}
10^{24 }  & (\mbox{\rm for} \lla {\cal A}_{CP} \rra)~,\\
10^{7 }  & (\mbox{\rm for} \lla {\cal P}_{L}^{-} \rra, \lla {\cal P}_{L}^{+} \rra)~,\\
10^{8}  & (\mbox{\rm for} \lla {\cal P}_{T}^{-} \rra, \lla {\cal P}_{T}^{+} \rra)~,\\
10^{12}  & (\mbox{\rm for} \lla {\cal P}_{N}^{-} \rra, \lla {\cal P}_{N}^{+} \rra)~,\\
\end{array} \right. \nnb \eea

\item for $\overline{B} \rar \overline{K}_0^{*}(1430) \tau^+ \tau^-$ decay \bea N \sim \left\{
\begin{array}{llll}
10^{27 }  & (\mbox{\rm for} \lla {\cal A}_{CP} \rra)~,\\
10^{10 }  & (\mbox{\rm for} \lla {\cal P}_{L}^{-} \rra, \lla {\cal P}_{L}^{+} \rra)~,\\
10^{10}  & (\mbox{\rm for} \lla {\cal P}_{T}^{-} \rra, \lla {\cal P}_{T}^{+} \rra)~,\\
10^{13}  & (\mbox{\rm for} \lla {\cal P}_{N}^{-} \rra, \lla {\cal P}_{N}^{+} \rra)~.\\
\end{array} \right. \nnb \eea

\end{itemize}

\section{Summary}
In short, in this paper by considering the theoretical
and experimental uncertainties in the SM,  we have presented a full
analysis related to the CP violating effects and single lepton polarization asymmetries for
$\overline{B}\rightarrow \overline{K}_{0}^*(1430) \ell^+\ell^-$ decay
 in  model III of 2HDM. At the same time we have compared the results of  both $\mu$ and $\tau$ channels to  each other. Also, the minimum required
number of events for measuring each asymmetry  has been obtained and compared with
those in LHC experiments,
 containing ATLAS, CMS and LHCb,  ($\sim 10^{12}$ per year)  or expected to be produced at  the Super-LHC experiments ( supposed to be $\sim 10^{13}$ per year). In conclusion, the following results
have been obtained:
\par
i) For the $\mu$ channel of single lepton polarization asymmetries (${\cal P}_{i}^{\mp}(q^2) i=L, N, T$ ) only the results obtained from case A differ from the SM expectations.  This fact indicates that these asymmetries are quite sensitive to the reduction of $|\lambda_{tt}\lambda_{bb}|$. Also, the decrease of the mass of $H^0$ and  simultaneously the increase  of the mass of $H^{\pm}$ can enhance the deviations from the SM predictions. Based on the above explanations in all single lepton polarization asymmetries the most deviations from the SM values happen in the case A of mass set 3.   On the other hand, for the $\mu$ channel of  CP violating asymmetry (${\cal A}_{CP}(q^2)$) the results obtained from all cases are different from that of the SM somehow the biggest deviation from the SM anticipation occurs in case C.  This fact indicates that this asymmetry is quite sensitive to the enhancement of $|\lambda_{tt}\lambda_{bb}|$.  Also, while this asymmetry is quite insensitive to the variation of mass of $H^0$, the deviations from the SM prediction increase by  decreasing the mass of $H^{\pm}$.   Based on the above explanations in CP violating asymmetry the most deviations from the SM value happen in the case C of mass sets 2 and 4. Paying attention to the minimum required number of events for detecting each asymmetry it is inferred while all single lepton polarization asymmetries are detectable at LHC,  CP violating asymmetry is not measurable in neither  LHC nor SLHC.

ii) For the $\tau$ channel of ${\cal P}_{T}^{-}(q^2)$  any sensitivity to the 2HDM parameters is not seen and for the $\tau$ channel of other single lepton polarization asymmetries (${\cal P}_{T}^{+}(q^2), {\cal P}_{i}^{\mp}(q^2) i=L, N$) only the results obtained from case A differ from the SM expectations. This fact indicates that these asymmetries are quite sensitive to the reduction of $|\lambda_{tt}\lambda_{bb}|$. Also, the decrease of the mass of $H^0$ and  simultaneously the increase  of the mass of $H^{\pm}$ can enhance the deviations from the SM predictions. Based on the above explanations in all single lepton polarization asymmetries except that ${\cal P}_{T}^{-}$ the most deviations from the SM values happen in the case A of mass set 3. On the other hand,  for the $\tau$ channel of CP violating asymmetry (${\cal A}_{CP}(q^2)$) the results obtained from all cases are different from that of the SM somehow the biggest deviation from the SM anticipation occurs in case C.  This fact indicates that this asymmetry is quite sensitive to the enhancement of $|\lambda_{tt}\lambda_{bb}|$.  Also, while this asymmetry is quite insensitive to the variation of mass of $H^0$, the deviations from the SM prediction increase by  decreasing the mass of $H^{\pm}$. Paying attention to the minimum required number of events for detecting each asymmetry it is inferred that ${\cal P}_{L}^{\mp}$ and ${\cal P}_{T}^{\mp}$ are detectable at LHC, ${\cal P}_{N}^{\mp}$ is measurable at SLHC and CP violating asymmetry is not detectable in neither  LHC nor SLHC.
\par

iii) For  $\mu$ channel, in ${ \la {\cal P}_{L}^{\mp}\ra}$ the results of  cases B and C for all mass sets don't lie between the limits of SM prediction. The maximum deviations of these asymmetries from the calculated values of SM happen in the case C of mass sets 2 and 4 which are $-3.2\% $ SM. In ${ \la {\cal P}_{N}^{\mp}\ra}$ the results of  case A for all mass sets don't lie between the limits of calculated SM prediction. The most deviations from the zero predictions of SM happen in the case A of mass set 3 which are $\mp0.004 $. In ${ \la {\cal P}_{T}^{\mp}\ra}$ the results of  case A for  mass sets 1, 3 and 4 don't lie between the limits of  SM prediction. The most deviations of ${ \la {\cal P}_{T}^{-}\ra}$ and ${ \la {\cal P}_{T}^{+}\ra}$ from the calculated values of SM happen in the case A of mass set 3 which are $+24\% $ SM and $-22\% $ SM, respectively. In ${ \la {\cal A}_{CP}\ra}$ the results of  all cases for all mass sets don't lie between the limits of SM prediction. The most deviations from the zero prediction of SM happen in the case C of mass sets 2 and 4 which are +0.005.
\par
iv) For  $\tau$ channel, in ${ \la {\cal P}_{L}^{\mp}\ra}$ the results of  case A for all mass sets don't lie between the limits of SM prediction. The most deviations of ${ \la {\cal P}_{L}^{-}\ra}$ and ${ \la {\cal P}_{L}^{+}\ra}$ from the obtained values in SM happen in the case A of mass set 3 which are $-4.9 $ times that of SM and $+6.6 $ times that of SM, respectively. In ${ \la {\cal P}_{N}^{\mp}\ra}$ the results of  cases A and B for all mass sets don't lie between the limits of SM prediction. The most deviations from the zero predictions of SM happen in the case A of mass set 3 which are $\mp0.024 $.  In ${ \la {\cal P}_{T}^{\mp}\ra}$, the results of all cases and   mass sets  lie between the limits of SM predictions  although the most deviation of ${ \la {\cal P}_{T}^{+}\ra}$ from the calculated value of SM is $-15\% $ SM.  In ${ \la {\cal A}_{CP}\ra}$, the results of  all cases and mass sets don't lie between the limits of SM prediction. The most deviations from the zero prediction of SM happen in the case C of mass sets 2 and 4 which are +0.004.
\par
v) By comparing the asymmetries of two channels it is understood that firstly the  ${\la{\cal A}_{CP}\ra}$ and ${ \la {\cal P}_{T}^{\mp}\ra}$ of  $\mu$ channel are more sensitive to the presence of new Higgs bosons than those of $\tau$ channel and secondly the ${ \la {\cal P}_{L}^{\mp}\ra}$ and ${ \la {\cal P}_{N}^{\mp}\ra}$ of $\tau$ channel show more dependency to the existence  of new Higgs bosons than those of $\mu$ channel.
\par Finally, it is worthwhile  mentioning that although  the muon polarization
is measured for stationary muons,  such experiments are very hard  to perform  in the near future. The tau
polarization can be studied by investigating the decay products of tau.
The measurement of tau polarization in this respect is
 easier than the polarization of muon.

\section{Acknowledgment}
The authors would like to thank V. Bashiry for his useful
discussions. Support of Research  Council of Shiraz University is
gratefully acknowledged.

\newpage

\begin{table}[t]
\begin{center}
\caption{The averaged  CP violation and single lepton polarization  asymmetries for
$\overline{B}\rightarrow \overline{K}_0^*(1430) \,\mu^{+}\mu^{-}$ in SM and 2HDM for   the mass sets 1 and 2 of Higgs bosons and the three cases A ($\theta=\pi/2$, $|\lambda_{tt}|=0.03$ and $|\lambda_{bb}|=100$), B ($\theta=\pi/2$, $|\lambda_{tt}|=0.15$ and $|\lambda_{bb}|=50$) and C ($\theta=\pi/2$, $|\lambda_{tt}|=0.3$ and $|\lambda_{bb}|=30$). The
errors shown for each asymmetry are due to the theoretical and
experimental uncertainties. The first ones are related to the
theoretical uncertainties and the second ones are due to experimental
uncertainties. The theoretical uncertainties come from the hadronic uncertainties
related to the form factors and the experimental uncertainties
originate from the mass of quarks and hadrons and Wolfenstein
parameters.}\label{masssetBK0smu12}
\begin{tabular}{clllllll}
\hline\hline
& \rm{SM} & \rm {Case A } & \rm{Case B }& \rm{Case C }& \rm {Case A }& \rm {Case B }& \rm {Case C }\\
&  & \rm { (Set 1)} & \rm{ (Set1)}& \rm{ (Set1)}& \rm { (Set 2)}& \rm { (Set 2)}& \rm { (Set 2)}\\
\hline

$\rm{ \la {\cal A}_{CP}\ra}$ &$\phantom{-}0.000^{+0.000+0.000}_{-0.000-0.000} $ & $+0.001$ & $+0.004$ & $+0.004$ & $+0.002$ & $+0.004$ & $+0.005$ \\
$ \rm{ \la {\cal P}_{L}^{-}\ra}$ & $ -0.952^{+0.002+0.001}_{-0.002-0.001}$ & $-0.945 $ & $-0.934$& $-0.929$& $-0.945$& $-0.928$ & $-0.922$\\
$\rm{ \la {\cal P}_{T}^{-}\ra}$ & $-0.158^{+0.009+0.002}_{-0.012-0.002}$& $-0.179$ & $-0.156$ & $-0.154$& $-0.170$& $-0.154$& $-0.153$ \\
$\rm{ \la {\cal P}_{N}^{-}\ra}$ &$\phantom{-}0.000^{+0.000+0.000}_{-0.000-0.000} $ & $-0.002$ & $-0.000$ & $-0.000$& $-0.001$ & $-0.000$ & $-0.000$ \\
$ \rm{ \la {\cal P}_{L}^{+}\ra}$ & $ +0.952^{+0.002+0.001}_{-0.002-0.001}$ & $+0.950$& $+0.934$& $+0.930$& $+0.948$ & $+0.929$& $+0.922$\\
$\rm{ \la {\cal P}_{T}^{+}\ra}$ & $+0.158^{+0.009+0.002}_{-0.012-0.002}$& $+0.140$ & $+0.154$& $+0.154$& $+0.149$& $+0.154$&$+0.153$ \\
$\rm{ \la {\cal P}_{N}^{+}\ra}$ &$\phantom{-}0.000^{+0.000+0.000}_{-0.000-0.000} $ & $+0.002$ & $+0.000$ & $+0.000$& $+0.001$ & $+0.000$ & $+0.000$ \\
 \hline\hline
\end{tabular}
\end{center}
\end{table}

\begin{table}[t]
\begin{center}
\caption{The same as TABLE \ref{masssetBK0smu12} but for the mass sets 3 and 4 of Higgs bosons.}\label{masssetBK0smu34}
\begin{tabular}{clllllll}
\hline\hline
& \rm{SM} & \rm {Case A } & \rm{Case B }& \rm{Case C }& \rm {Case A }& \rm {Case B }& \rm {Case C }\\
&  & \rm { (Set 3)} & \rm{ (Set3)}& \rm{ (Set3)}& \rm { (Set 4)}& \rm { (Set 4)}& \rm { (Set 4)}\\
\hline

$\rm{ \la {\cal A}_{CP}\ra}$ &$\phantom{-}0.000^{+0.000+0.000}_{-0.000-0.000} $ & $+0.001$ & $+0.004$ & $+0.004$ & $+0.002$ & $+0.004$ & $+0.005$ \\
$ \rm{ \la {\cal P}_{L}^{-}\ra}$ & $ -0.952^{+0.002+0.001}_{-0.002-0.001}$ & $-0.942 $ & $-0.934$& $-0.929$& $-0.943$& $-0.928$ & $-0.922$\\
$\rm{ \la {\cal P}_{T}^{-}\ra}$ & $-0.158^{+0.009+0.002}_{-0.012-0.002}$& $-0.196$ & $-0.156$ & $-0.155$& $-0.183$& $-0.155$& $-0.153$ \\
$\rm{ \la {\cal P}_{N}^{-}\ra}$ &$\phantom{-}0.000^{+0.000+0.000}_{-0.000-0.000} $ & $-0.004$ & $-0.000$ & $-0.000$& $-0.003$ & $-0.000$ & $-0.000$ \\
$ \rm{ \la {\cal P}_{L}^{+}\ra}$ & $ +0.952^{+0.002+0.001}_{-0.002-0.001}$ & $+0.952$& $+0.935$& $+0.930$& $+0.950$ & $+0.929$& $+0.922$\\
$\rm{ \la {\cal P}_{T}^{+}\ra}$ & $+0.158^{+0.009+0.002}_{-0.012-0.002}$& $+0.123$ & $+0.154$& $+0.154$& $+0.136$& $+0.153$&$+0.153$ \\
$\rm{ \la {\cal P}_{N}^{+}\ra}$ &$\phantom{-}0.000^{+0.000+0.000}_{-0.000-0.000} $ & $+0.004$ & $+0.000$ & $+0.000$& $+0.003$ & $+0.000$ & $+0.000$ \\
\hline\hline
\end{tabular}
\end{center}
\end{table}

\begin{table}[t]
\begin{center}
\caption{The same as TABLE \ref{masssetBK0smu12} except for $\overline{B}\to \overline{K}_0^*(1430) \tau^+\tau^-$.}\label{masssetBK0stau12}
\begin{tabular}{clllllll}
\hline\hline
& \rm{SM} & \rm {Case A } & \rm{Case B }& \rm{Case C }& \rm {Case A }& \rm {Case B }& \rm {Case C }\\
&  & \rm { (Set 1)} & \rm{ (Set1)}& \rm{ (Set1)}& \rm { (Set 2)}& \rm { (Set 2)}& \rm { (Set 2)}\\
\hline

$\rm{ \la {\cal A}_{CP}\ra}$ &$\phantom{-}0.000^{+0.000+0.000}_{-0.000-0.000} $ & $+0.001$ & $+0.003$ & $+0.003$ & $+0.001$ & $+0.003$ & $+0.004$ \\
$ \rm{ \la {\cal P}_{L}^{-}\ra}$ & $ -0.066^{+0.030+0.011}_{-0.077-0.013}$ & $+0.147 $ & $-0.056$& $-0.066$& $+0.057$ & $-0.060$&$-0.063$\\
$ \rm{ \la {\cal P}_{T}^{-}\ra}$ & $ -0.628^{+0.123+0.010}_{-0.127-0.017}$ & $-0.619$ & $-0.619$ & $-0.612$& $-0.616 $ & $-0.617$& $-0.609$\\
$\rm{ \la {\cal P}_{N}^{-}\ra}$ &$\phantom{-}0.000^{+0.000+0.000}_{-0.000-0.000} $ & $-0.013$ & $-0.001$ & $-0.000$ & $-0.008$ & $-0.001$ & $-0.000$ \\
$ \rm{ \la {\cal P}_{L}^{+}\ra}$ & $ +0.066^{+0.030+0.011}_{-0.077-0.013}$ & $+0.266 $ & $+0.073$& $+0.066$& $+0.176$ & $+0.069$&$+0.065$\\
$ \rm{ \la {\cal P}_{T}^{+}\ra}$ & $ +0.628^{+0.123+0.010}_{-0.127-0.017}$ & $+0.579$ & $+0.617$ & $+0.612$& $+0.593 $ & $+0.616$& $+0.609$\\
$\rm{ \la {\cal P}_{N}^{+}\ra}$ &$\phantom{-}0.000^{+0.000+0.000}_{-0.000-0.000} $ & $+0.013$ & $+0.001$ & $+0.000$ & $+0.008$ & $+0.001$ & $+0.000$ \\

 \hline\hline
\end{tabular}
\end{center}
\end{table}

\begin{table}[t]
\begin{center}
\caption{The same as TABLE \ref{masssetBK0stau12}  but for the mass sets 3 and 4 of Higgs bosons.}\label{masssetBK0stau34}
\begin{tabular}{clllllll}
\hline\hline
& \rm{SM} & \rm {Case A } & \rm{Case B }& \rm{Case C }& \rm {Case A }& \rm {Case B }& \rm {Case C }\\
&  & \rm { (Set 3)} & \rm{ (Set3)}& \rm{ (Set3)}& \rm { (Set 4)}& \rm { (Set 4)}& \rm { (Set 4)}\\
\hline

$\rm{ \la {\cal A}_{CP}\ra}$ &$\phantom{-}0.000^{+0.000+0.000}_{-0.000-0.000} $ & $+0.001$ & $+0.003$ & $+0.003$ & $+0.001$ & $+0.003$ & $+0.004$ \\
$ \rm{ \la {\cal P}_{L}^{-}\ra}$ & $ -0.066^{+0.030+0.011}_{-0.077-0.013}$ & $+0.322 $ & $-0.049$& $-0.060$& $+0.197$ & $-0.054$&$-0.061$\\
$ \rm{ \la {\cal P}_{T}^{-}\ra}$ & $ -0.628^{+0.123+0.010}_{-0.127-0.017}$ & $-0.611$ & $-0.620$ & $-0.613$& $-0.617 $ & $-0.617$& $-0.609$\\
$\rm{ \la {\cal P}_{N}^{-}\ra}$ &$\phantom{-}0.000^{+0.000+0.000}_{-0.000-0.000} $ & $-0.024$ & $-0.002$ & $-0.000$ & $-0.017$ & $-0.001$ & $-0.000$ \\
$ \rm{ \la {\cal P}_{L}^{+}\ra}$ & $ +0.066^{+0.030+0.011}_{-0.077-0.013}$ & $+0.436 $ & $+0.081$& $+0.068$& $+0.314$ & $+0.075$&$+0.067$\\
$ \rm{ \la {\cal P}_{T}^{+}\ra}$ & $ +0.628^{+0.123+0.010}_{-0.127-0.017}$ & $+0.537$ & $+0.617$ & $+0.612$& $+0.567 $ & $+0.615$& $+0.609$\\
$\rm{ \la {\cal P}_{N}^{+}\ra}$ &$\phantom{-}0.000^{+0.000+0.000}_{-0.000-0.000} $ & $+0.024$ & $+0.002$ & $+0.000$ & $+0.017$ & $+0.001$ & $+0.000$ \\
 \hline\hline
\end{tabular}
\end{center}
\end{table}

\begin{figure}[ht]
  \centering
  \setlength{\fboxrule}{2pt}
        \centering
                     \includegraphics[height=1.7in]{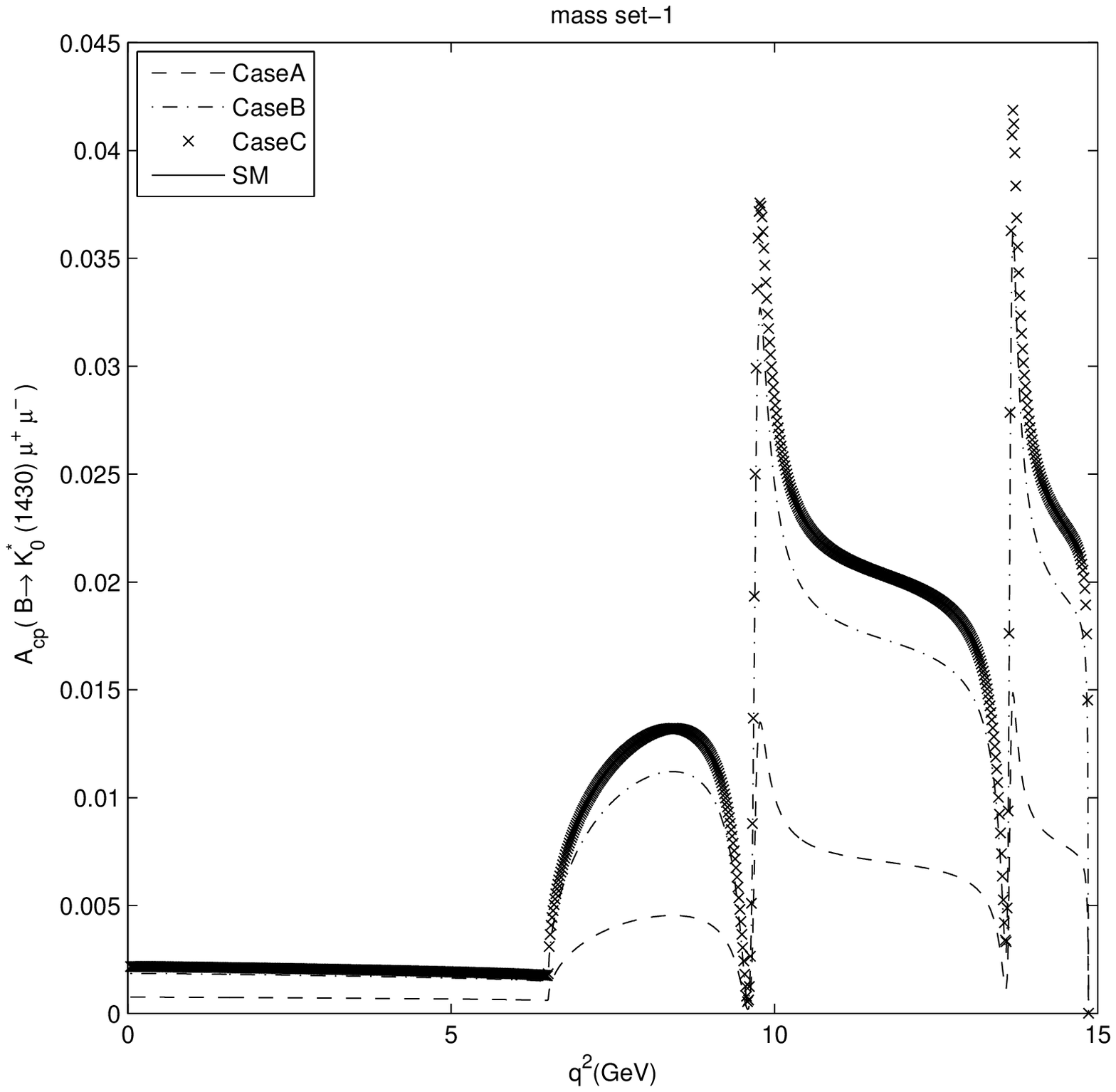}~
             \includegraphics[height=1.7in]{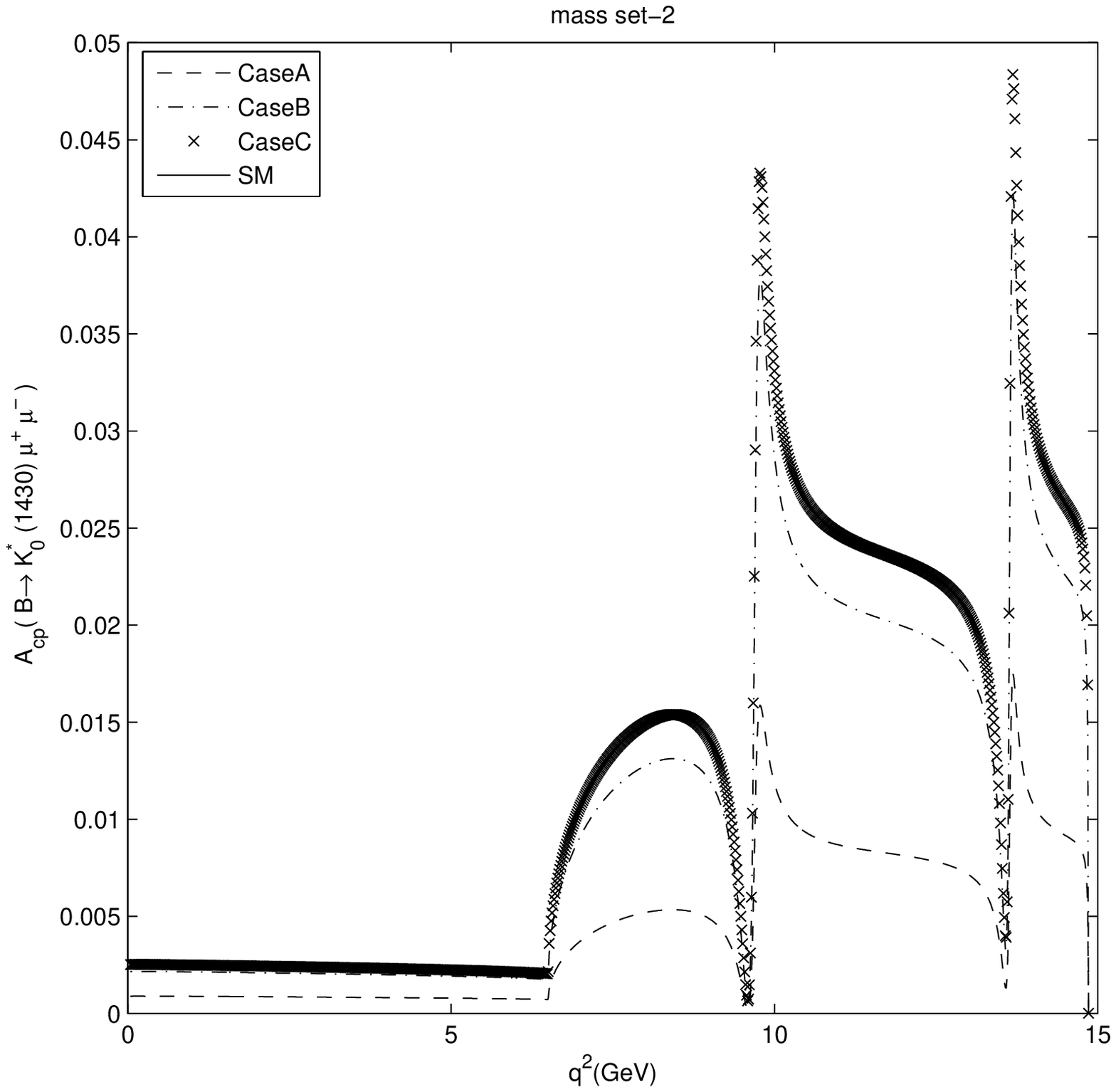}
             \includegraphics[height=1.7in]{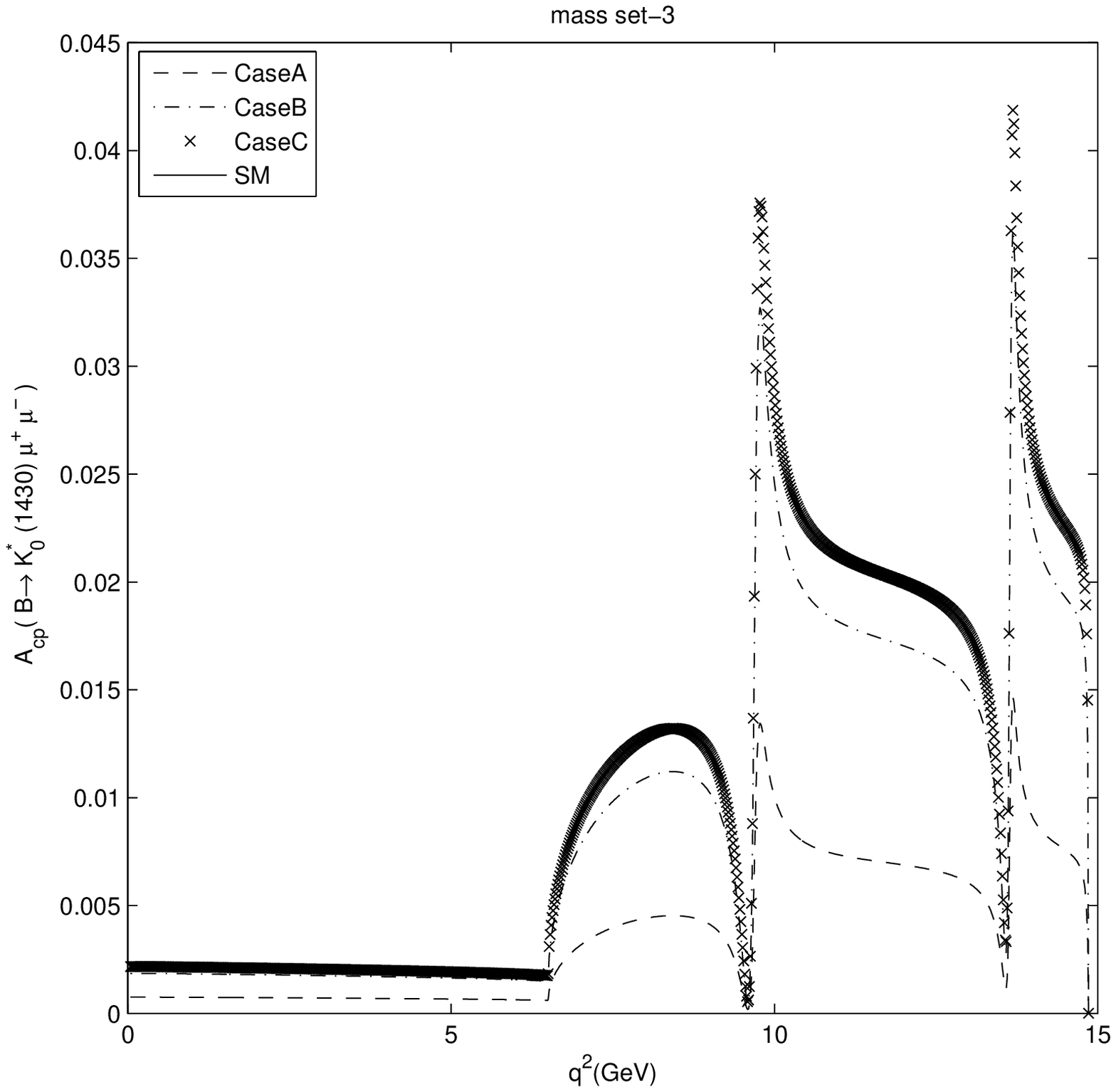}~
             \includegraphics[height=1.7in]{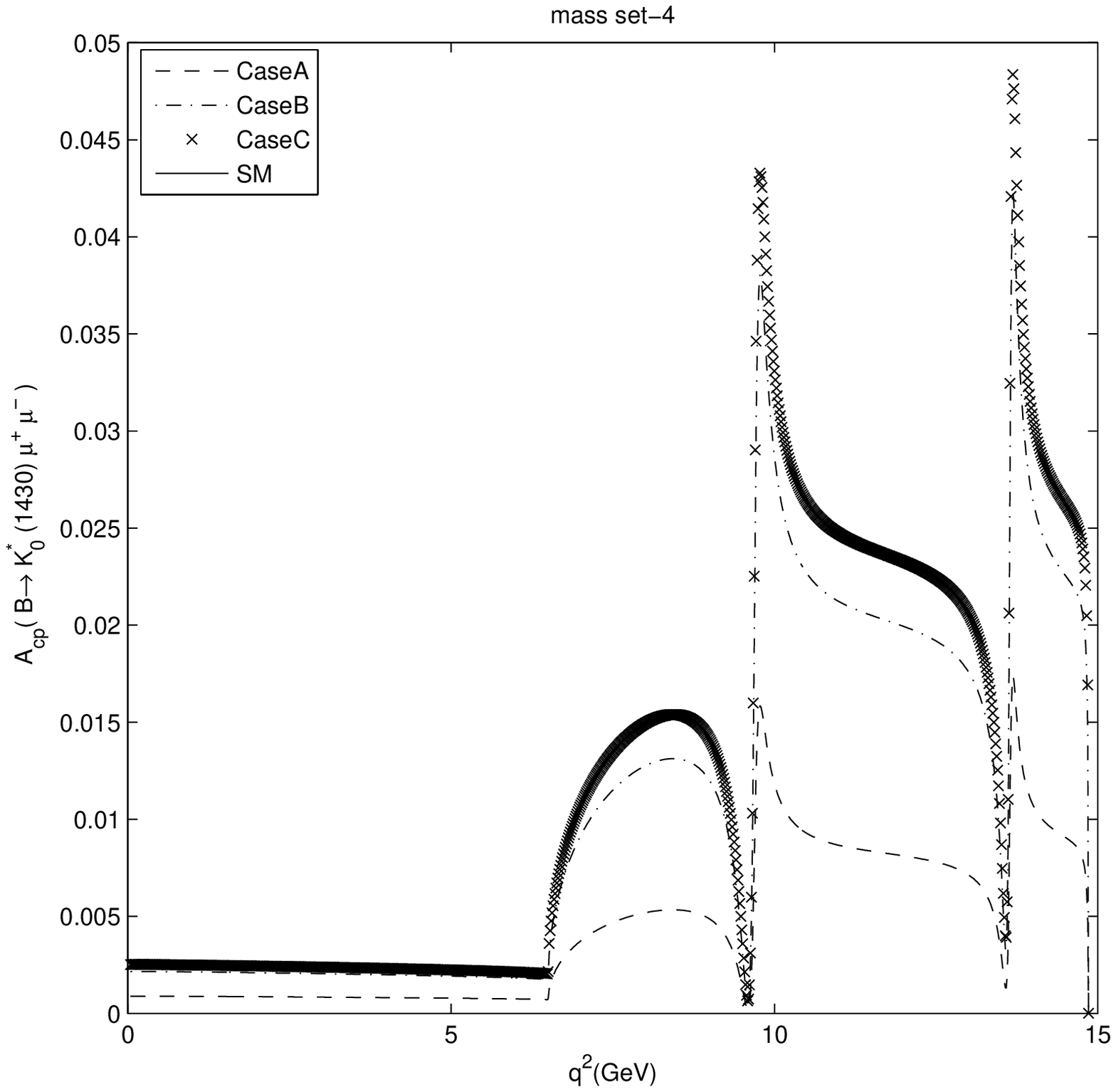}
               \caption{The dependence of the $ {\cal A}_{CP}$ polarization  on $q^2$  and the three typical cases of 2HDM, i.e.
               cases A, B and C and SM  for  the $\mu$  channel of  $\overline{B}\to\overline{K}_0^{*}$ transition for the  mass sets 1, 2, 3  and 4. } \label{ACPmKstar}
    \end{figure}
    \begin{figure}[ht]
  \centering
  \setlength{\fboxrule}{2pt}
        \centering
                     \includegraphics[height=1.7in]{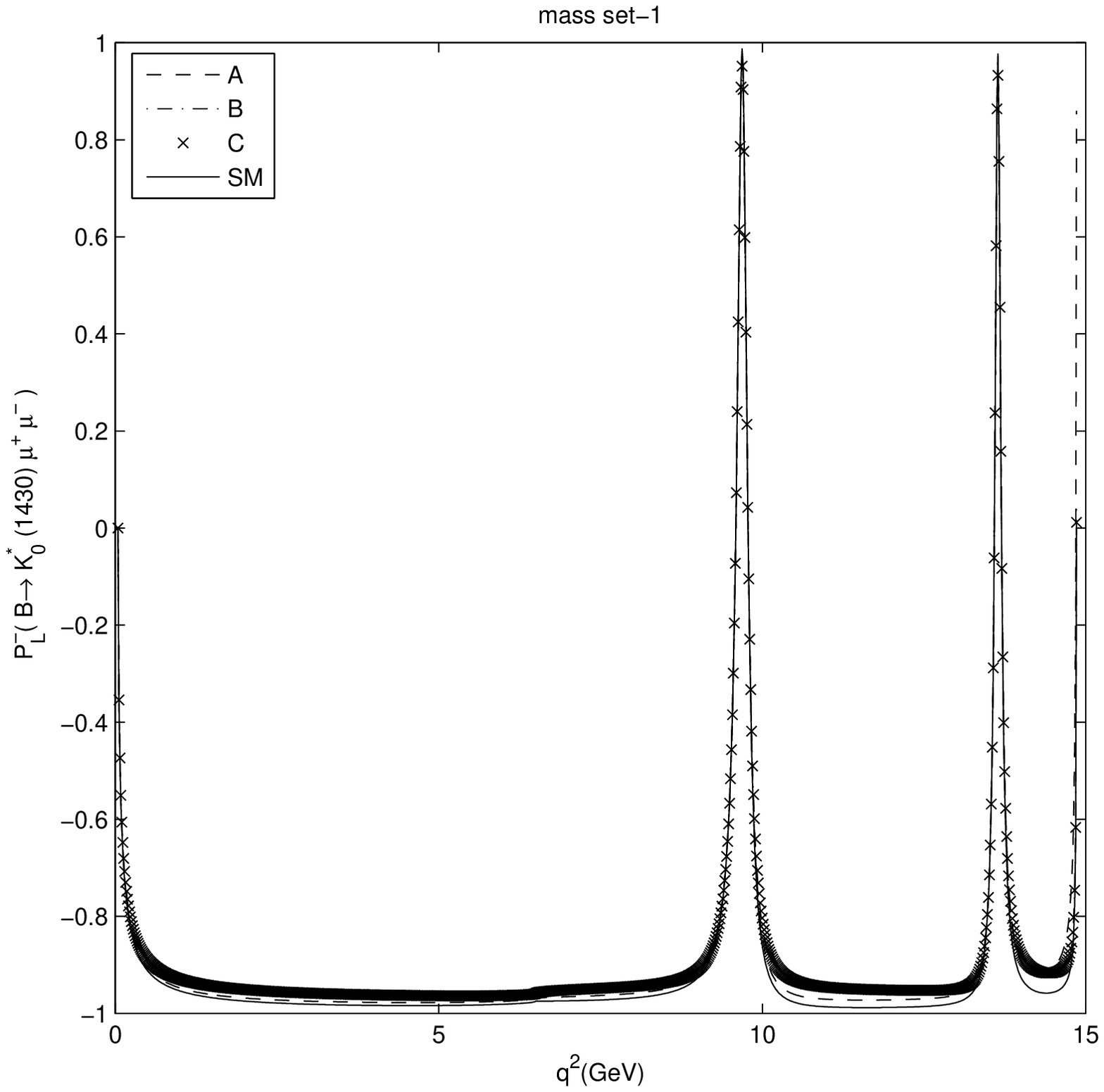}~
             \includegraphics[height=1.7in]{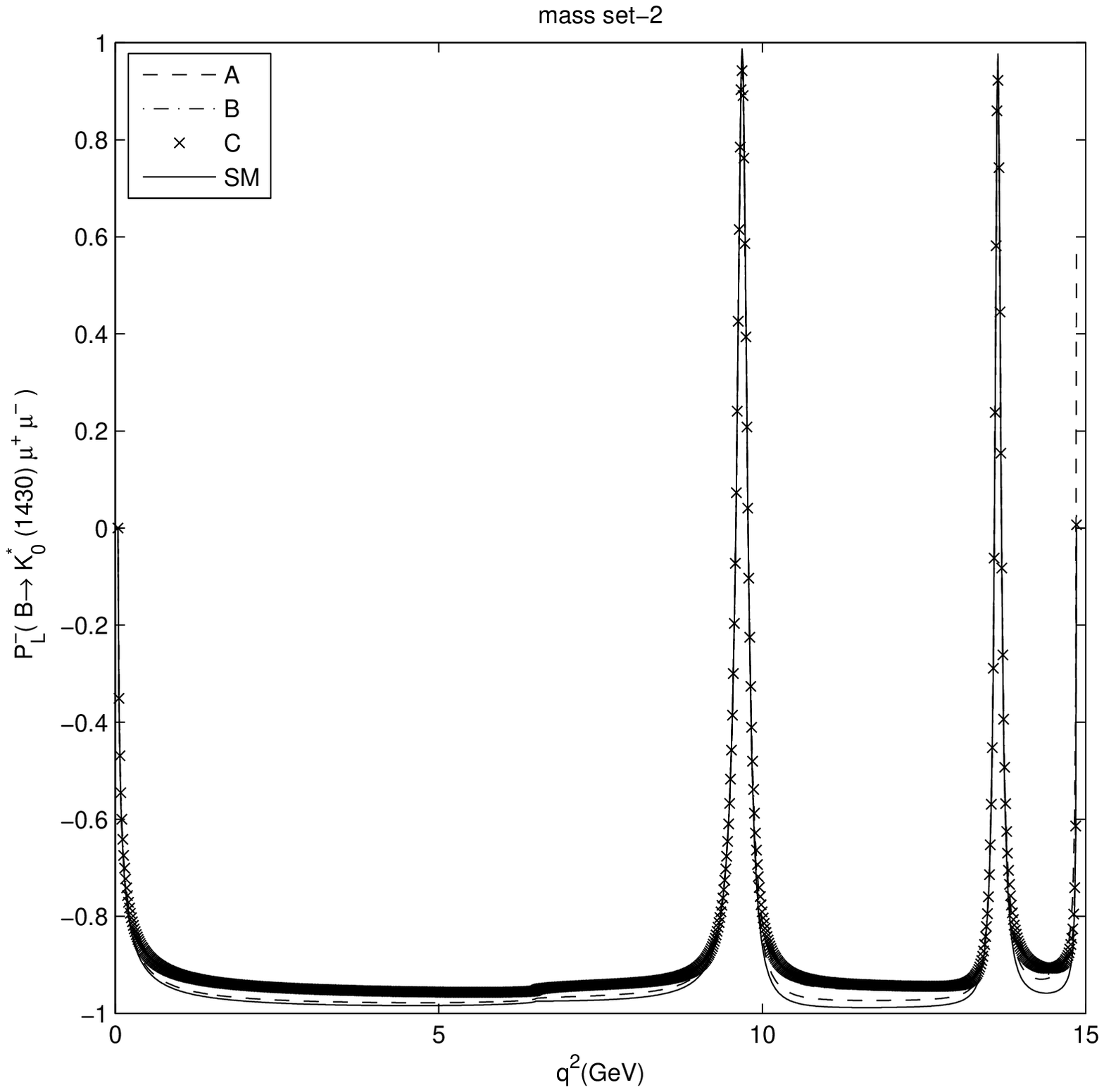}
             \includegraphics[height=1.7in]{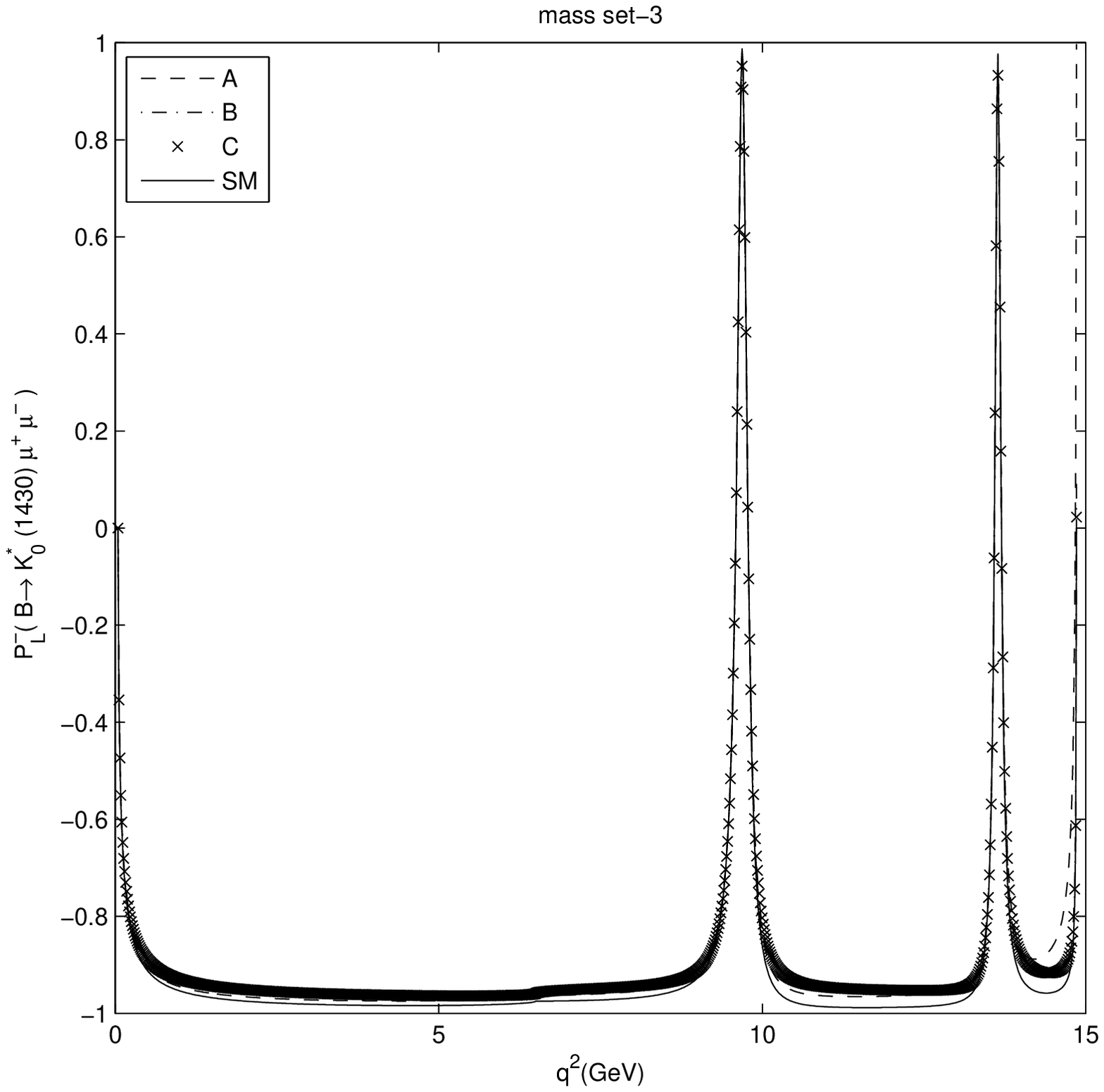}~
             \includegraphics[height=1.7in]{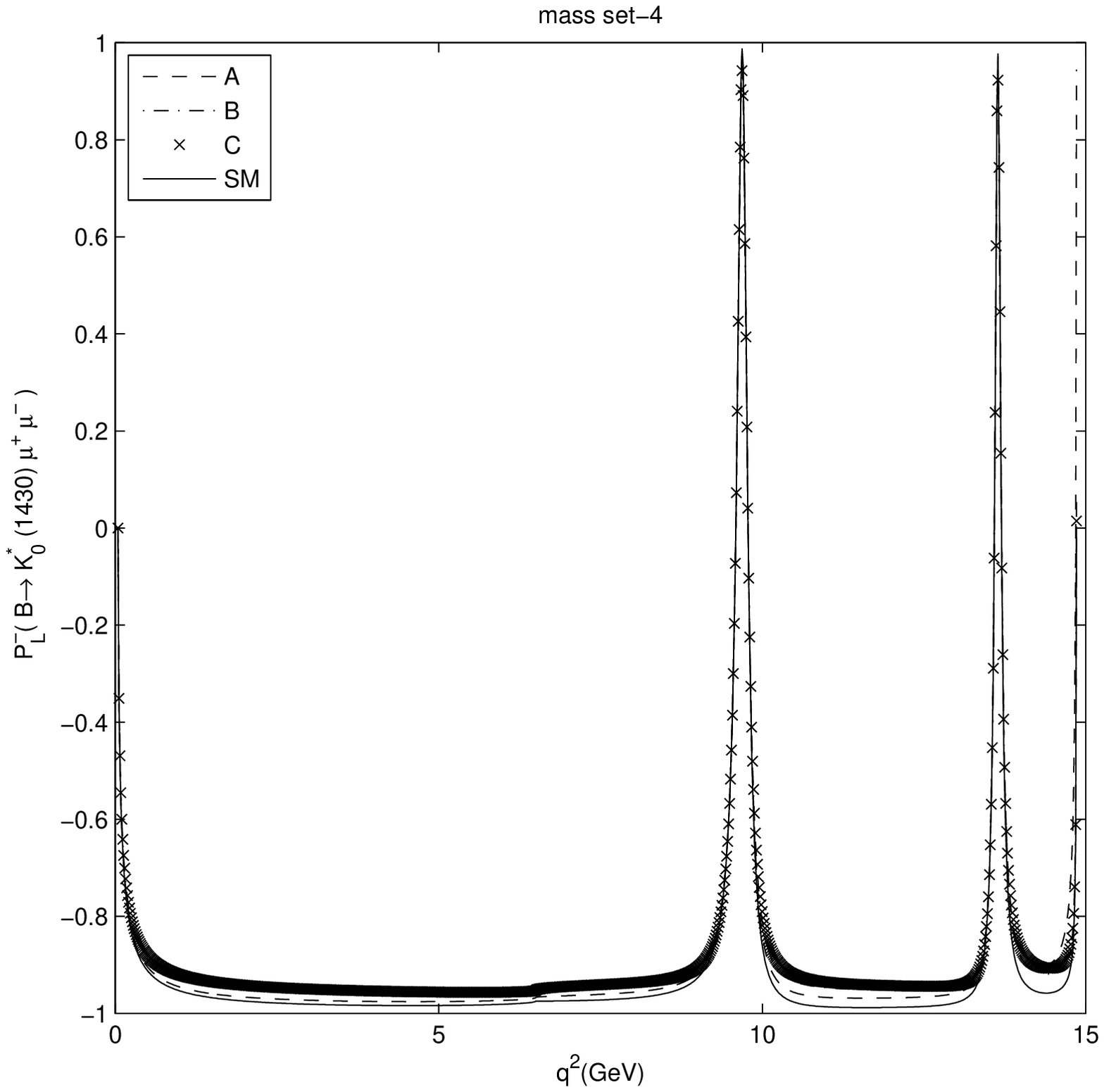}
               \caption{The dependence of the $ {\cal P}_{L}^{-}$ polarization  on $q^2$  and the three typical cases of 2HDM, i.e.
               cases A, B and C and SM  for  the $\mu$  channel of  $\overline{B}\to\overline{K}_0^{*}$ transition for the  mass sets 1, 2, 3  and 4. } \label{PLmmKstar}
    \end{figure}
        \begin{figure}[ht]
  \centering
  \setlength{\fboxrule}{2pt}
        \centering
                     \includegraphics[height=1.7in]{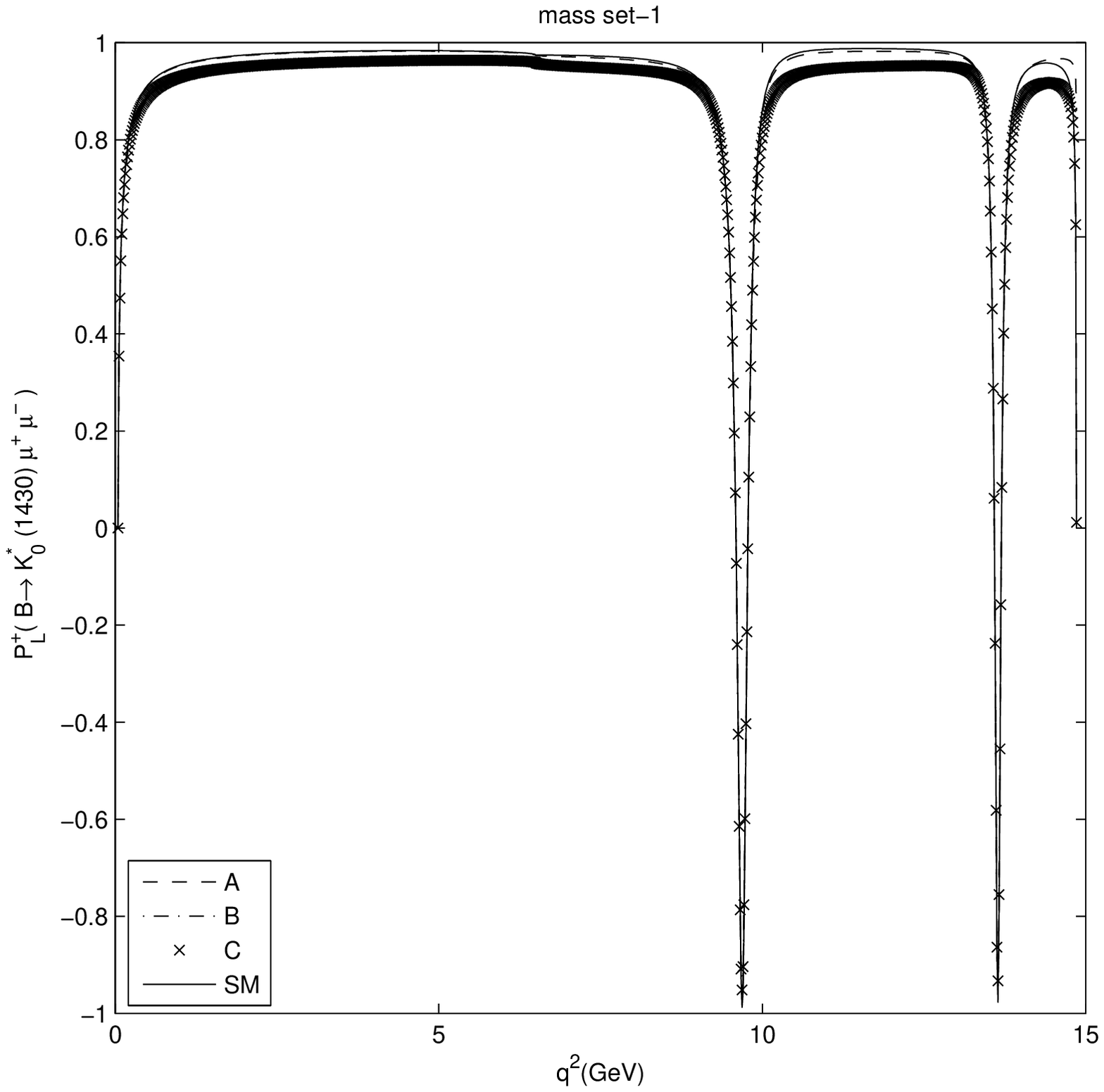}~
             \includegraphics[height=1.7in]{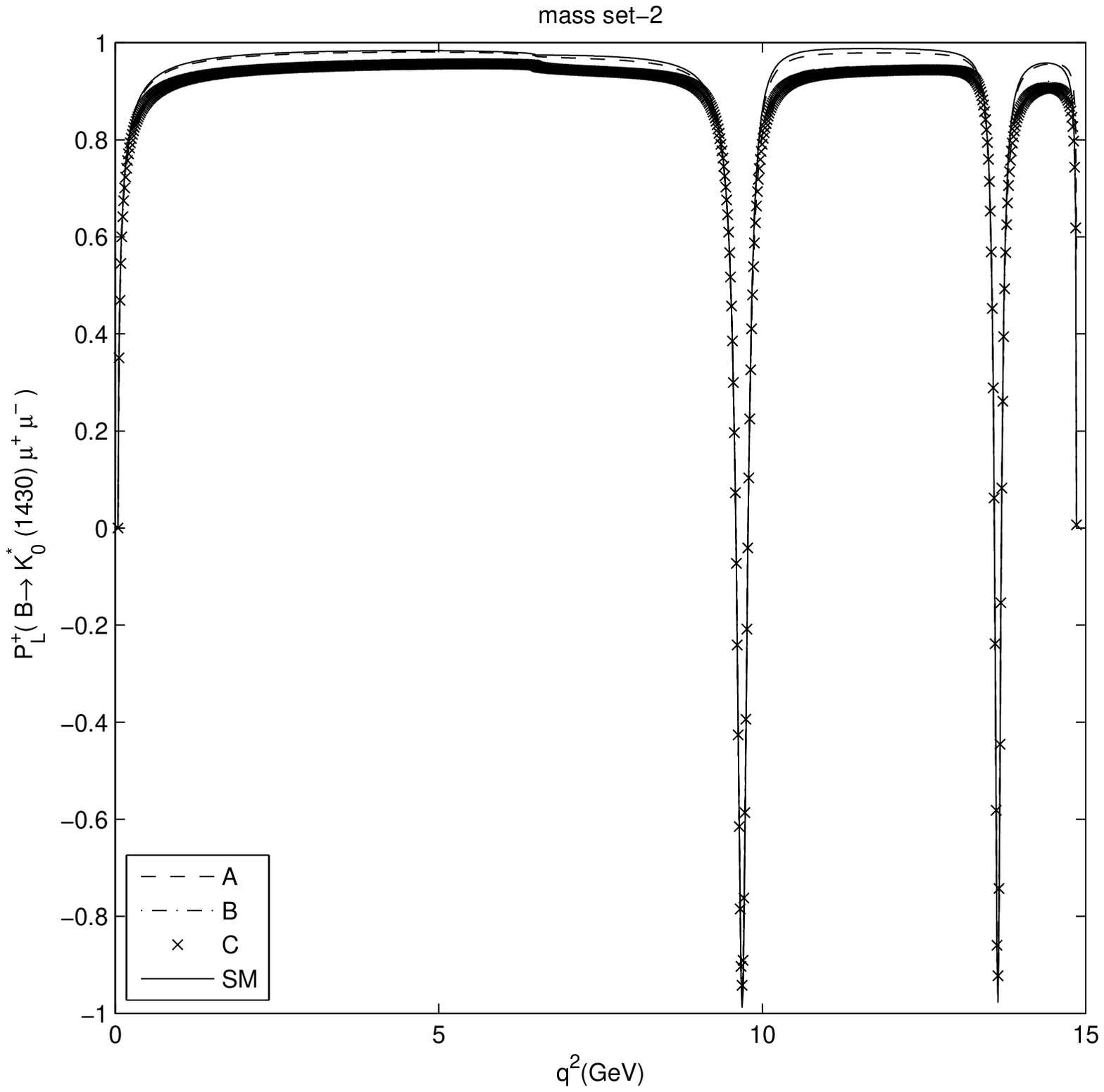}
             \includegraphics[height=1.7in]{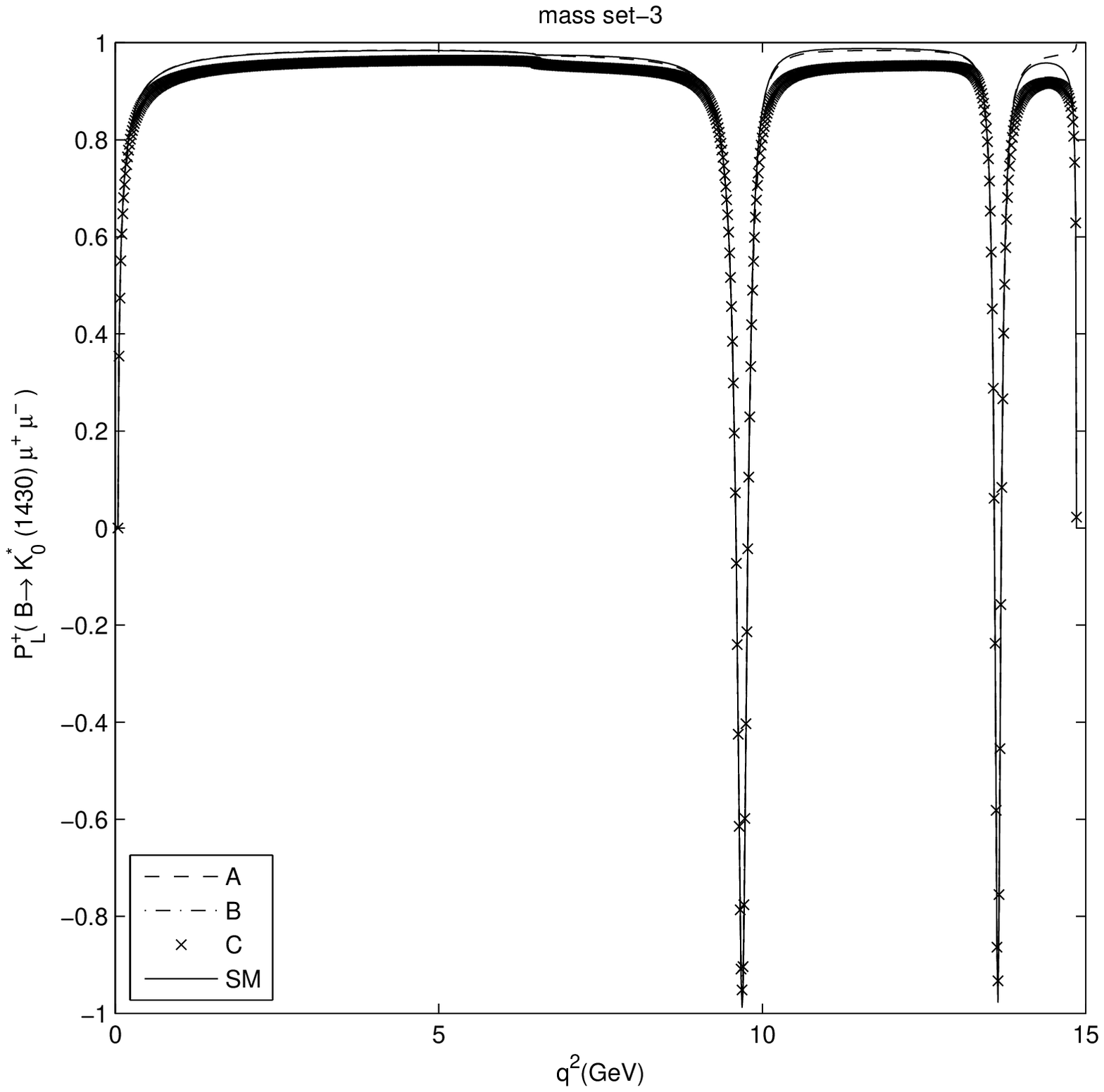}~
             \includegraphics[height=1.7in]{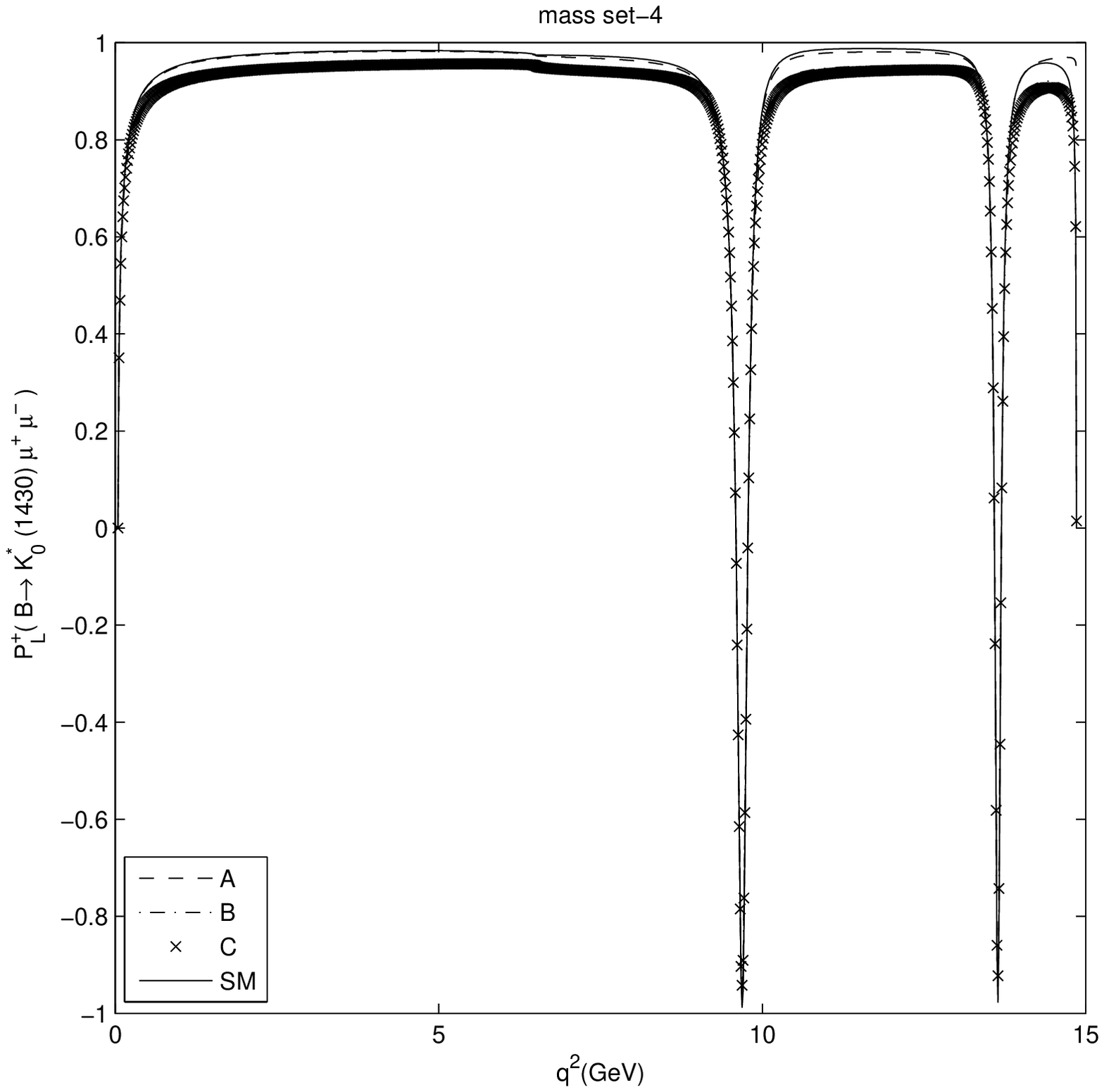}
               \caption{The dependence of the $ {\cal P}_{L}^{+}$ polarization  on $q^2$  and the three typical cases of 2HDM, i.e.
               cases A, B and C and SM  for  the $\mu$  channel of  $\overline{B}\to\overline{K}_0^{*}$ transition for the  mass sets 1, 2, 3  and 4. } \label{PLpmKstar}
    \end{figure}
      \begin{figure}[ht]
  \centering
  \setlength{\fboxrule}{2pt}
        \centering
                     \includegraphics[height=1.7in]{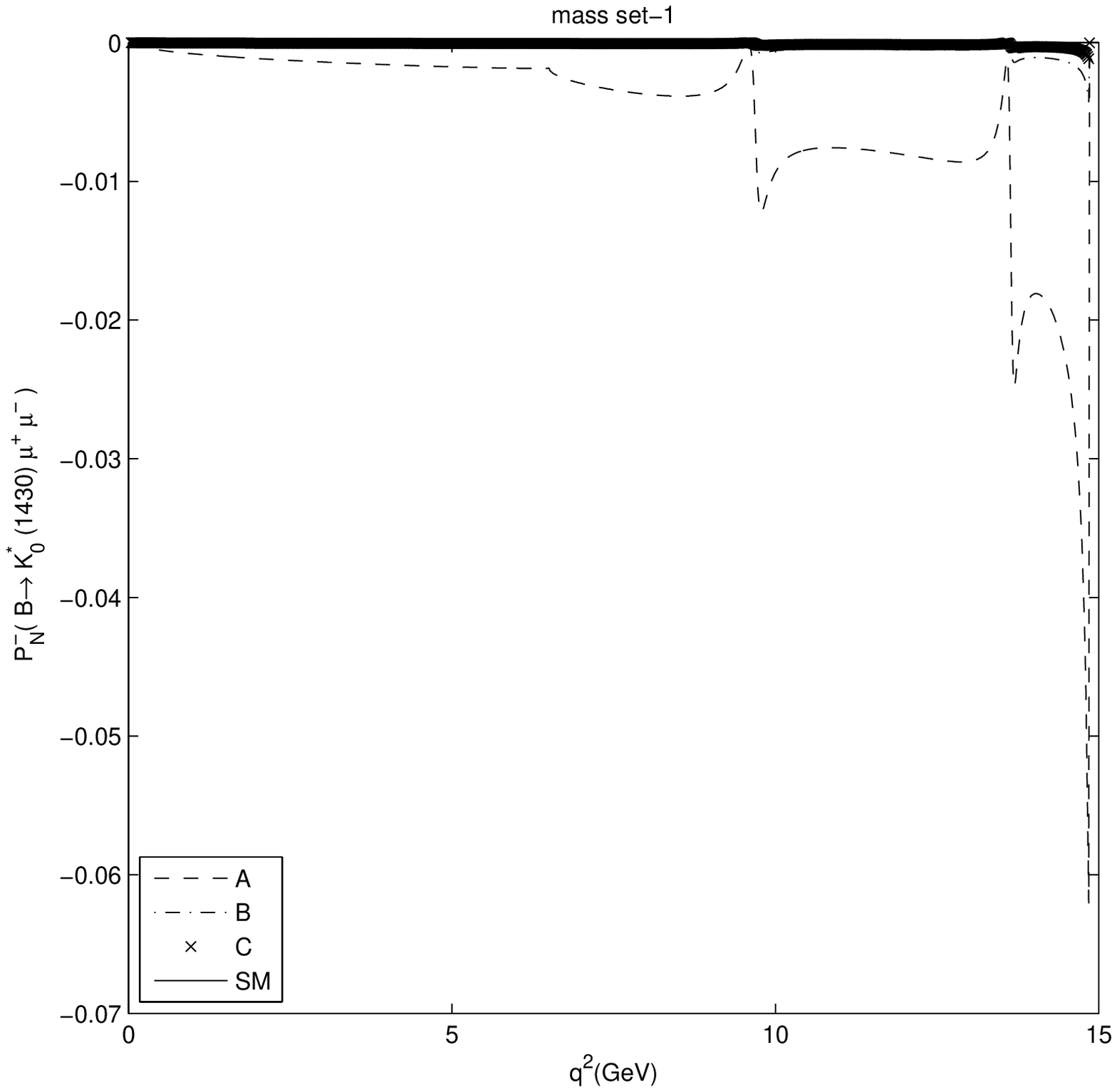}~
             \includegraphics[height=1.7in]{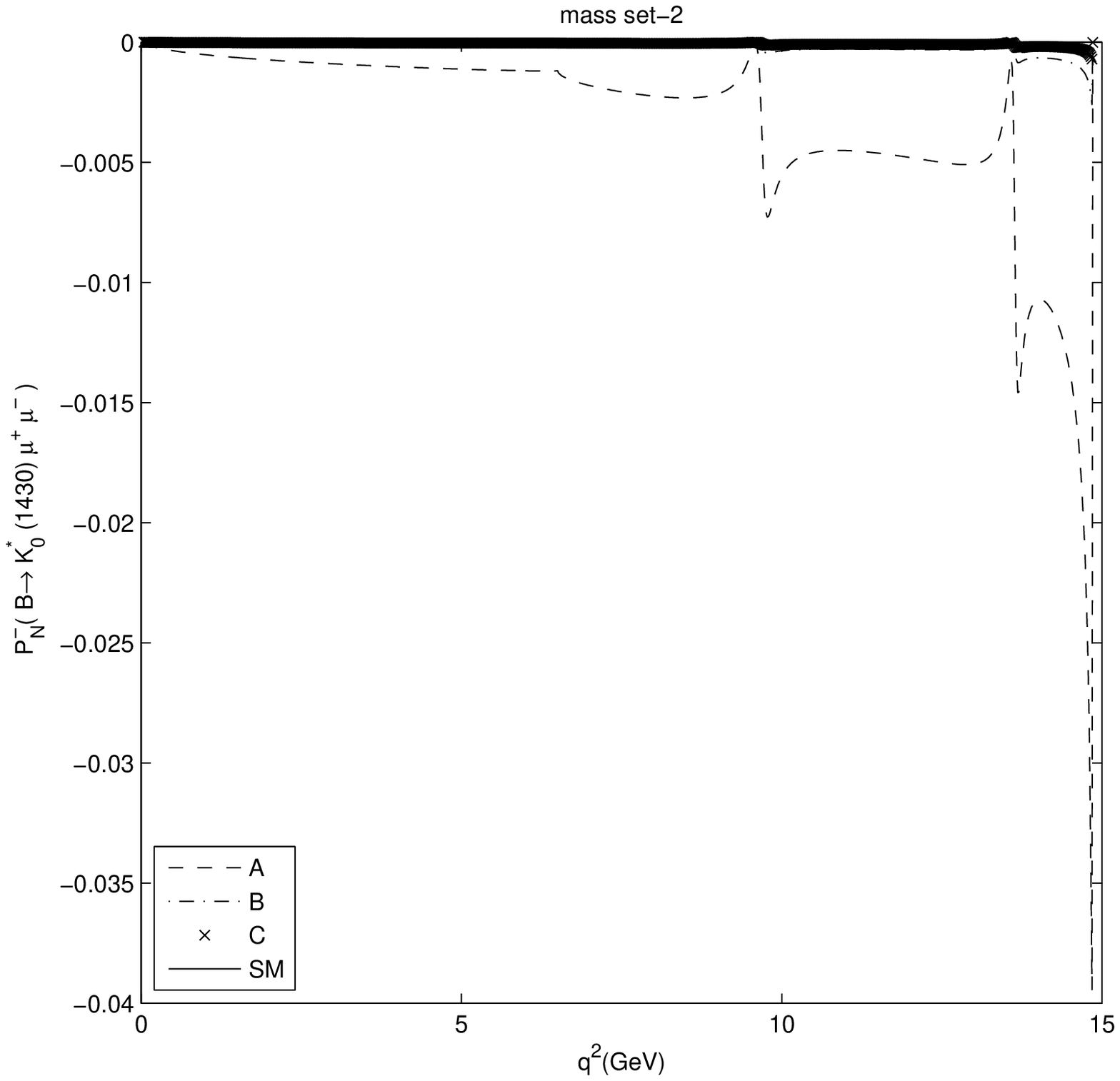}
             \includegraphics[height=1.7in]{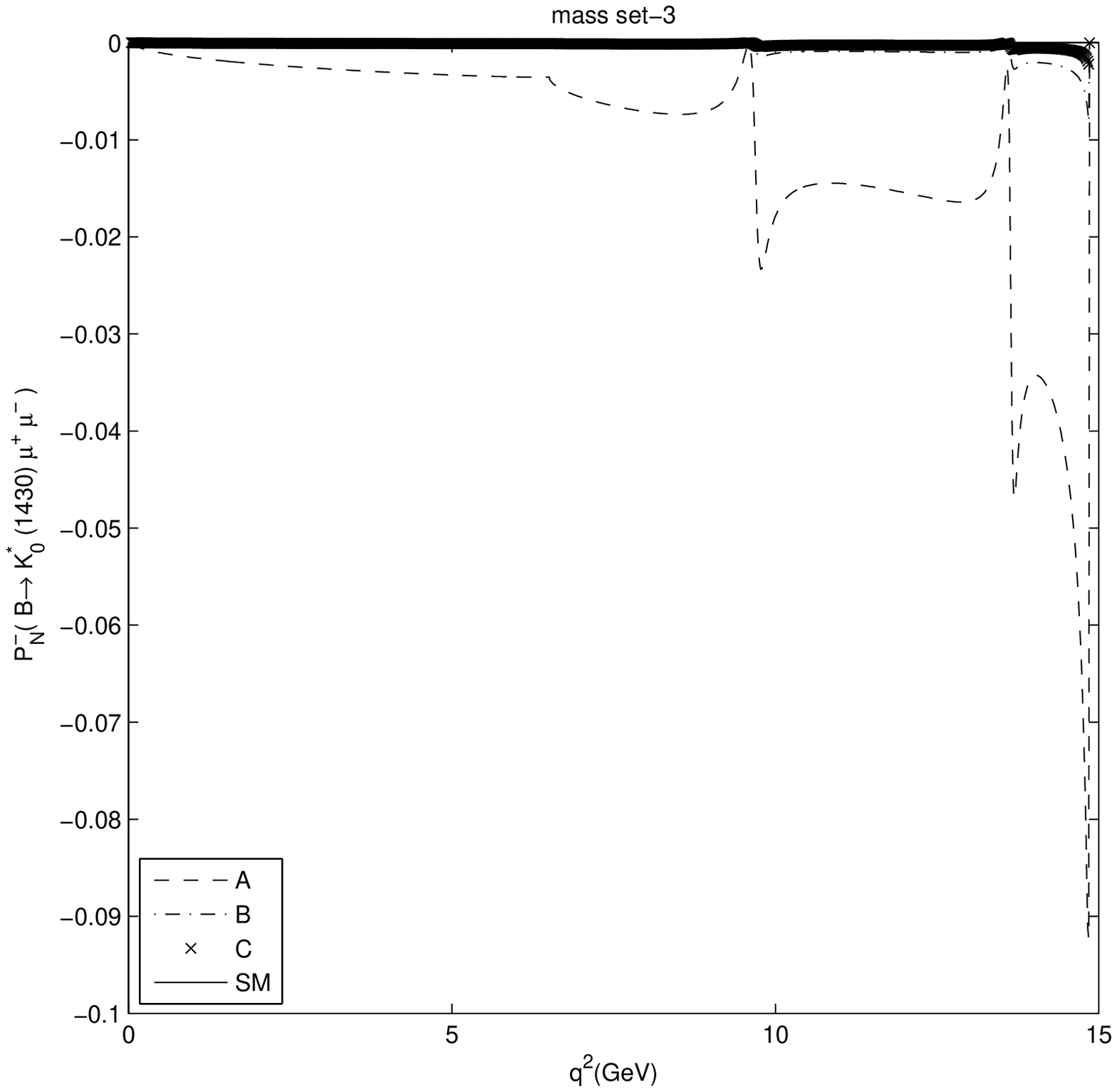}~
             \includegraphics[height=1.7in]{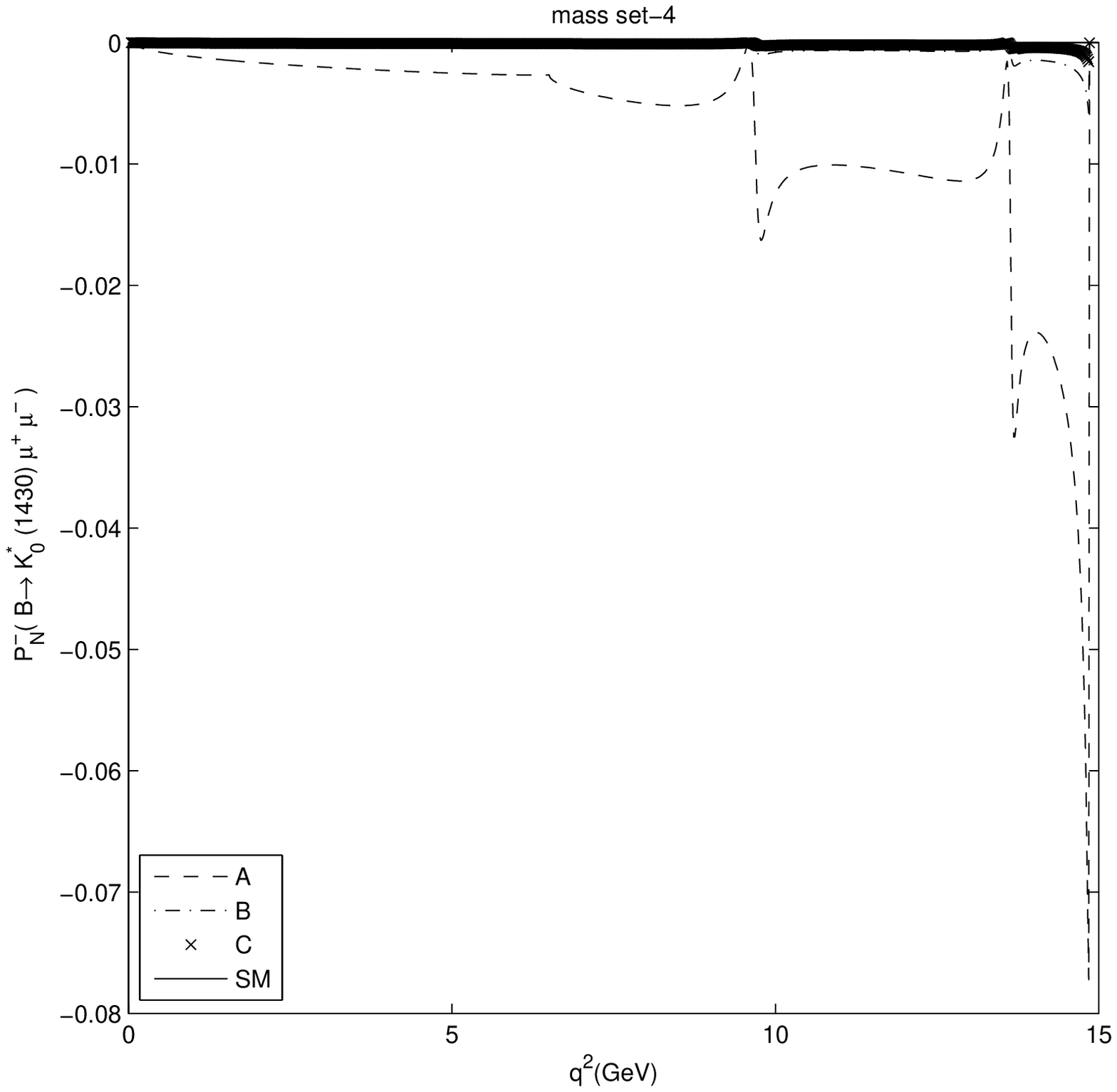}
               \caption{The dependence of the $ {\cal P}_{N}^{-}$ polarization  on $q^2$  and the three typical cases of 2HDM, i.e.
               cases A, B and C and SM  for  the $\mu$  channel of  $\overline{B}\to\overline{K}_0^{*}$ transition for the  mass sets 1, 2, 3  and 4. } \label{PNmmKstar}
    \end{figure}
          \begin{figure}[ht]
  \centering
  \setlength{\fboxrule}{2pt}
        \centering
                     \includegraphics[height=1.7in]{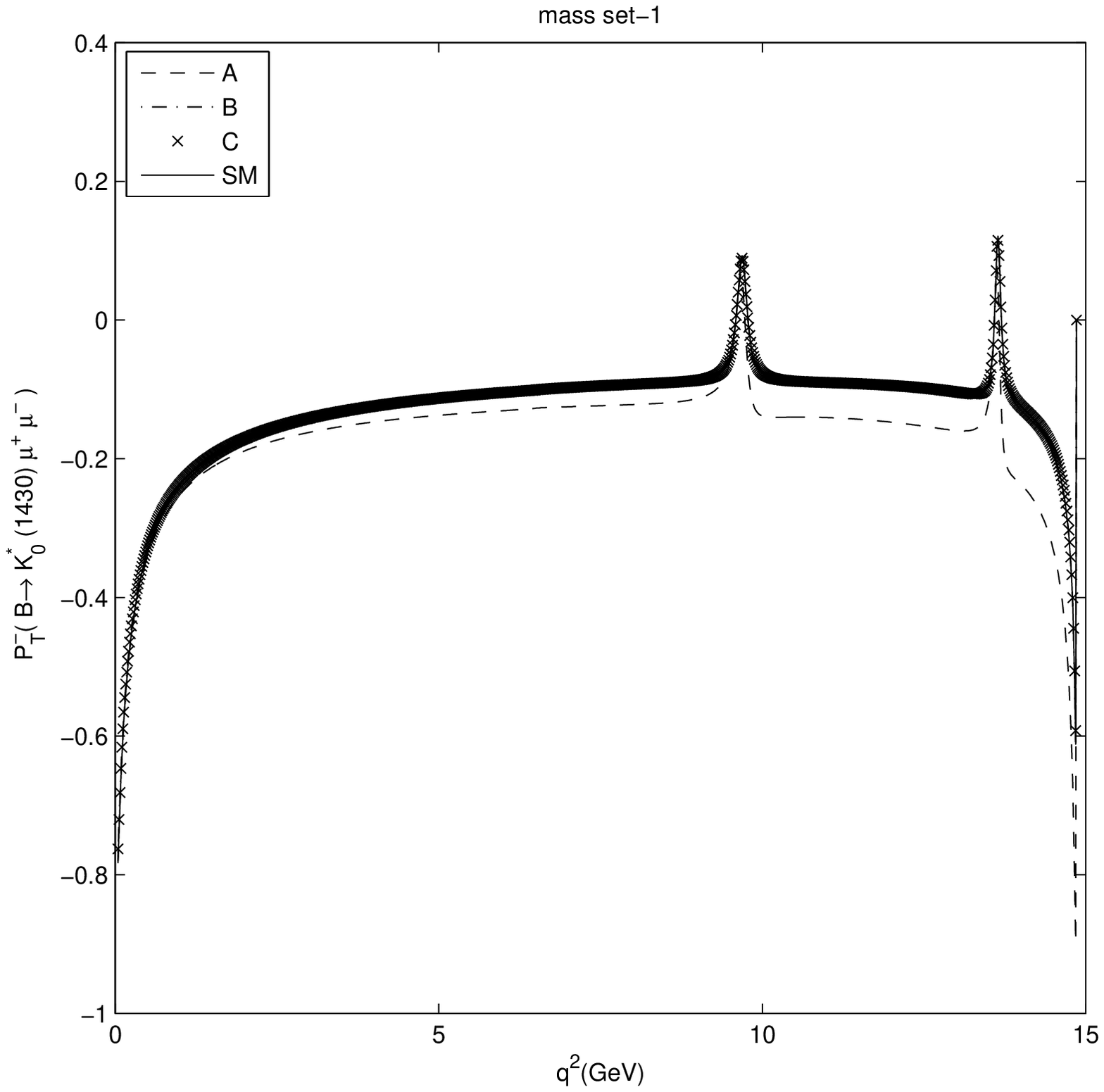}~
             \includegraphics[height=1.7in]{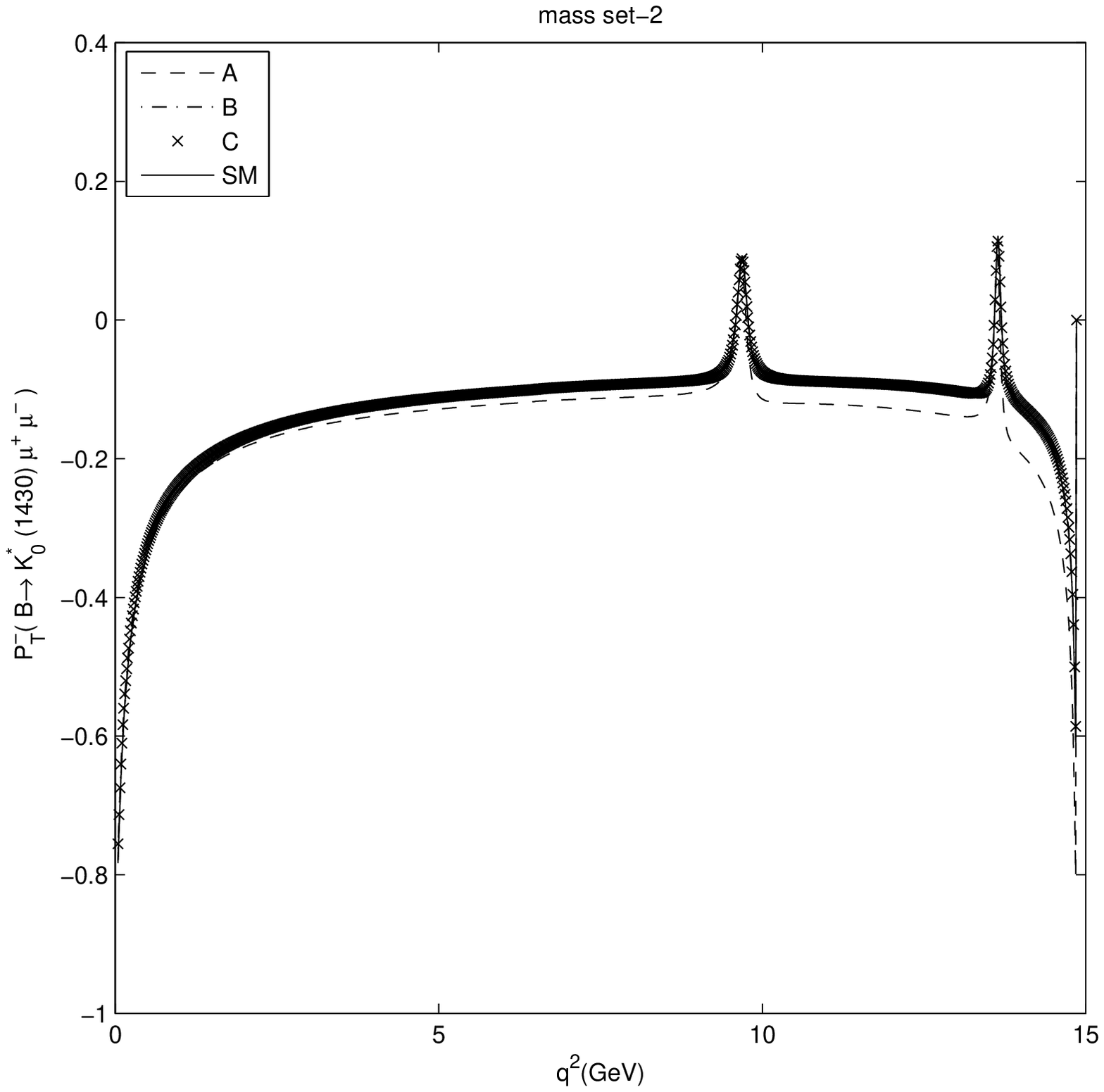}
             \includegraphics[height=1.7in]{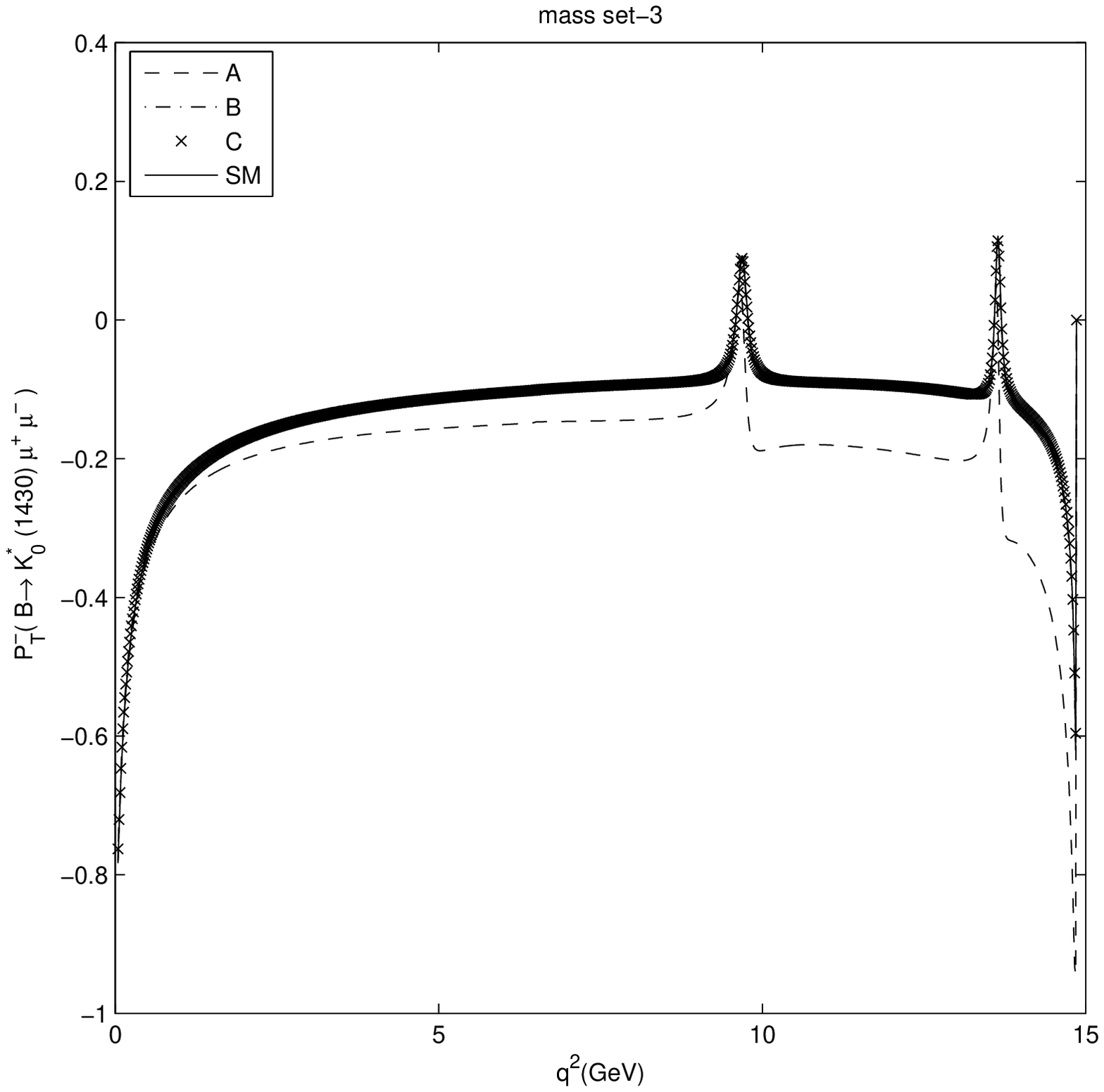}~
             \includegraphics[height=1.7in]{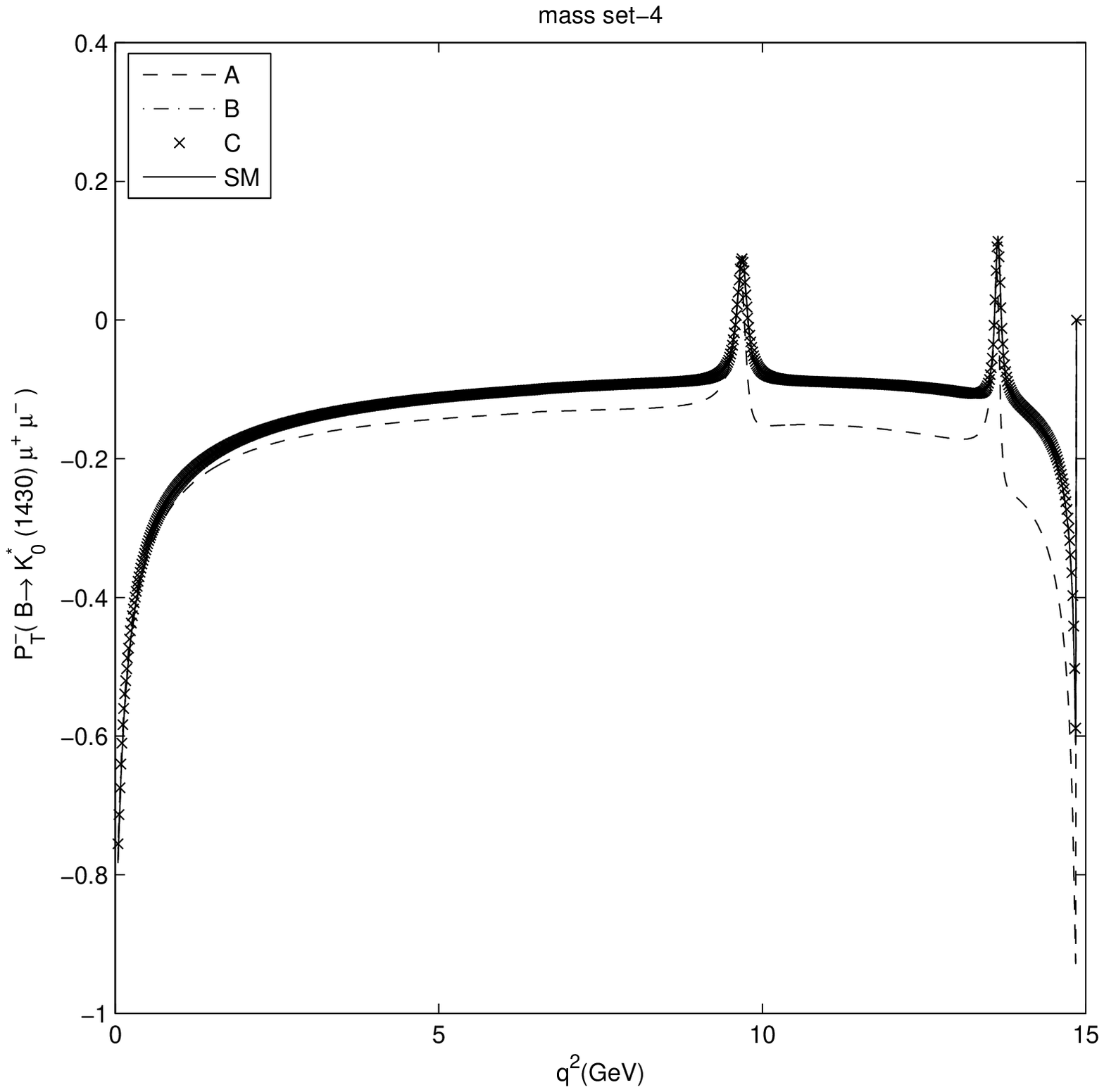}
               \caption{The dependence of the $ {\cal P}_{T}^{-}$ polarization  on $q^2$  and the three typical cases of 2HDM, i.e.
               cases A, B and C and SM  for  the $\mu$  channel of  $\overline{B}\to\overline{K}_0^{*}$ transition for the  mass sets 1, 2, 3  and 4. } \label{PTmmKstar}
    \end{figure}
    \begin{figure}[ht]
  \centering
  \setlength{\fboxrule}{2pt}
        \centering
                     \includegraphics[height=1.7in]{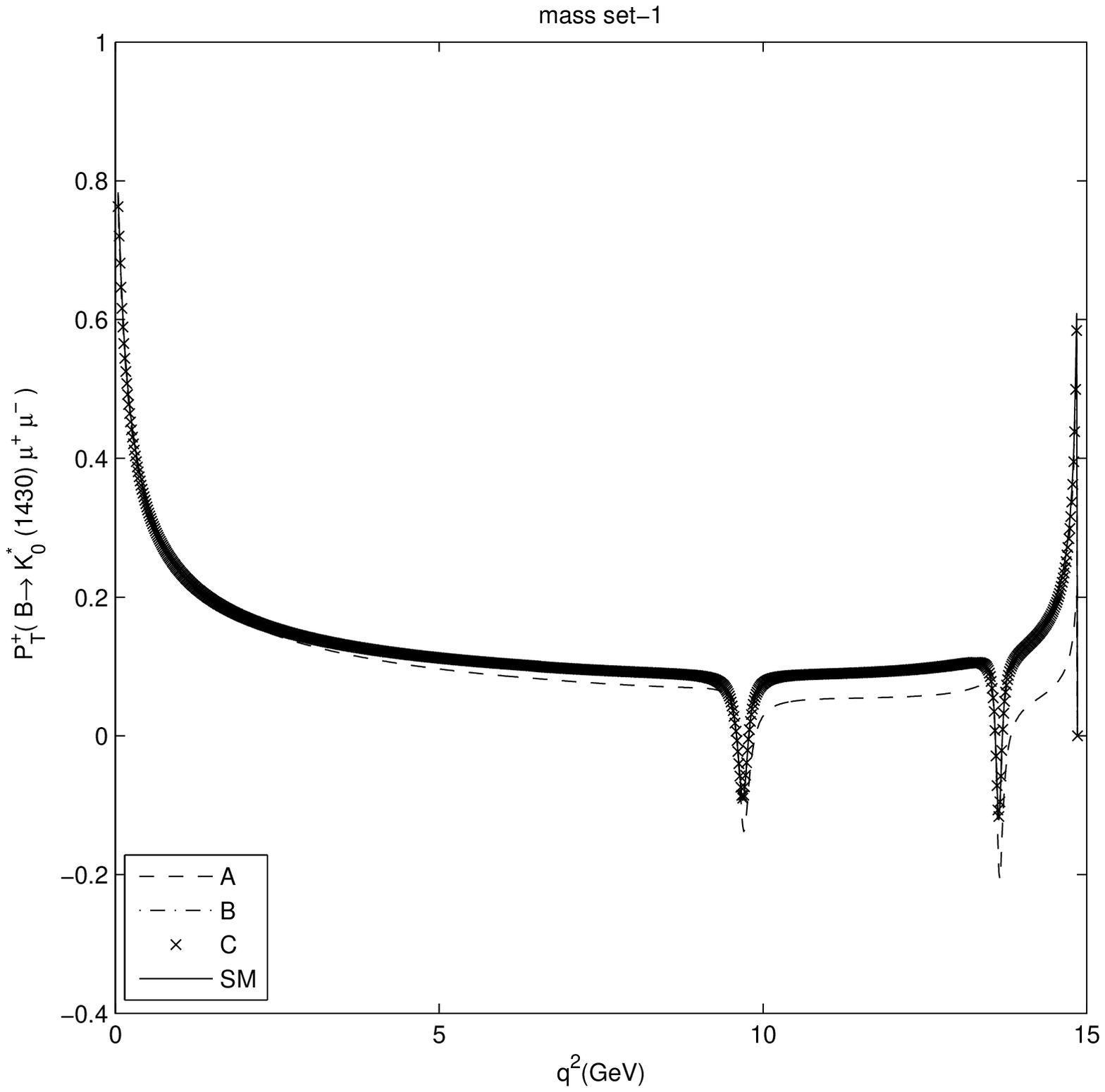}~
             \includegraphics[height=1.7in]{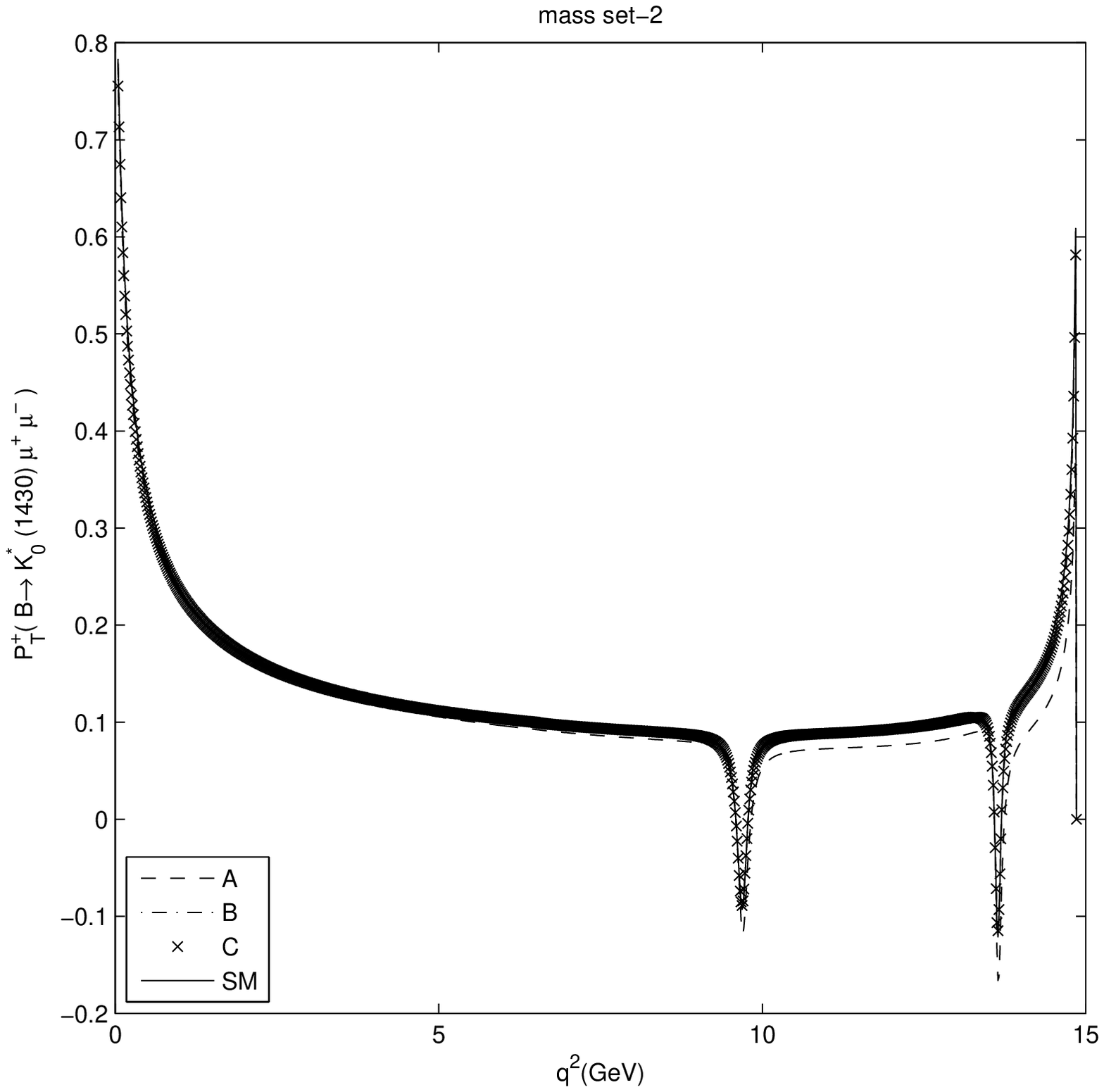}
             \includegraphics[height=1.7in]{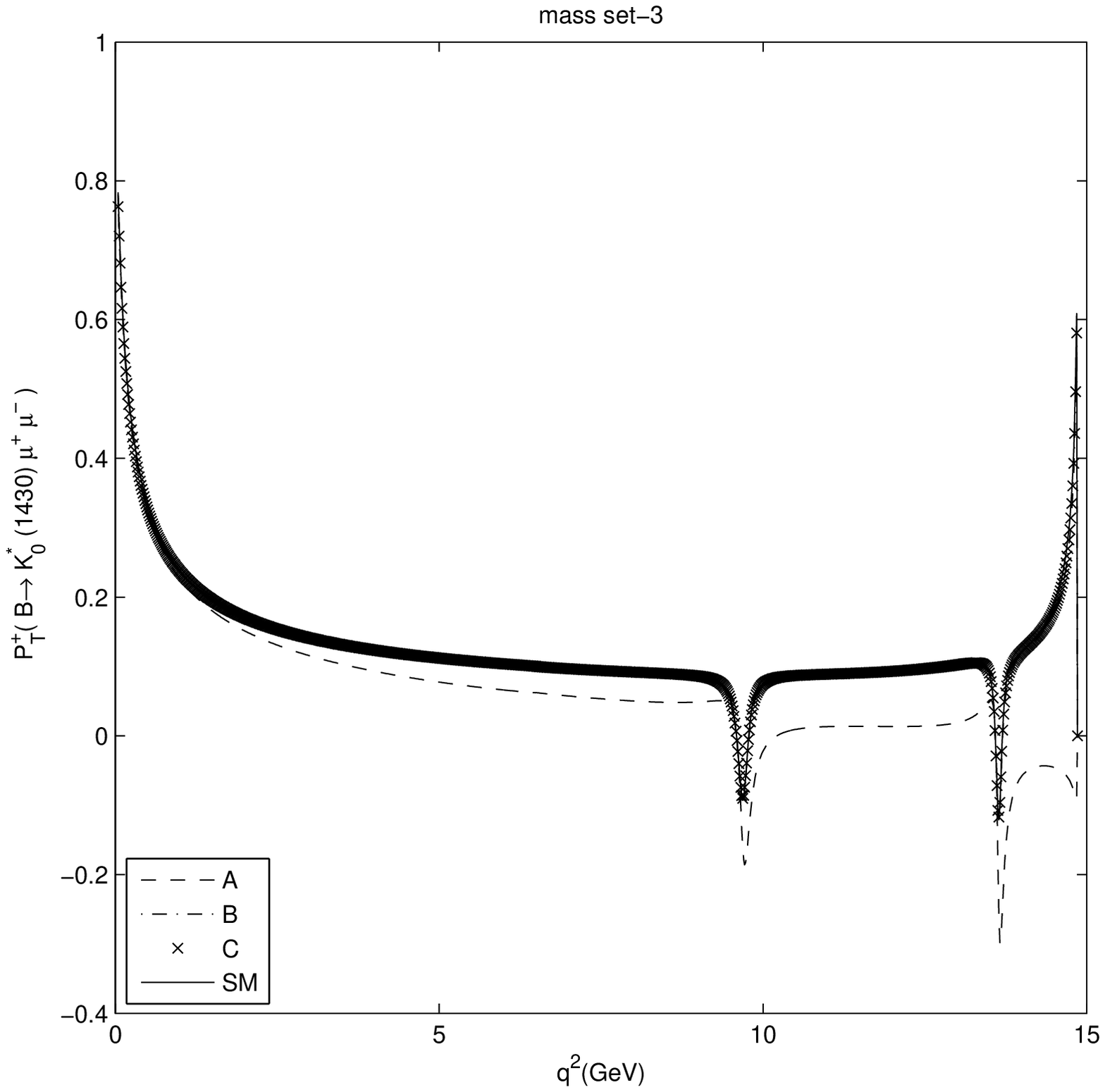}~
             \includegraphics[height=1.7in]{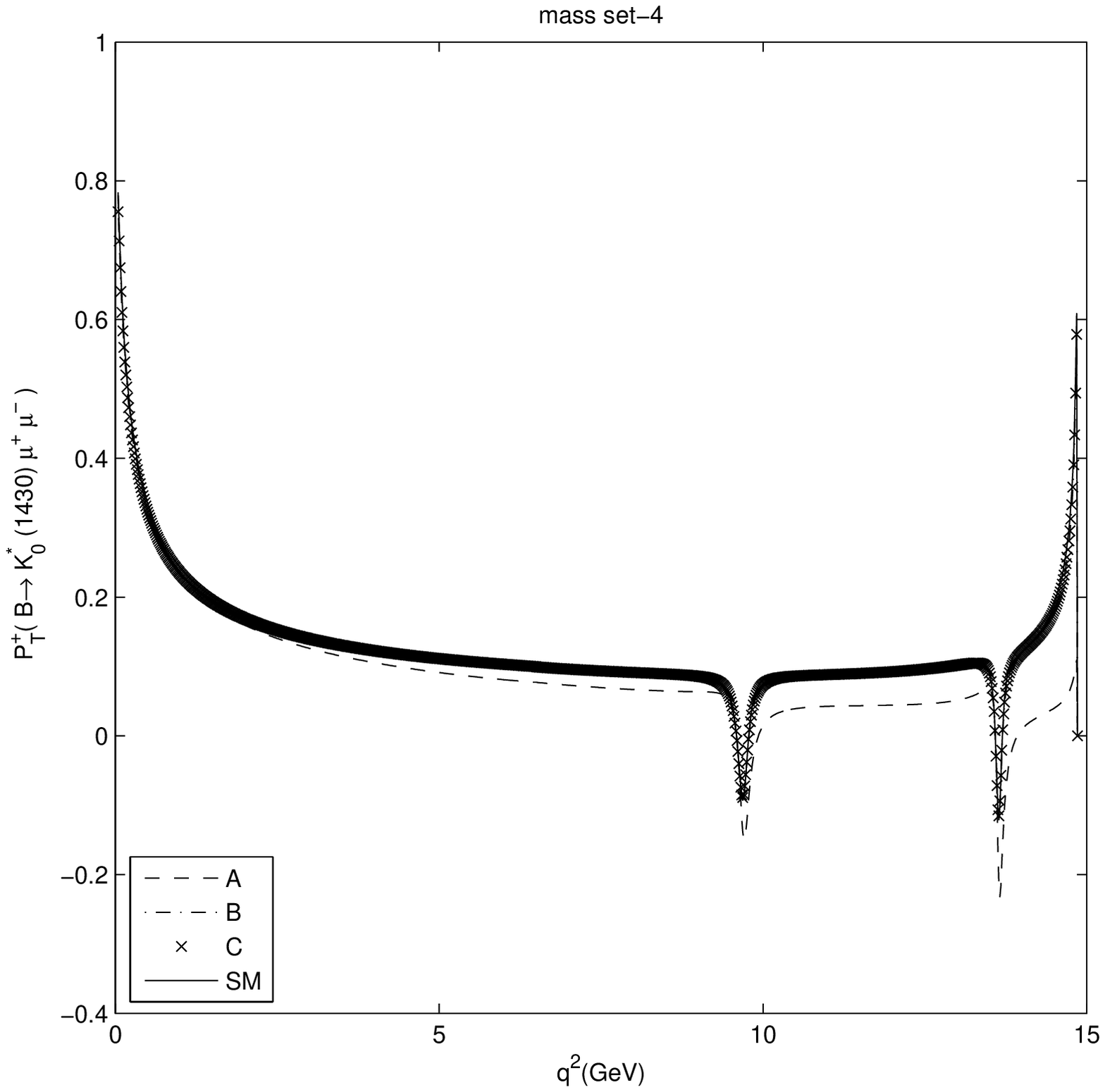}
               \caption{The dependence of the $ {\cal P}_{T}^{+}$ polarization  on $q^2$  and the three typical cases of 2HDM, i.e.
               cases A, B and C and SM  for  the $\mu$  channel of  $\overline{B}\to\overline{K}_0^{*}$ transition for the  mass sets 1, 2, 3  and 4. } \label{PTpmKstar}
    \end{figure}
    \begin{figure}[ht]
  \centering
  \setlength{\fboxrule}{2pt}
        \centering
                     \includegraphics[height=1.7in]{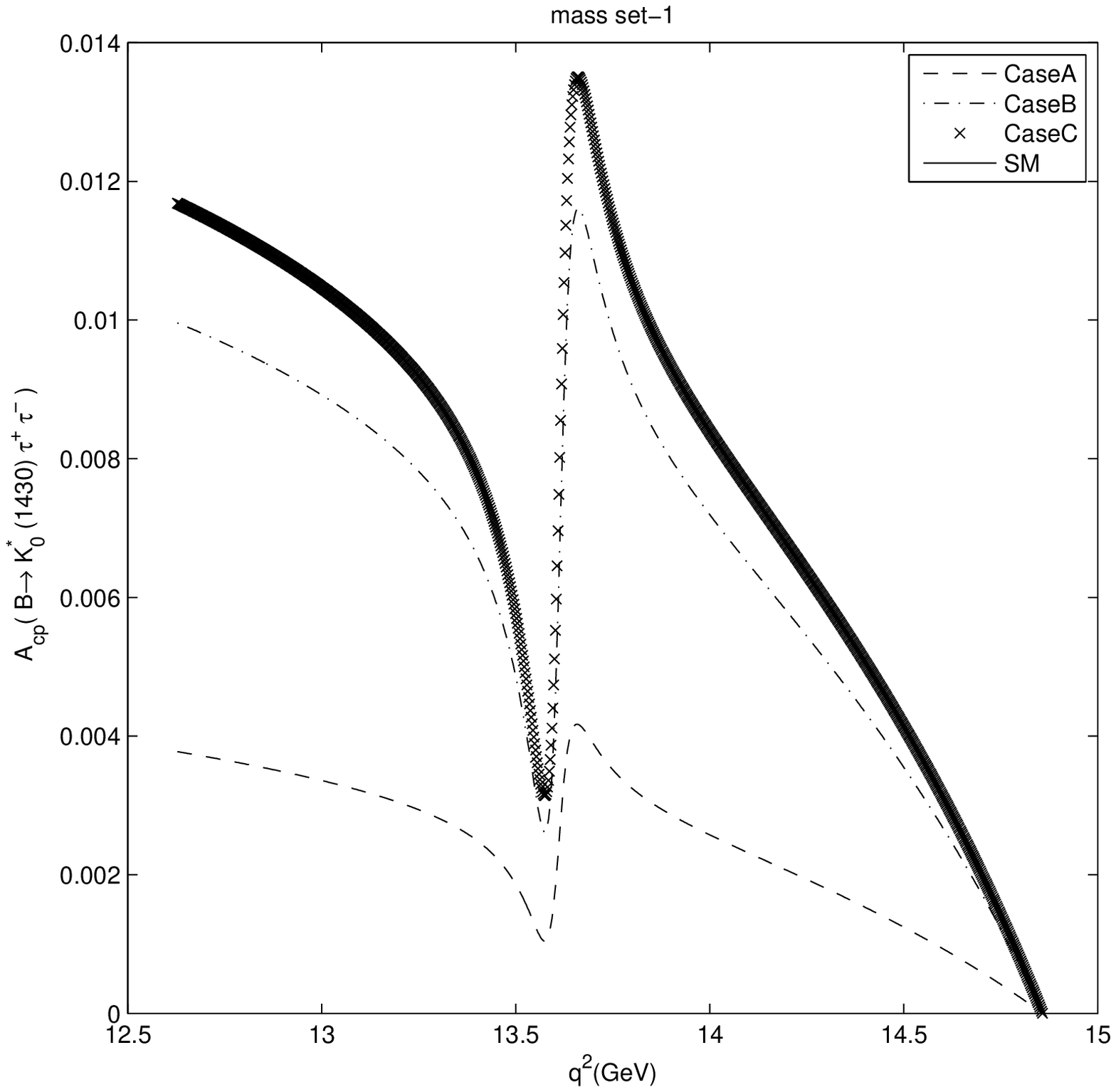}~
             \includegraphics[height=1.7in]{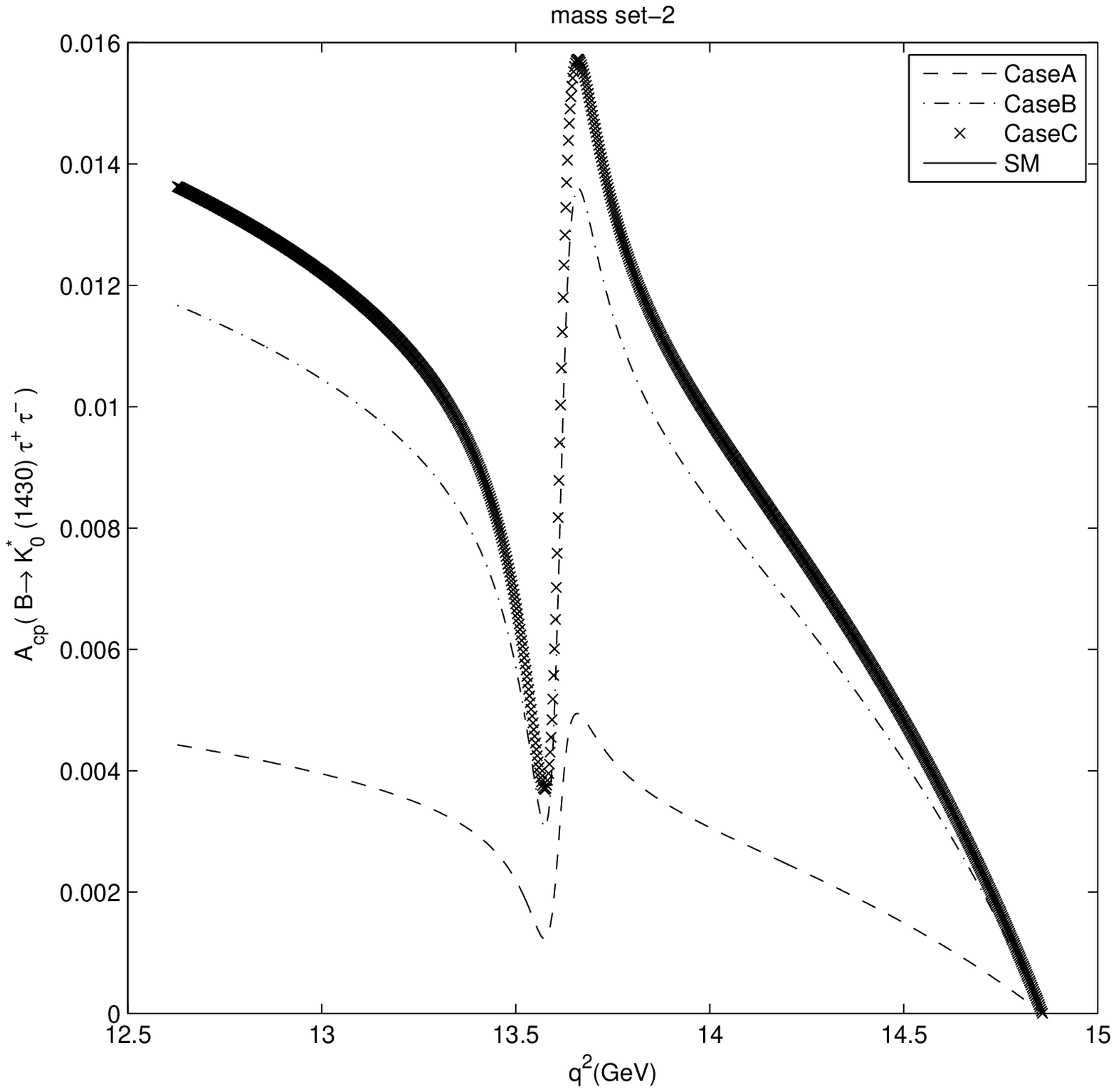}
             \includegraphics[height=1.7in]{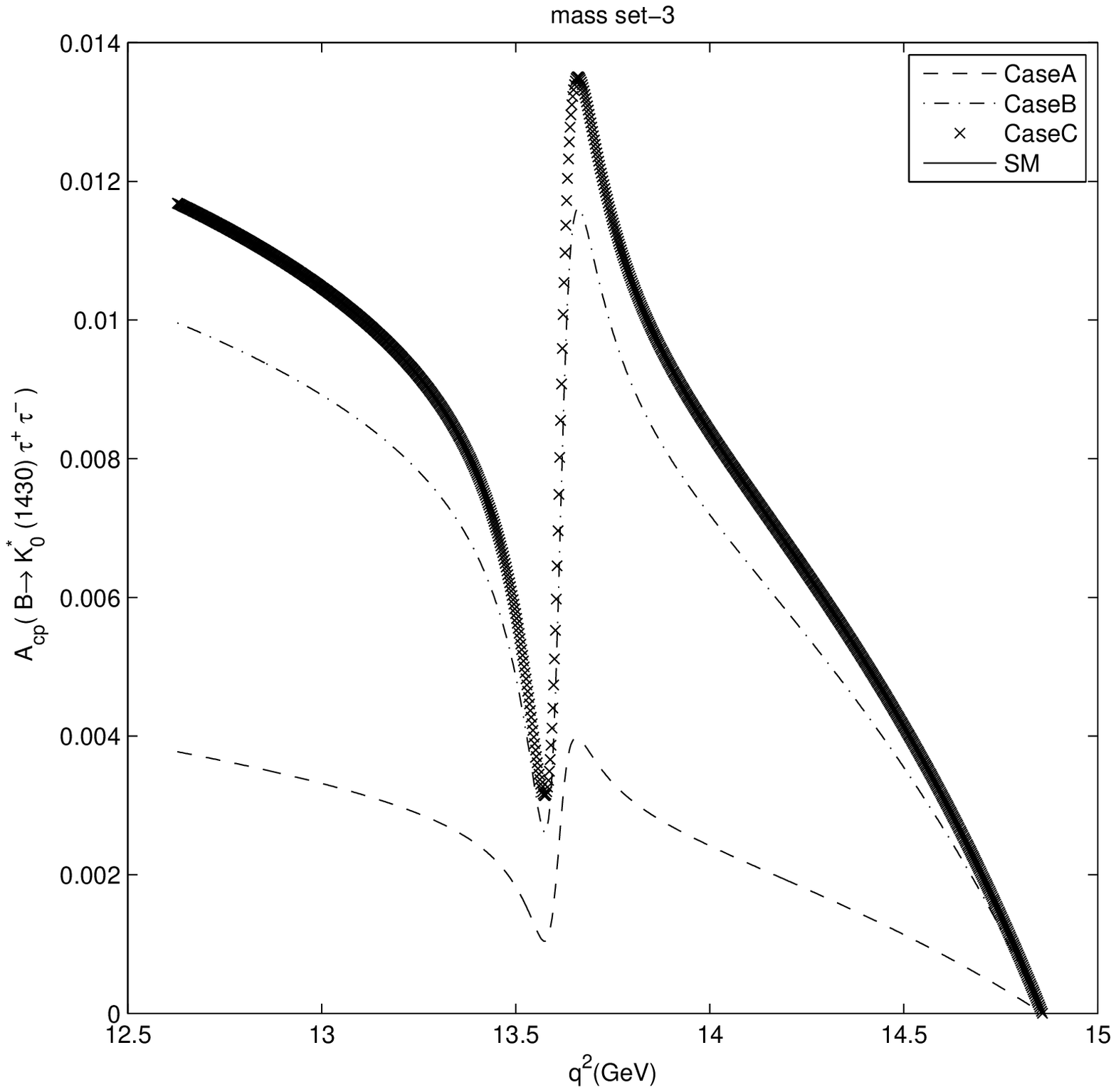}~
             \includegraphics[height=1.7in]{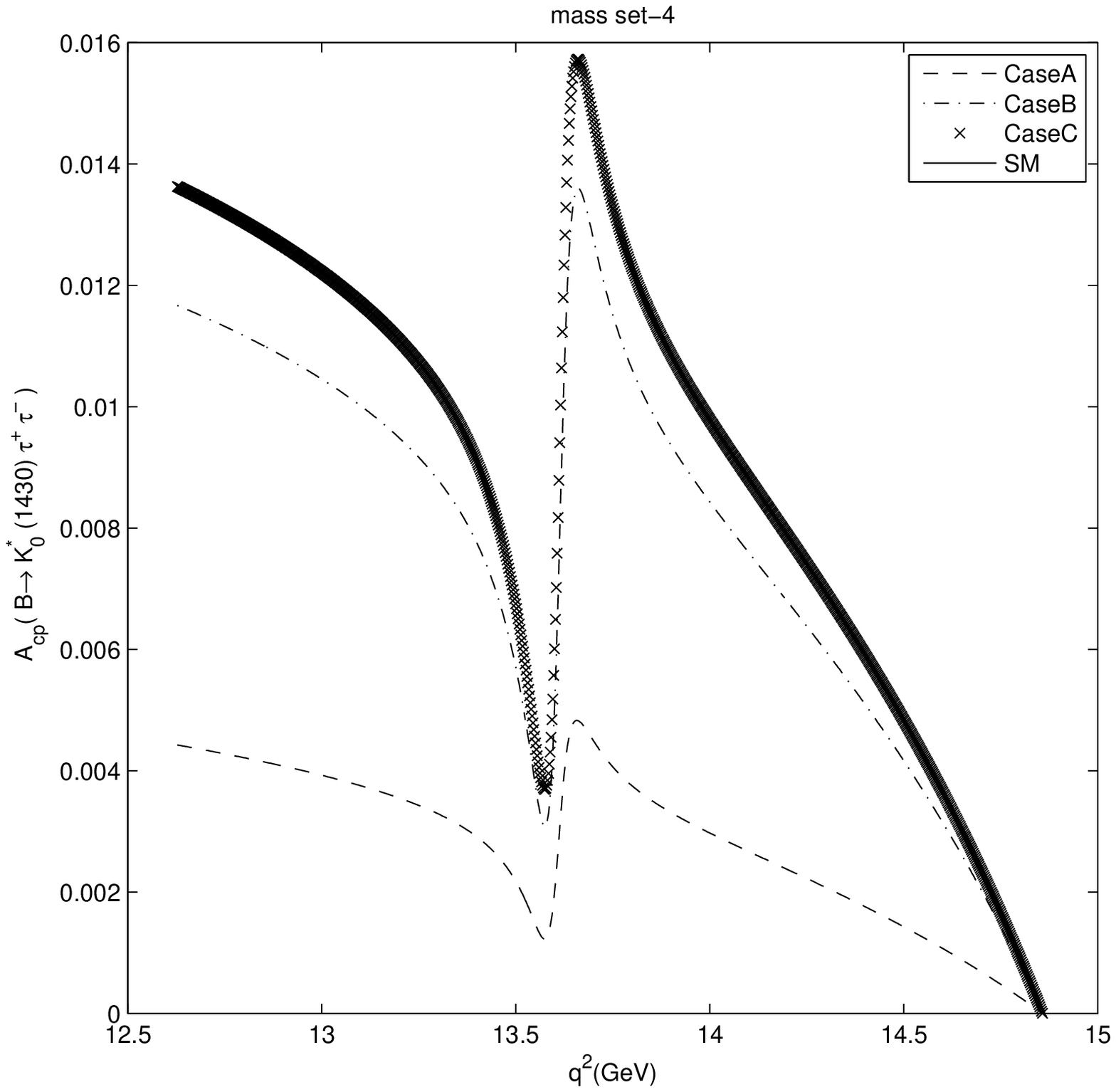}
               \caption{The dependence of the $ {\cal A}_{CP}$ polarization  on $q^2$  and the three typical cases of 2HDM, i.e.
               cases A, B and C and SM  for  the $\tau$  channel of  $\overline{B}\to\overline{K}_0^{*}$ transition for the  mass sets 1, 2, 3  and 4. } \label{ACPtKstar}
    \end{figure}
    \begin{figure}[ht]
  \centering
  \setlength{\fboxrule}{2pt}
        \centering
                     \includegraphics[height=1.7in]{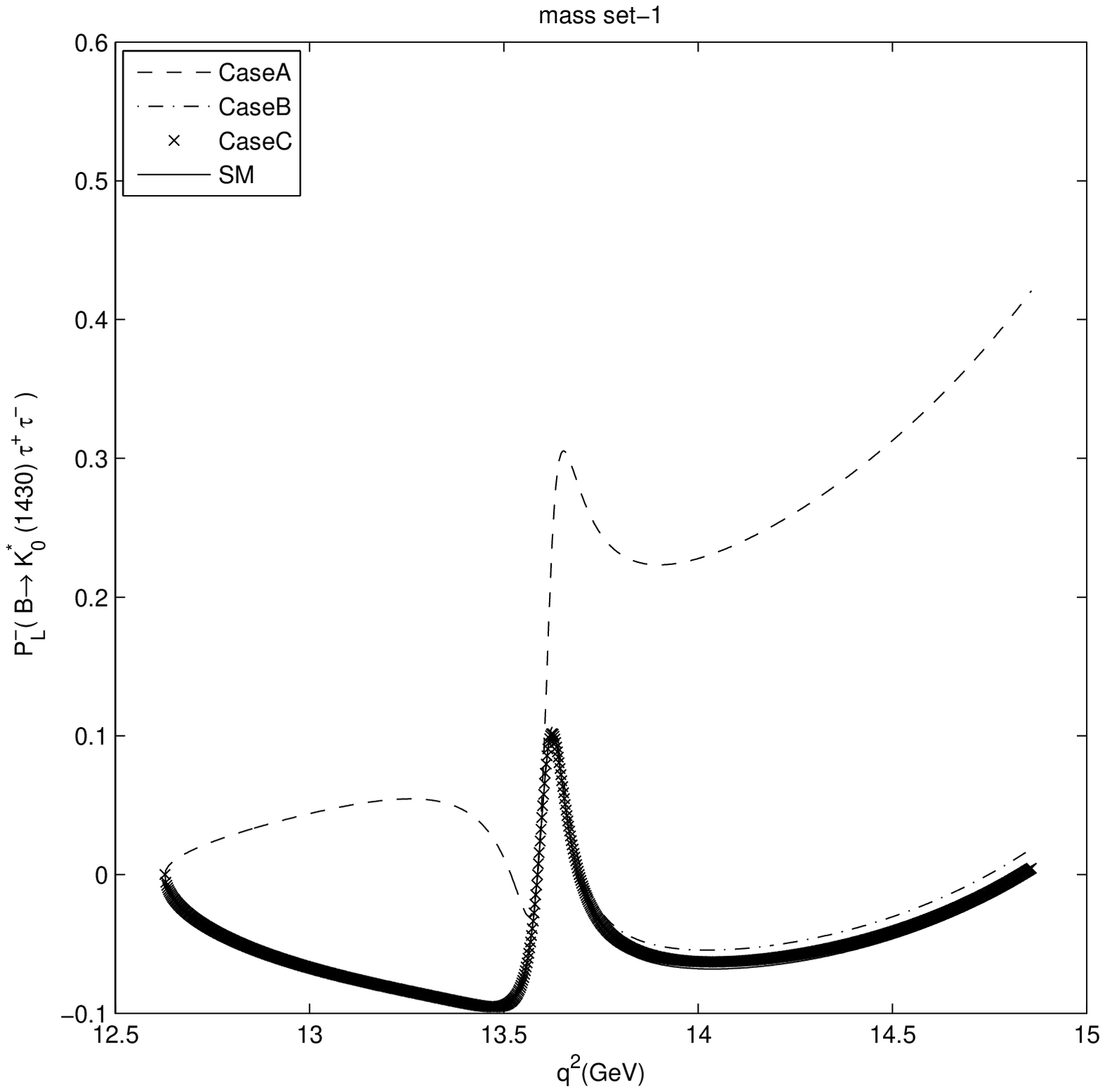}~
             \includegraphics[height=1.7in]{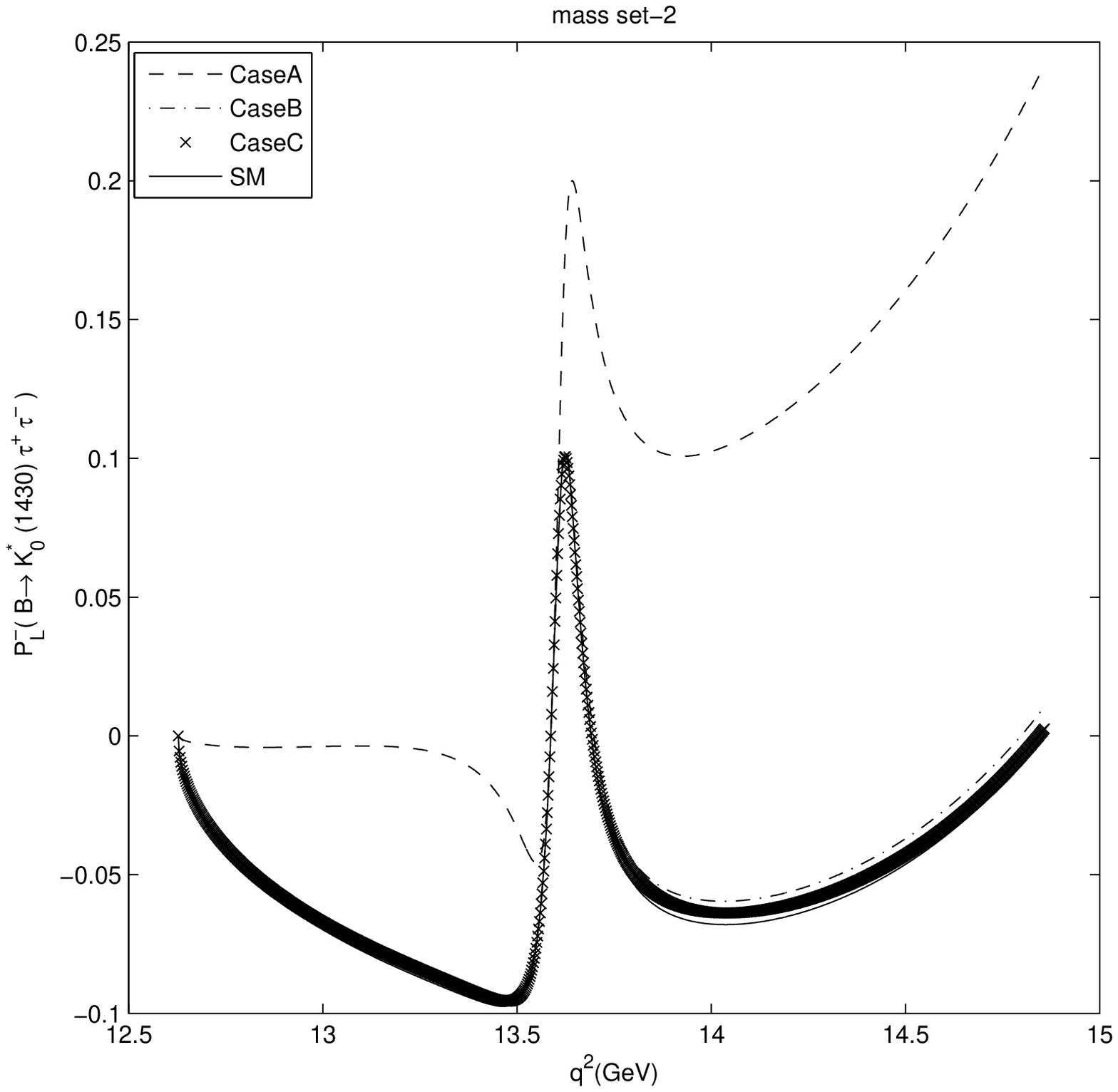}
             \includegraphics[height=1.7in]{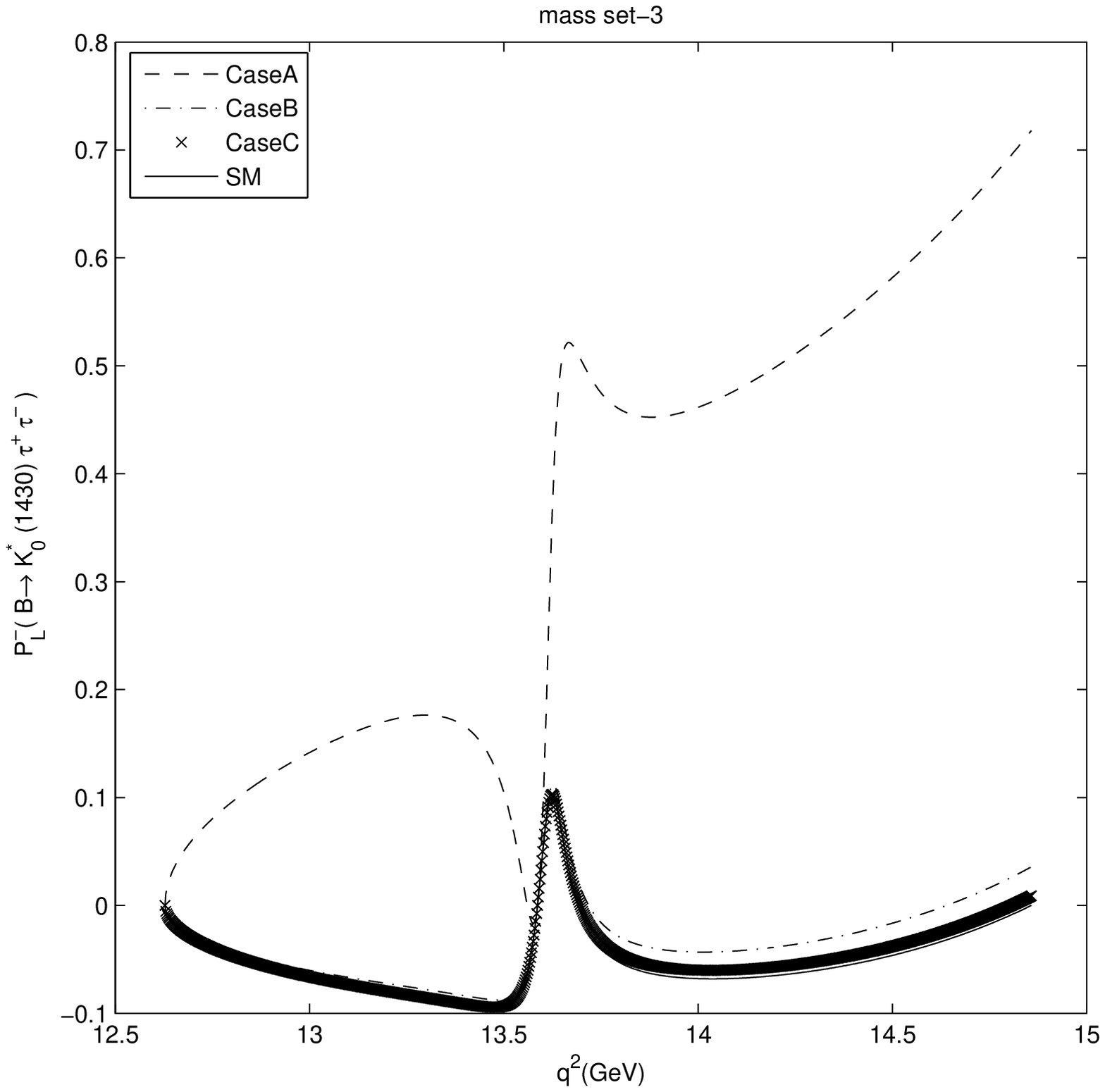}~
             \includegraphics[height=1.7in]{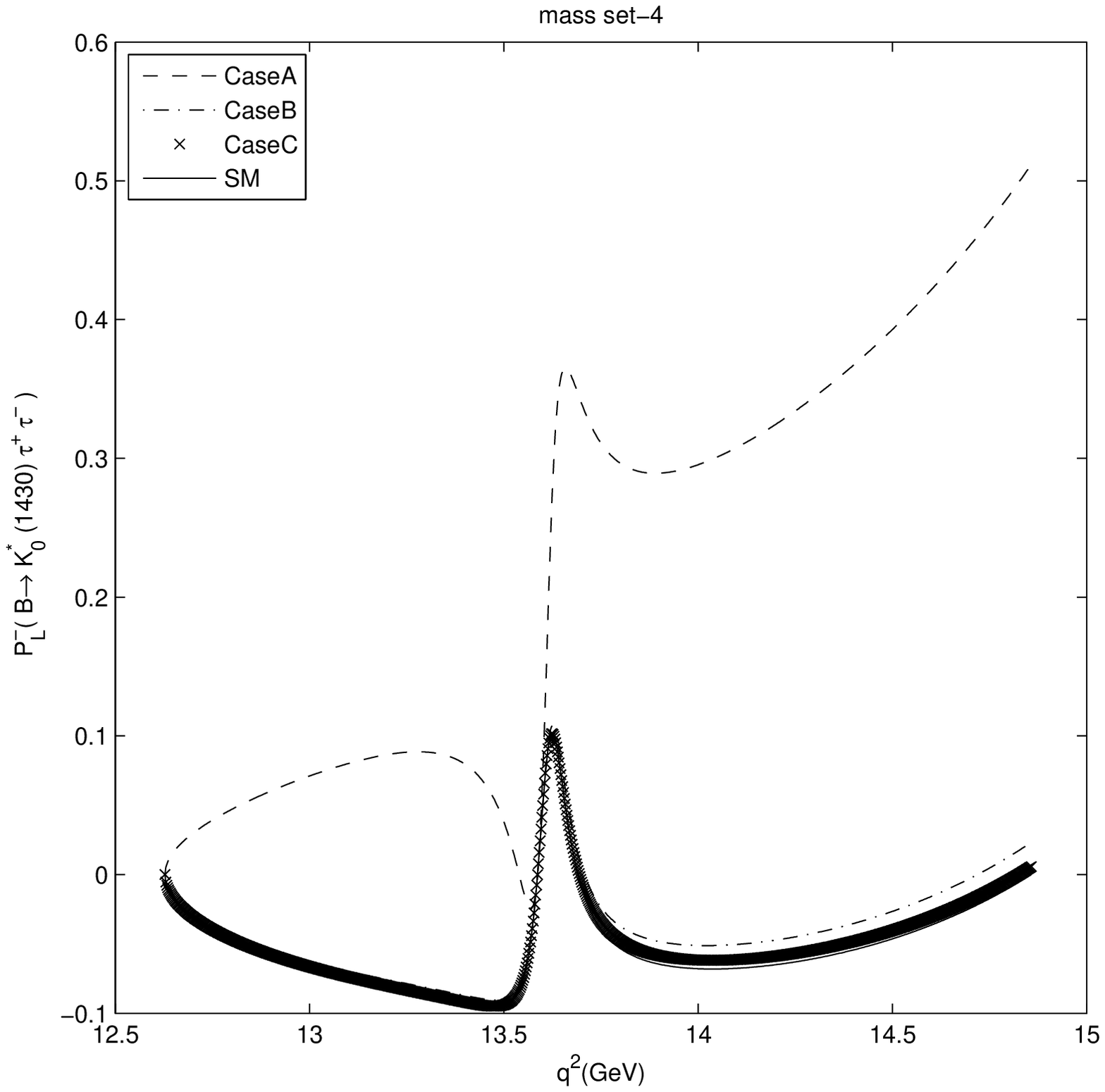}
               \caption{The dependence of the $ {\cal P}_{L}^{-}$ polarization  on $q^2$  and the three typical cases of 2HDM, i.e.
               cases A, B and C and SM  for  the $\tau$  channel of  $\overline{B}\to\overline{K}_0^{*}$ transition for the  mass sets 1, 2, 3  and 4. } \label{PLmtKstar}
    \end{figure}
    \begin{figure}[ht]
  \centering
  \setlength{\fboxrule}{2pt}
        \centering
                     \includegraphics[height=1.7in]{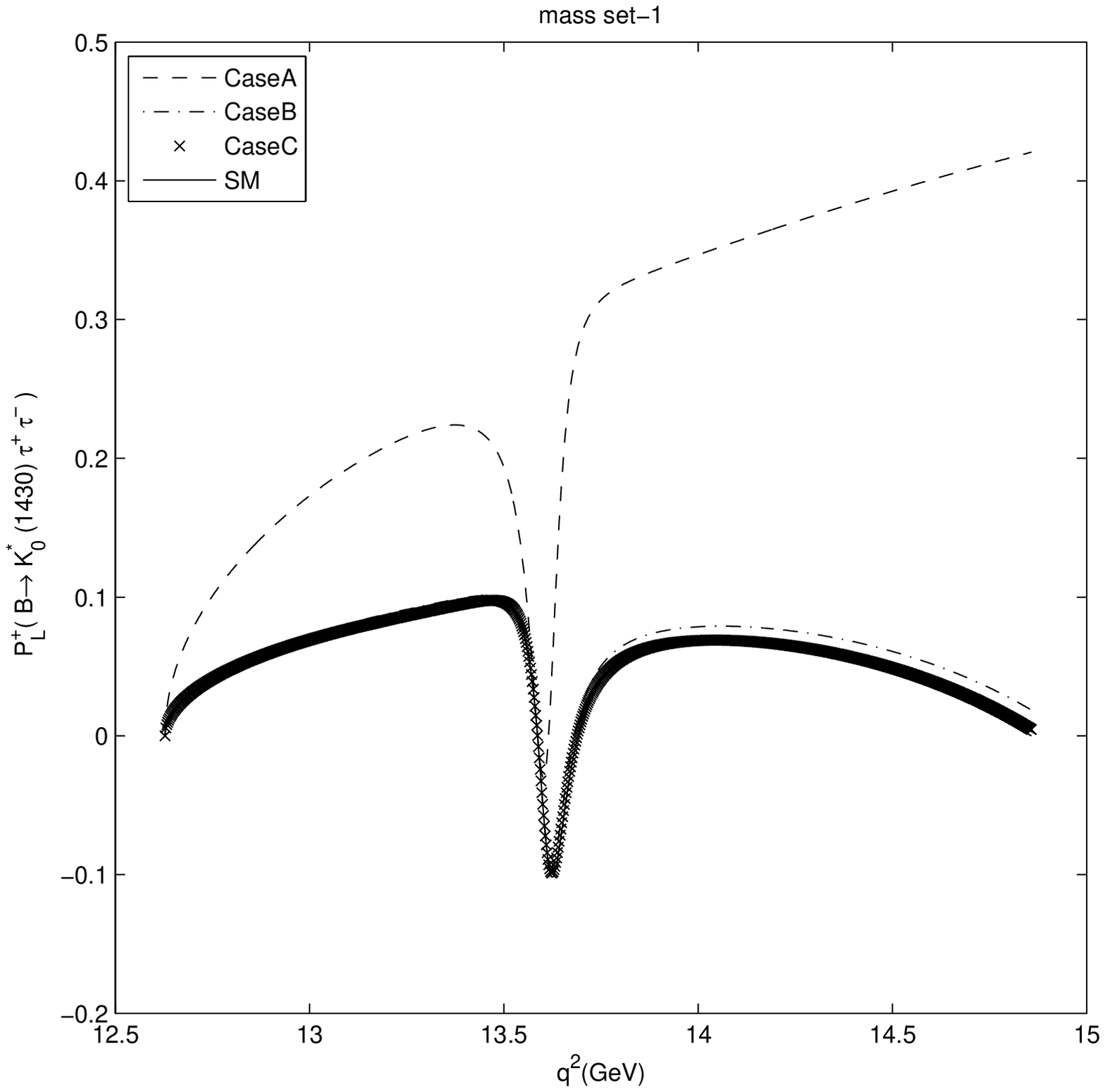}~
             \includegraphics[height=1.7in]{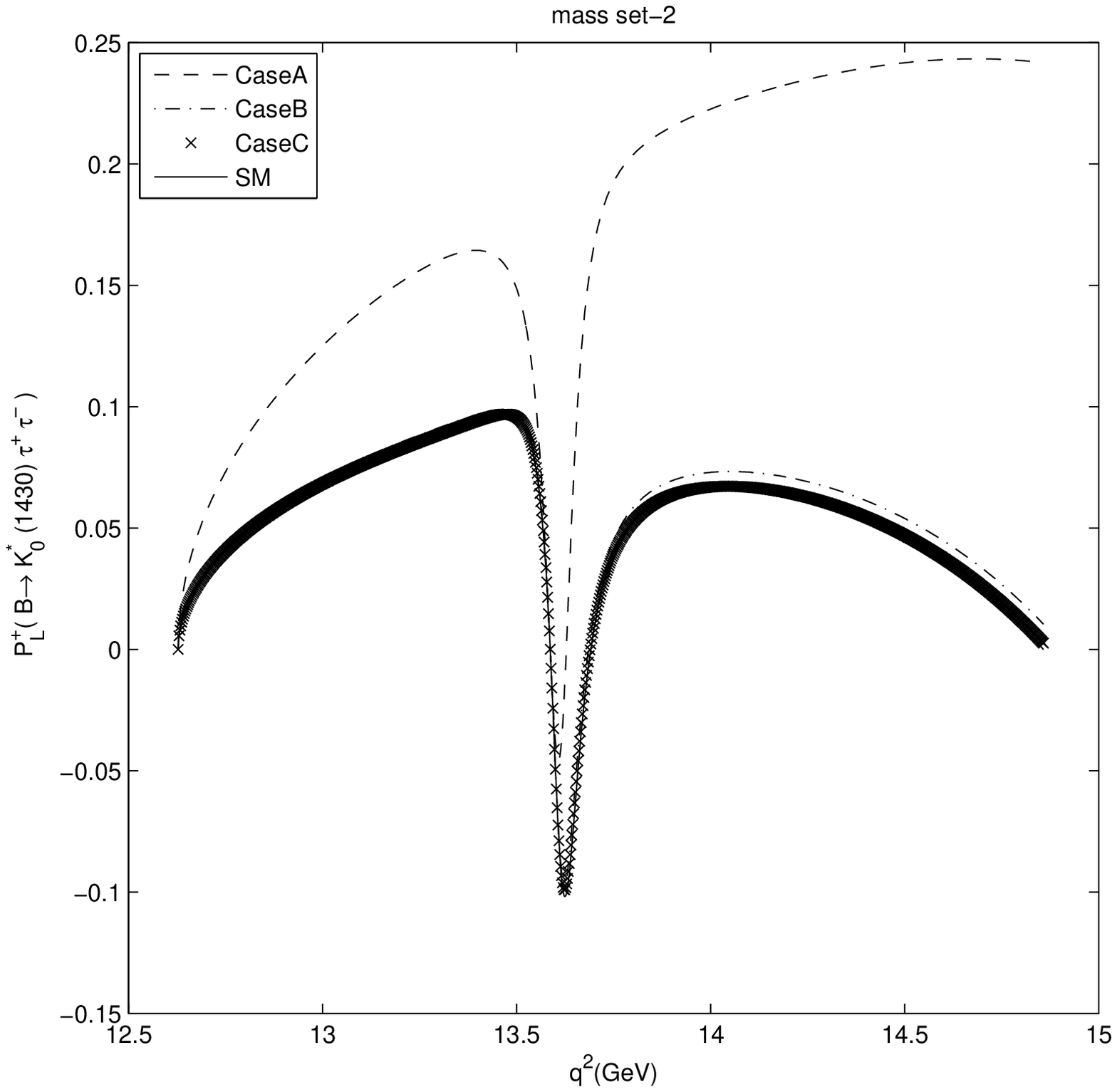}
             \includegraphics[height=1.7in]{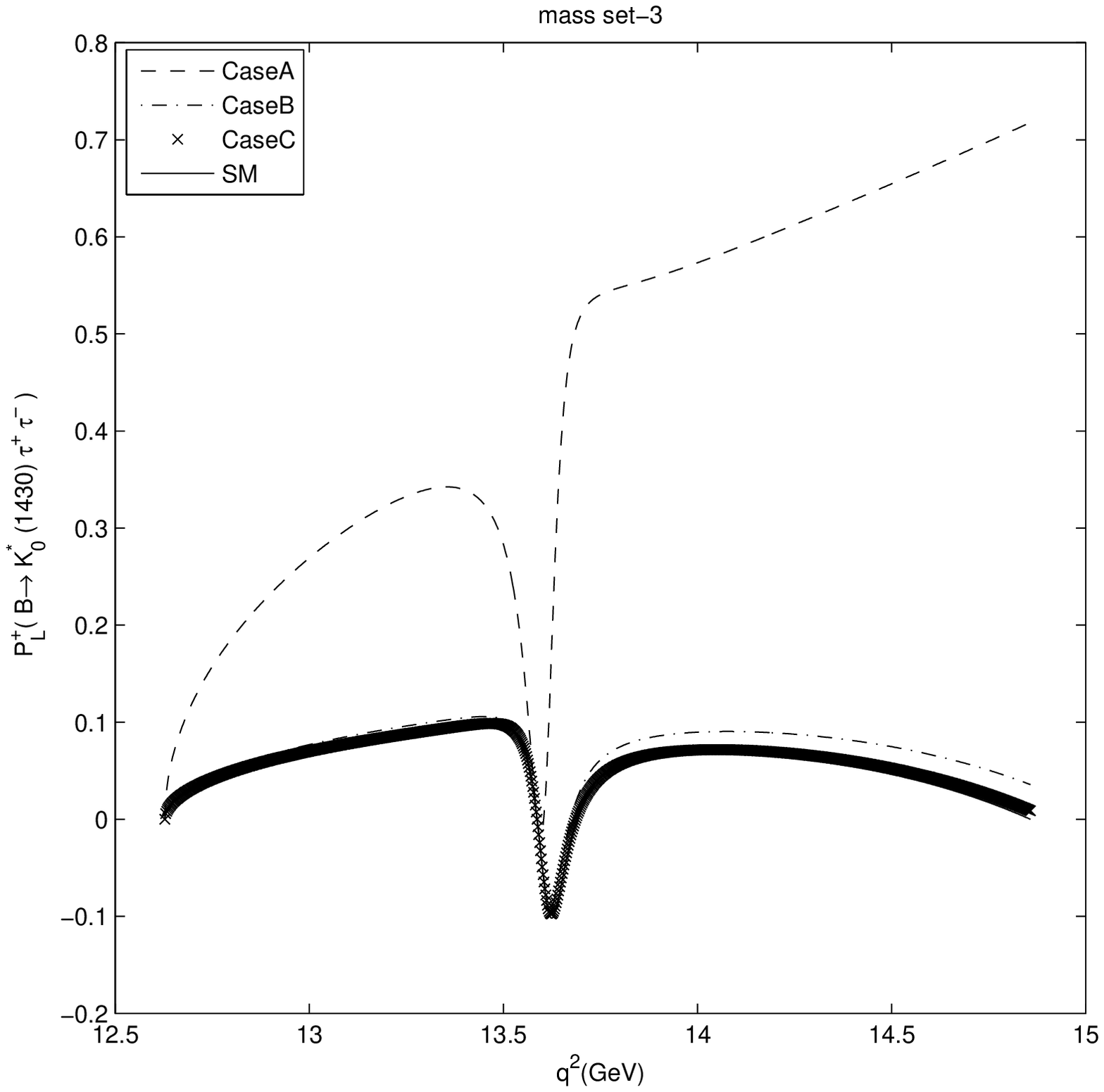}~
             \includegraphics[height=1.7in]{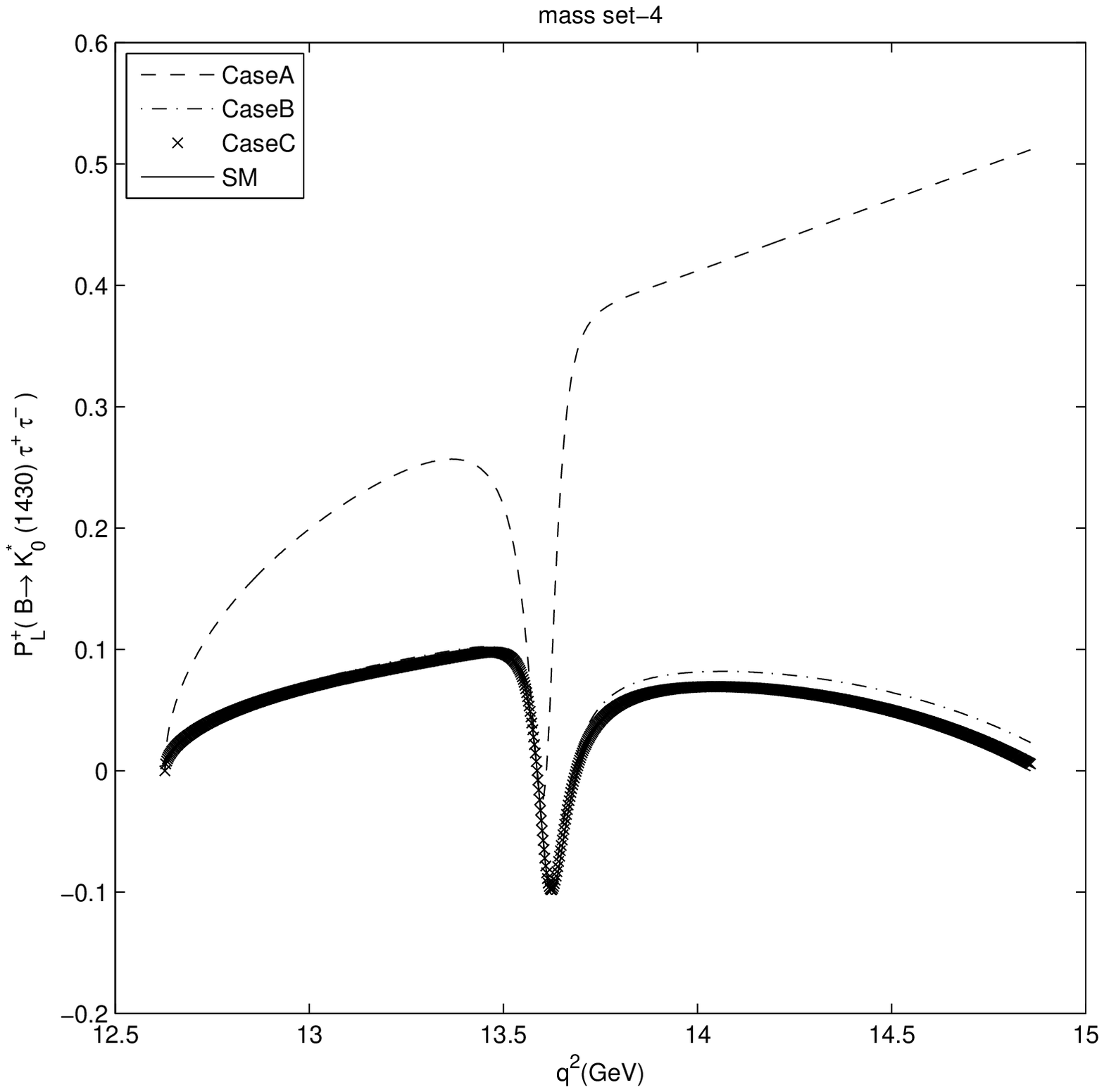}
               \caption{The dependence of the $ {\cal P}_{L}^{+}$ polarization  on $q^2$  and the three typical cases of 2HDM, i.e.
               cases A, B and C and SM  for  the $\tau$  channel of  $\overline{B}\to\overline{K}_0^{*}$ transition for the  mass sets 1, 2, 3  and 4. } \label{PLptKstar}
    \end{figure}
     \begin{figure}[ht]
  \centering
  \setlength{\fboxrule}{2pt}
        \centering
                     \includegraphics[height=1.7in]{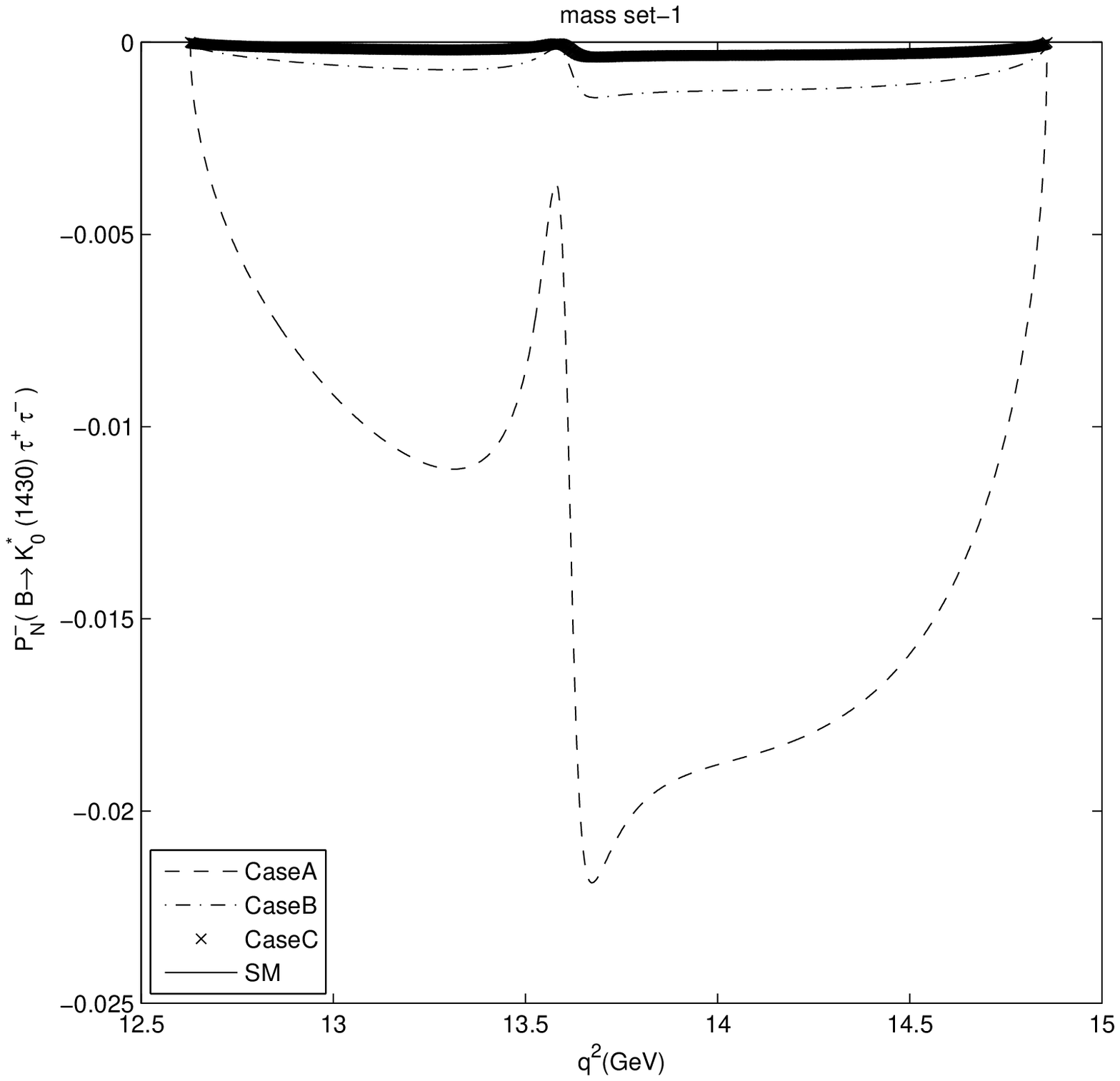}~
             \includegraphics[height=1.7in]{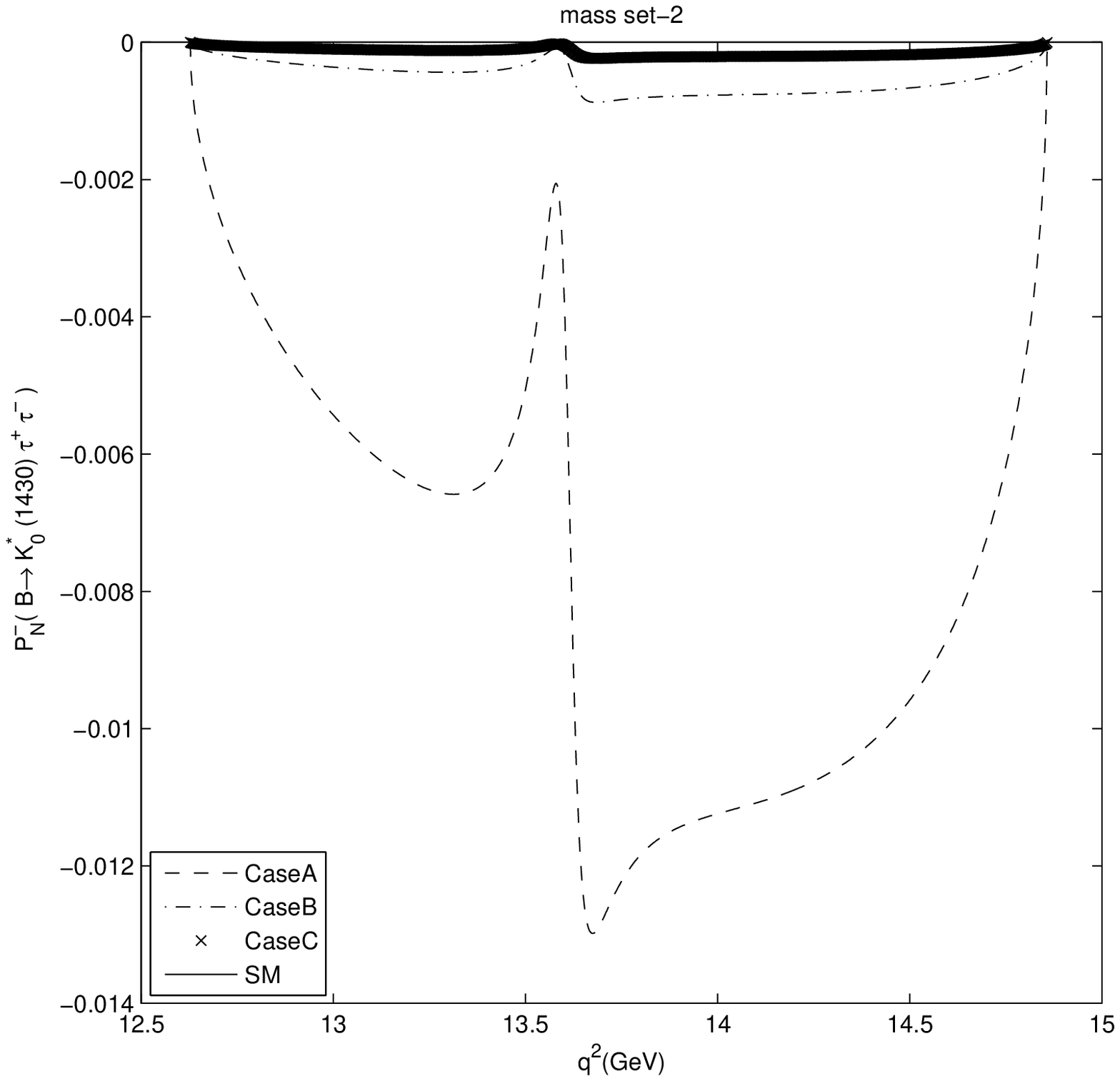}
             \includegraphics[height=1.7in]{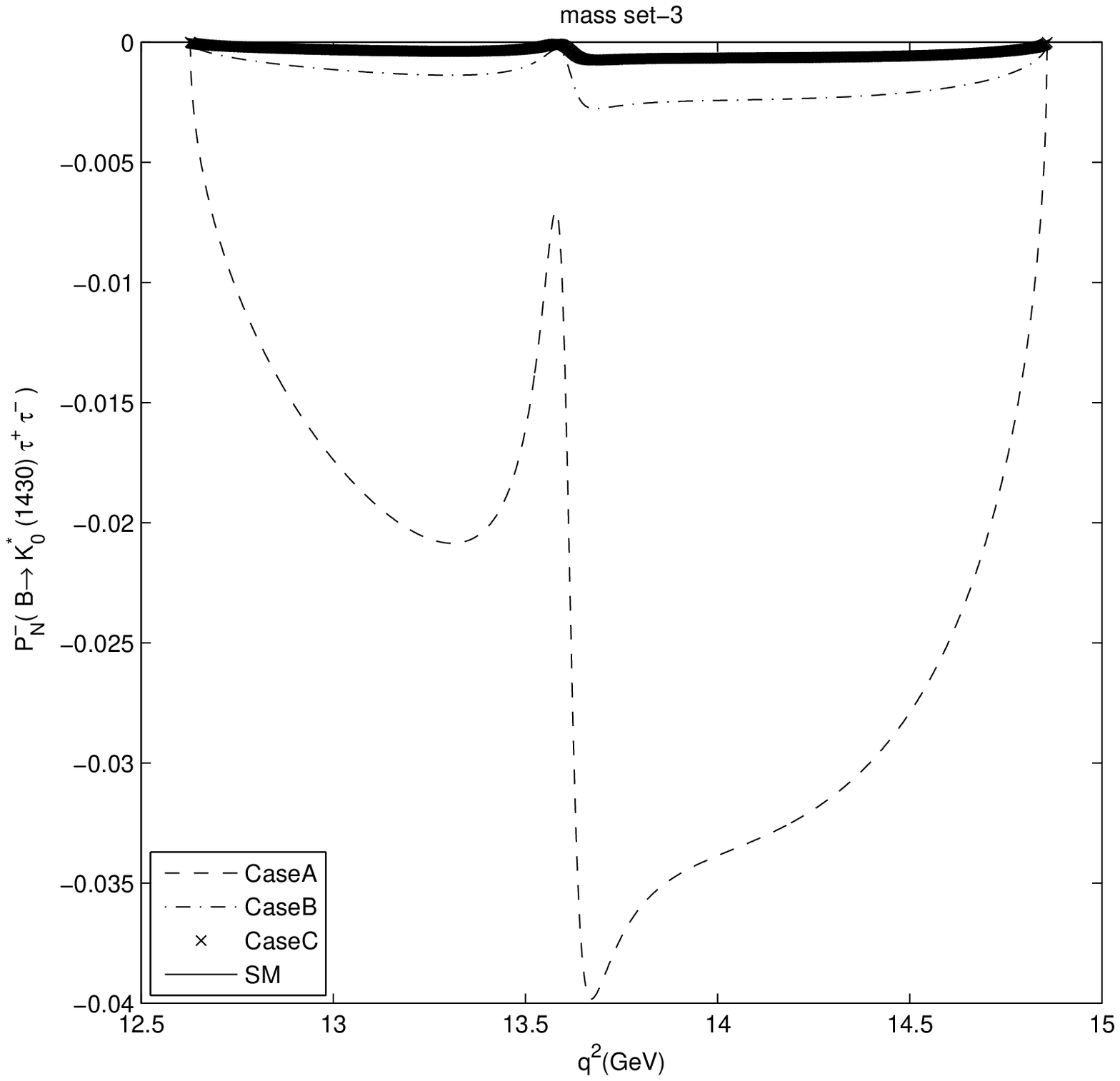}~
             \includegraphics[height=1.7in]{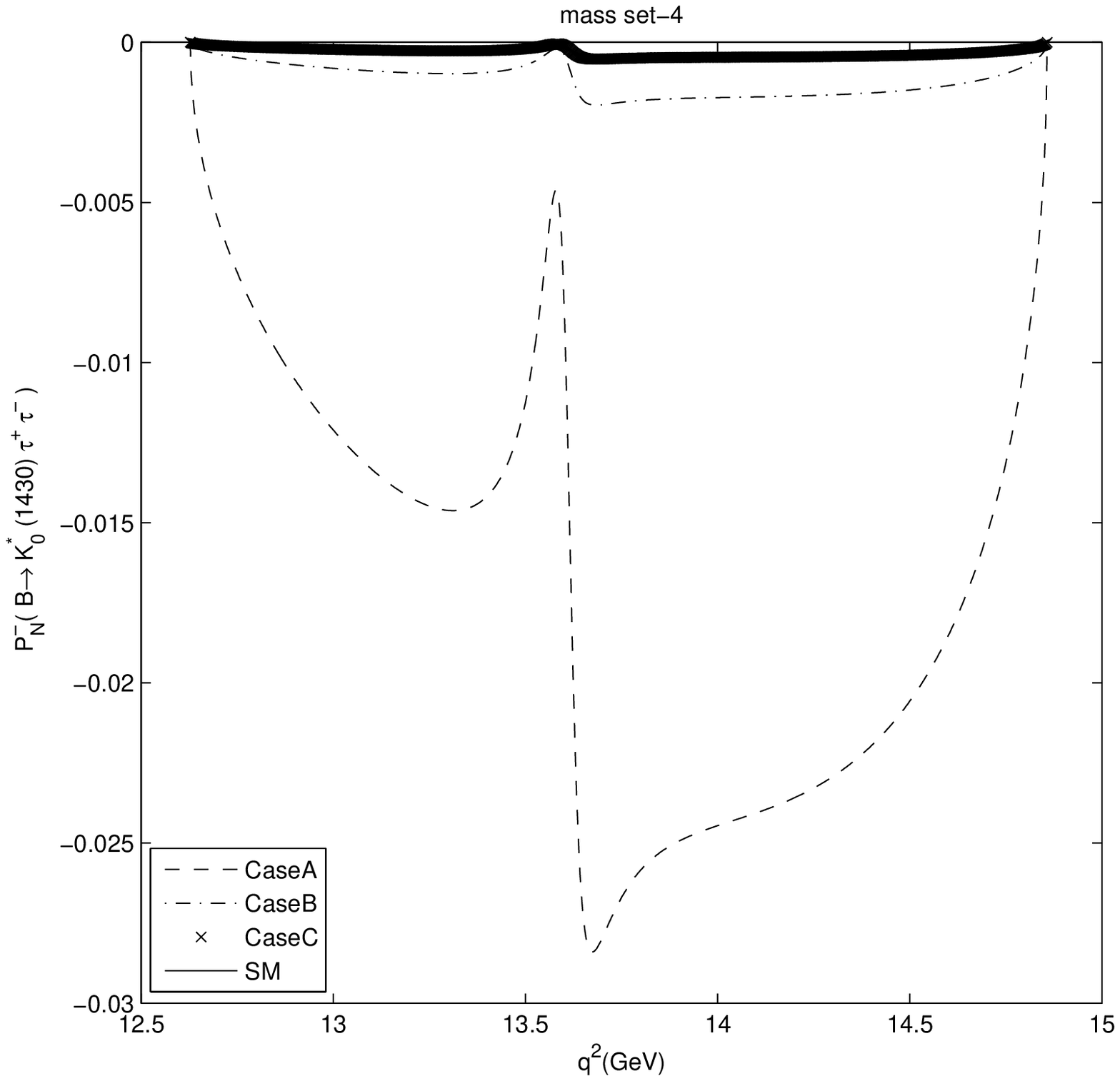}
               \caption{The dependence of the $ {\cal P}_{N}^{-}$ polarization  on $q^2$  and the three typical cases of 2HDM, i.e.
               cases A, B and C and SM  for  the $\tau$  channel of  $\overline{B}\to\overline{K}_0^{*}$ transition for the  mass sets 1, 2, 3  and 4. } \label{PNmtKstar}
    \end{figure}
    \begin{figure}[ht]
  \centering
  \setlength{\fboxrule}{2pt}
        \centering
                     \includegraphics[height=1.7in]{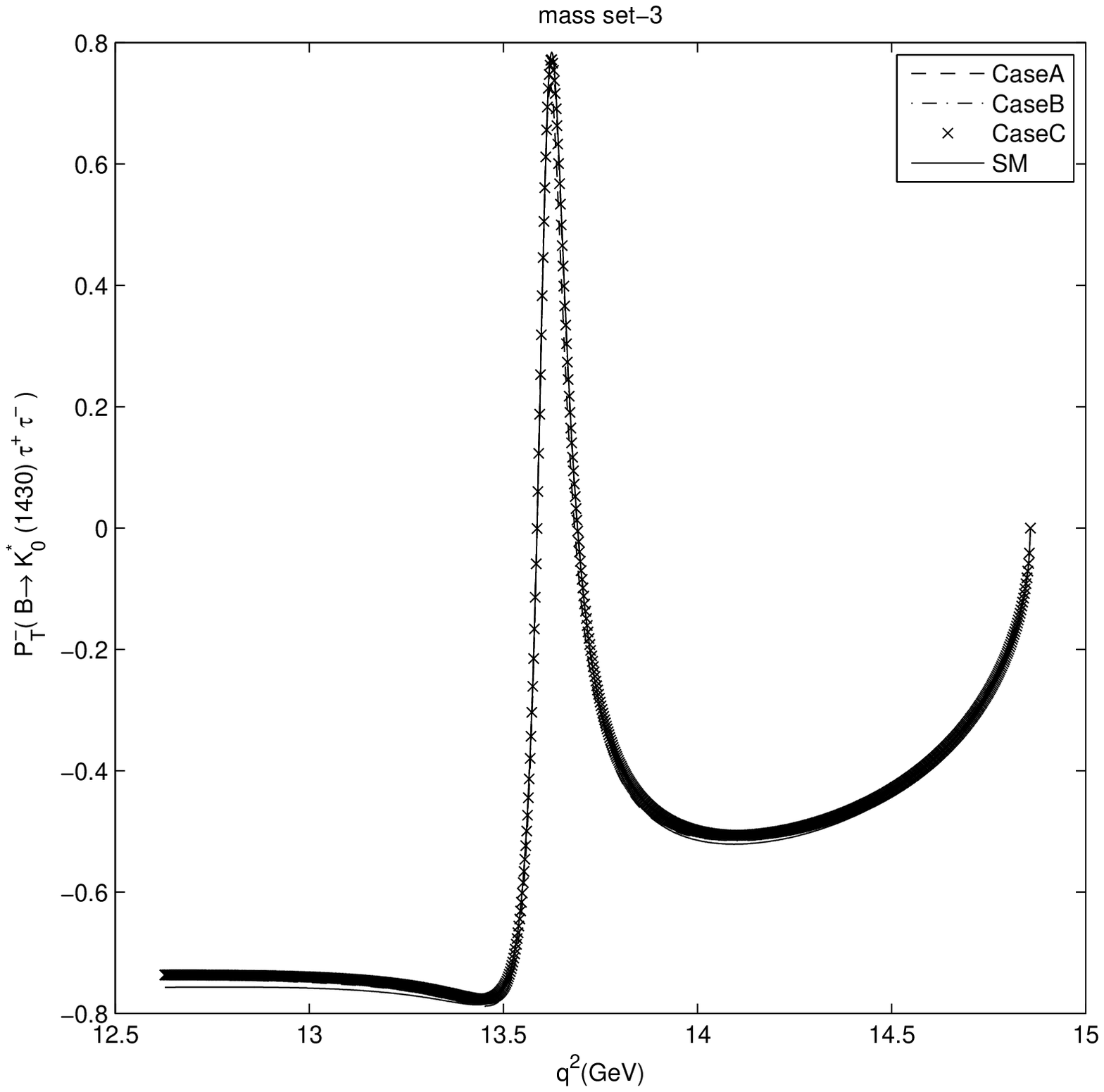}~
             \includegraphics[height=1.7in]{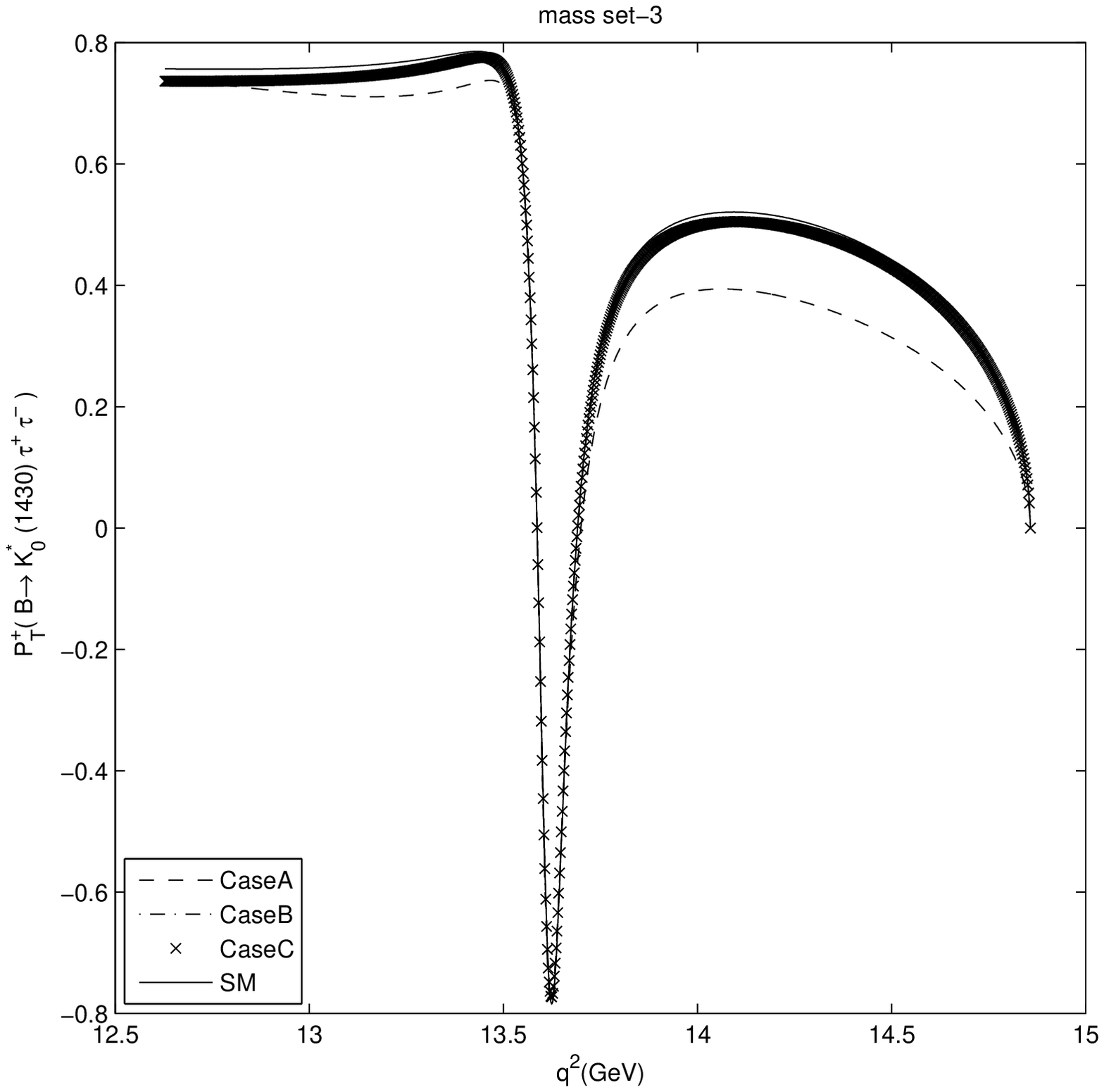}
               \caption{The dependence of the $ {\cal P}_{T}^{\mp}$ polarizations  on $q^2$  and the three typical cases of 2HDM, i.e.
               cases A, B and C and SM  for  the $\tau$  channel of  $\overline{B}\to\overline{K}_0^{*}$ transition for the  mass set 3. } \label{PTmptKstar}
    \end{figure}
    \end{document}